%% file: ms.tex
\documentclass[twocolumn,times]{aastex61}

\usepackage{comment}
\usepackage{graphicx}

\received{May 11, 2017}
\revised{November 14, 2017}
\accepted{November 26, 2017}
\submitjournal{AJ}

\shorttitle{885$\mu$m ALMA Survey of Stellar/Substellar Protoplanetary Disk Masses}
\shortauthors{Ward-Duong et al.}

\begin{document}

\title{The Taurus Boundary of Stellar/Substellar (TBOSS) Survey II. \\ Disk Masses from ALMA Continuum Observations}


\correspondingauthor{K. Ward-Duong}
\email{kwardduo@asu.edu}

\author[0000-0002-4479-8291]{K. Ward-Duong}
\affiliation{School of Earth and Space Exploration, Arizona State University, P.O. Box 871404, Tempe, AZ 85287, USA}

\author{J. Patience}
\affiliation{School of Earth and Space Exploration, Arizona State University, P.O. Box 871404, Tempe, AZ 85287, USA}

\author[0000-0003-4641-2003]{J. Bulger}
\affiliation{Subaru Telescope, NAOJ, 650 N. A‘ohoku Pl., Hilo, HI 96720, USA}
\affiliation{Institute for Astronomy Maui, University of Hawaii, 34 Ohia Ku St., Pukalani, HI, 96768, USA}

\author[0000-0001-5688-187X]{G. van der Plas}
\affiliation{DAS, Universidad de Chile, Santiago, Chile}
\affiliation{Millennium Nucleus, ``Protoplanetary Disks'', El Observatorio 1515, Las Condes, Santiago, Chile}
\affiliation{Univ. Grenoble Alpes, Institut de Plan\'{e}tologie et d`Astrophysique de Grenoble (IPAG, UMR 5274), 38000 Grenoble, France}

\author{F. M\'{e}nard}
\affiliation{Univ. Grenoble Alpes, Institut de Plan\'{e}tologie et d`Astrophysique de Grenoble (IPAG, UMR 5274), 38000 Grenoble, France}

\author[0000-0001-5907-5179]{C. Pinte}
\affiliation{Univ. Grenoble Alpes, Institut de Plan\'{e}tologie et d`Astrophysique de Grenoble (IPAG, UMR 5274), 38000 Grenoble, France}

\author[0000-0003-4393-9520]{A. P. Jackson}
\affiliation{Centre for Planetary Sciences, University of Toronto, 1265 Military Trail, Toronto, Ontario, M1C 1A4, Canada}
\affiliation{School of Earth and Space Exploration, Arizona State University, P.O. Box 871404, Tempe, AZ 85287, USA}

\author{G. Bryden}
\affiliation{Jet Propulsion Laboratory, California Institute of Technology, 4800 Oak Grove Drive, Pasadena, CA 91109-8099, USA}

\author[0000-0001-8292-1943]{N. J. Turner}
\affiliation{Jet Propulsion Laboratory, California Institute of Technology, 4800 Oak Grove Drive, Pasadena, CA 91109-8099, USA}

\author{P. Harvey}
\affiliation{Univ. of Texas at Austin, Austin, TX 78712, USA}

\author[0000-0001-5073-2849]{A. Hales}
\affiliation{Atacama Large Millimeter/Submillimeter Array, Joint ALMA Observatory, Alonso de C\'{o}rdova 3107, Vitacura 763-0355, Santiago, Chile}
\affiliation{National Radio Astronomy Observatory, 520 Edgemont Road, Charlottesville, Virginia, 22903-2475, USA}

\author[0000-0002-4918-0247]{R. J. De Rosa}
\affiliation{Astronomy Department, University of California, Berkeley, CA 94720, USA}

\begin{abstract}

We report 885$\mu$m ALMA continuum flux densities for 24 Taurus members spanning the stellar/substellar boundary, with spectral types from M4 to M7.75. Of the 24 systems, 22 are detected at levels ranging from 1.0--55.6 mJy. The two non-detections are transition disks, though other transition disks in the sample are detected. Converting ALMA continuum measurements to masses using standard scaling laws and radiative transfer modeling yields dust mass estimates ranging from $\sim$0.3--20M$_{\oplus}$. The dust mass shows a declining trend with central object mass when combined with results from submillimeter surveys of more massive Taurus members. The substellar disks appear as part of a continuous sequence and not a distinct population. Compared to older Upper Sco members with similar masses across the substellar limit, the Taurus disks are brighter and more massive. Both Taurus and Upper Sco populations are consistent with an approximately linear relationship in $M_{dust}$ to $M_{star}$, although derived power-law slopes depend strongly upon choices of stellar evolutionary model and dust temperature relation. The median disk around early M-stars in Taurus contains a comparable amount of mass in small solids as the average amount of heavy elements in \textit{Kepler} planetary systems on short-period orbits around M-dwarf stars, with an order of magnitude spread in disk dust mass about the median value. Assuming a gas:dust ratio of 100:1, only a small number of low-mass stars and brown dwarfs have a total disk mass amenable to giant planet formation, consistent with the low frequency of giant planets orbiting M-dwarfs.
\end{abstract}



\section{Introduction} \label{sec:intro}
Submillimeter and millimeter wavelength observations of protoplanetary disks provide views into the disk structure, composition, evolution, and dust grain properties within the nascent environments of planet formation \citep[see, e.g.,][]{andrewswilliams05, andrews07b, birnstiel10, ricci10b}. Given assumptions regarding disk temperature and spatial extent, and grain properties (e.g., opacity, emissivity and size distribution), measurements of sub-mm/mm disk flux density can be translated into dust masses of grains with sizes similar to the observation wavelength \citep{beckwith90}. 

By studying the properties of protoplanetary disks in star-forming regions with known ages, it is possible to use the abundance of dust and gas content within disks to trace disk evolution pathways and timescales. However, this is complicated by the dominant mode and scale of star formation, such as the environmental impacts of high-mass stellar populations, as within the Orion Molecular Cloud (OMC), or relatively quiescent low-mass environments, like the Taurus star-forming region. Measurements of disk evolution timescales and natal environments refine our understanding of formation mechanisms, and provide context for the history of the solar system, for which the meteoritic record and isotopic evidence offer important benchmarks on planetesimal growth timescales and indications of the Sun's formation environment \citep[cf.][]{macpherson95, russell06}. 

Previous surveys have examined stars with $M_{*} > 0.1M_{\odot}$ in a number of diverse star-forming regions, including: Taurus \citep{andrewswilliams05, andrews13}, IC348 \citep{lee11}, Upper Sco \citep{mathews12, carpenter14, gvdp16, barenfeld16}, Lupus \citep{ansdell16}, sigma Orionis \citep{ansdell17}, Chamaeleon~I \citep{pascucci16}, and Orion \citep{williams13, eisner16}. In particular, great emphasis has been placed on the Taurus star-forming region given its proximity ($\sim$140~pc) and canonically young age \citep[$\sim$1-2 Myr, although an older sub-population may extend up to 20 Myr;][]{daemgen15}, which enable detailed studies of its stellar population. Surveys of Taurus have demonstrated a correlation of increasing disk mass with stellar mass \citep{andrewswilliams05, andrews13}, suggesting that the mass of the disks in the Class II Taurus population ranges from $\sim$0.2\%-0.6\% of the host mass. With comparisons to regions at the older age of Upper Sco, studies have also shown trends of decreased dust mass for the same stellar masses at later ages \citep{carpenter14, gvdp16, barenfeld16}, and at mid-infrared wavelengths, disk studies of the low-mass stellar population with \textit{Spitzer} revealed longer-lived excess emission for lower-mass stellar hosts \citep{carpenter06}.

With studies largely focusing on stars with masses $>0.1M_{\odot}$, key questions remain as to whether similar disk mass relations and depletion timescales hold for lower-mass stars and substellar objects. As the lowest-mass stars ultimately become the bulk of the stellar population by number -- with M-dwarfs comprising $\sim$75\% of the neighboring field population \citep{henry06, lepine05a} -- their disk properties represent what may be the most common pathways of planet formation. For the Taurus star-forming region that is the subject of this study, previous surveys \citep[e.g.,][]{andrews13} have provided high detection rates around Class II solar-mass stars, but few detections in the M-star range ($0.1-0.6M_{\odot}$), and M-star disk detections are limited to the brightest subset of disks. To probe the full population of disks around low-mass stars and brown dwarfs in Taurus extending below the upper envelope of disk continuum emission, more sensitive observations are required and are the subject of this study. Furthermore, extending disk measurements across the hydrogen-burning limit is of significant interest as relatively little is yet known about the planet populations of the lowest-mass stars and brown dwarfs. Recent transiting planet searches have revealed intriguing systems of low-mass planets orbiting M-dwarf hosts, including potentially temperate planets around Proxima Centauri \citep[M5.5V; 0.12$M_{\odot}$,][]{anglada-escude16} and LHS1140 \citep[M4.5V; 0.15$M_{\odot}$,][]{dittman17}, and the seven planet system of TRAPPIST-1, an ultracool dwarf residing at the stellar/brown dwarf boundary \citep[M8V; 0.08$M_{\odot}$][]{gillon17}. To provide context for planet-hosting low-mass stars, investigations into protoplanetary disk hosts as younger analogues to systems like TRAPPIST-1 illustrate the early environments and physical processes relevant to low-mass systems, allowing us to ascertain how their conditions impact the formation of planets.

To understand the diversity and evolution of planet forming environments, and to enable a comparison with the detected exoplanet population, comprehensive studies of disk properties require a wide range of stellar host masses, ages, and star-forming environments. Constraining disk properties for the full population therefore requires traversing the substellar boundary, and necessitates sensitive observations in a lower luminosity regime. Long-wavelength observations of the dust content within low-mass stellar and substellar disks have become viable with facilities such as the IRAM~30m telescope, providing some of the initial explorations of brown dwarf disks \citep{scholz06}. The large-program Submillimeter Array (SMA) survey by \citet[][with a 3$\sigma$ sensitivity limit of 3 mJy]{andrews13}, enabled disk detections for many higher-mass ($>0.1M_{\odot}$) members of Taurus, but few detections of the brightest low-mass stellar and brown dwarf disks. Recently, studies using the Atacama Large Millimeter/submillimeter Array (ALMA) have enabled the measurement of disk properties for detected brown dwarf disks in three systems in Taurus \citep{ricci14}, seven systems in Upper Sco \citep{gvdp16}, and 11 systems in $\rho$ Ophiuchus \citep{testi16}, providing initial results regarding disk mass deficits for these lower-mass hosts. With the sensitivity of ALMA for sub-mm/mm detections of brown dwarf disks, large systematic surveys of disk populations bridging the gap across the sub-stellar boundary are now possible.

In this paper, we present new ALMA Cycle 1 885 $\mu$m continuum observations of 24 low mass stars and brown dwarfs in the Taurus star forming region, which were selected on the basis of previous \textit{Herschel} detections at 70$\mu$m and 160$\mu$m \citep{bulger14}. In Section~\ref{sec:sample}, we describe the sample and its selection from previous far-infrared Taurus surveys. Details of the ALMA observations and data reduction procedures are listed in Section~\ref{sec:observations}. Section~\ref{sec:data} provides the analysis methods to process the ALMA data and determine source flux densities, the results of which are given in Section~\ref{sec:results}. In Section~\ref{sec:discussion}, we describe the various methods used to estimate the dust masses of the disks and the central object masses of the host stars, and discuss these relations in terms of the feasibility and timescale of planet formation. The summary and conclusions are given in Section~\ref{sec:summary}.


\section{Sample}
\label{sec:sample}
\begin{figure}
    \centering
    \includegraphics[scale=0.6]{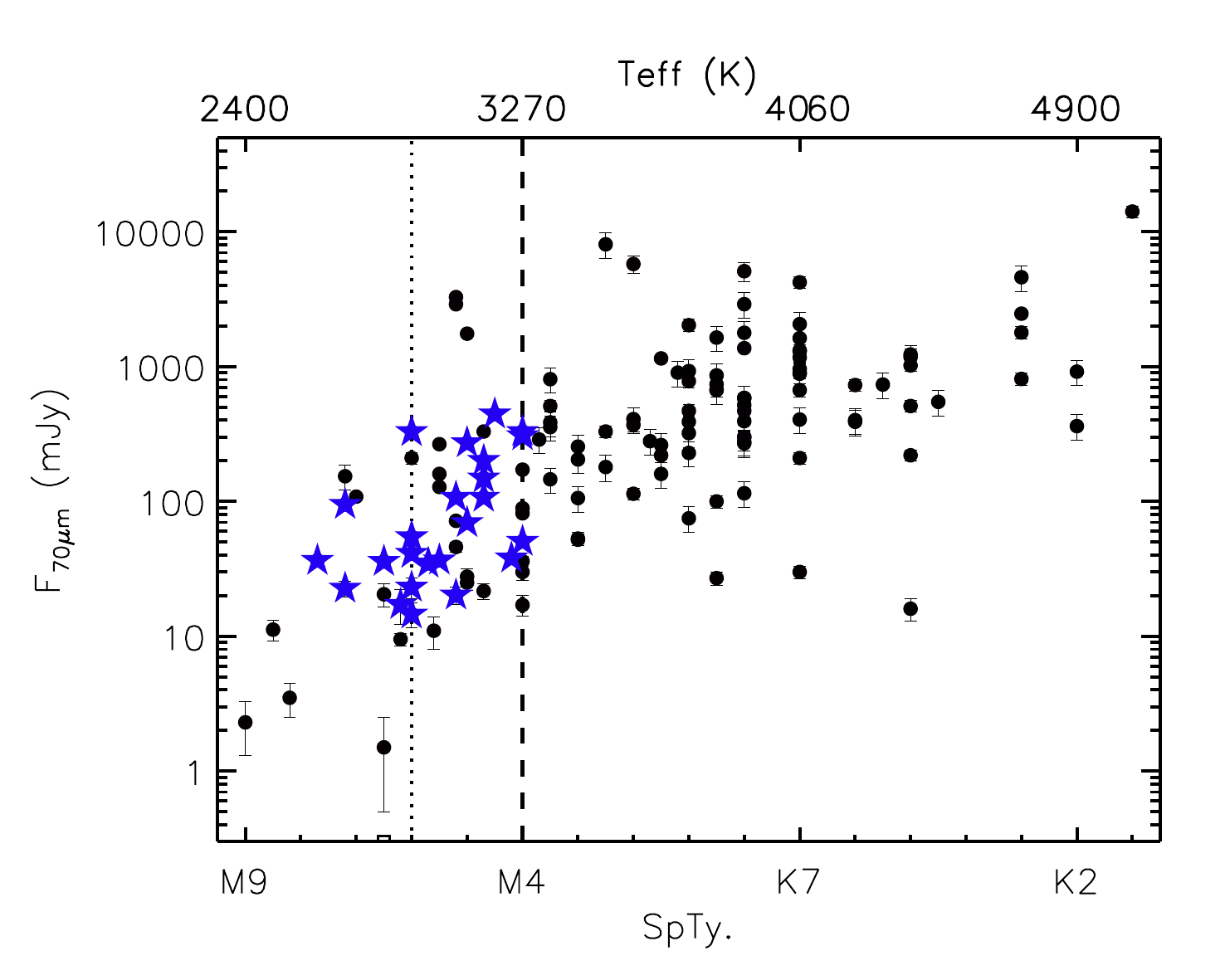}
    \caption{Flux at 70$\mu$m from \textit{Herschel} PACS or \textit{Spitzer} MIPS observations of Taurus members as a function of spectral type. Only detections are plotted. The ALMA sample is indicated with blue stars. The dashed vertical line denotes the earliest M4 spectral type of the TBOSS sample and the dotted line is the M6 spectral type near the stellar/substellar limit. The ALMA sample spans the range of 70$\mu$m fluxes rather than being limited to the upper envelope of brightest sources.}
    \label{fig:herschelcomp}
\end{figure}

\begin{figure}
    \centering
    \includegraphics[scale=0.73]{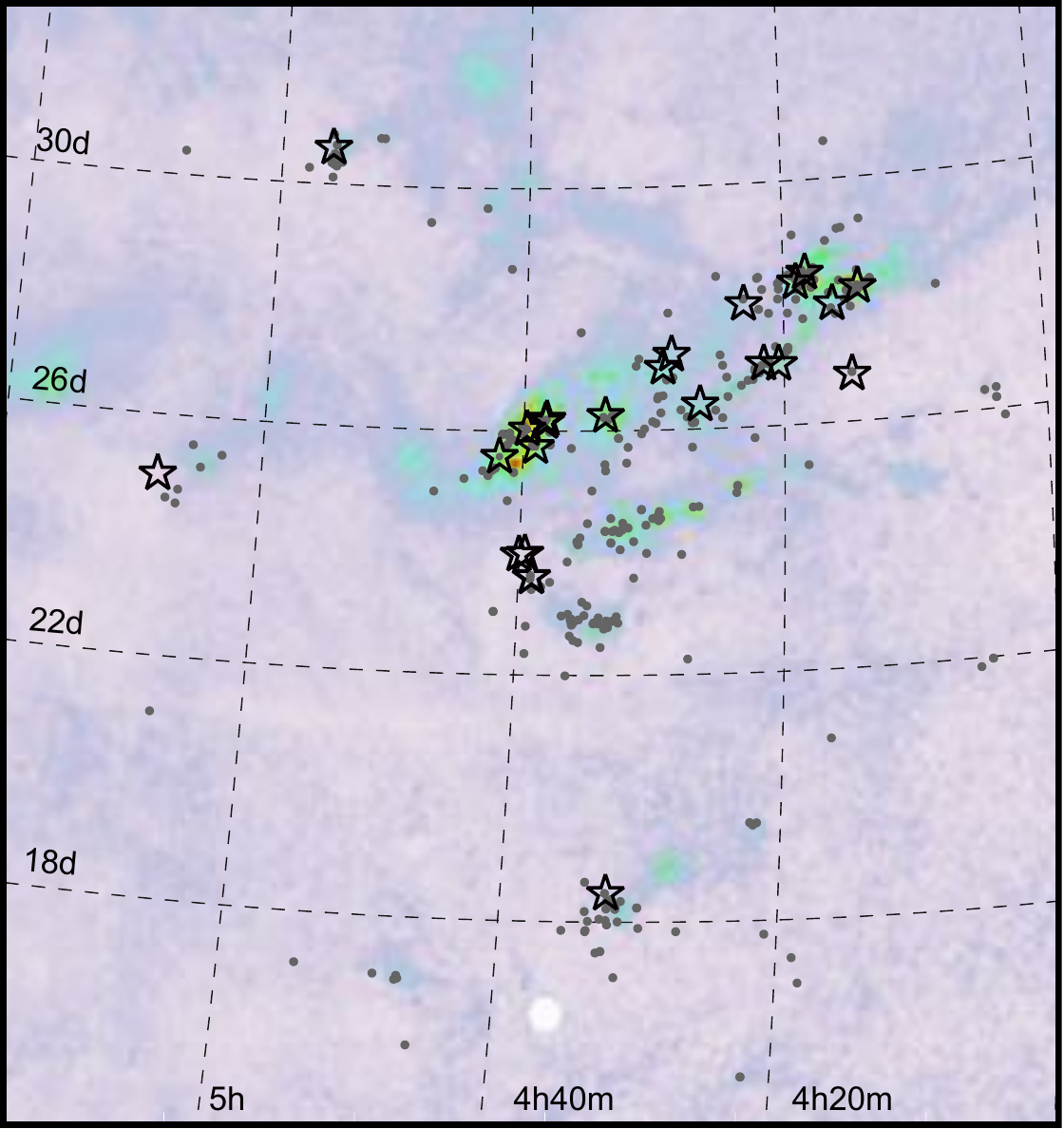}
    \caption{Spatial distribution of the ALMA sample (open star symbols) compared to the full TBOSS sample \mbox{\citep[grey circles]{bulger14}}, overlaid on the extinction map from \mbox{\citet{dobashi05}}. The ALMA sample covers many of the sub-regions in Taurus.}
    \label{fig:spatialdist}
\end{figure}

The ALMA target sample consists of 24 Taurus low mass stars and brown dwarfs with spectral types of M4-M7.75. The 24 targets represent a subset of \textit{Herschel}-detected members from the 153-object TBOSS (Taurus Boundary of Stellar/Substellar) sample \citep{bulger14} that is a 99\% complete sample of M4-L0 Taurus members covering Class I-III objects. Class I and Class III detections from the TBOSS survey were not considered for the ALMA study. As shown in Figure~\ref{fig:herschelcomp}, the observed targets span the full range of measured \textit{Herschel} PACS \citep{poglitsch10} fluxes so the sample is not biased to include only the brightest far-IR detections. Of the Class II M4-L0 members observed with \textit{Herschel}, 75\% were detected \citep{bulger14}\footnote{OT1\_jpatienc\_1 }, making the \textit{Herschel}-detection criterion representative of the majority of the lowest mass Class II Taurus objects. Table~\ref{tab:sample} lists the basic information for the ALMA Taurus targets, and the spatial distribution of the sample is mapped in Figure~\ref{fig:spatialdist} along with the full TBOSS sample. While not a selection criterion, the sample includes seven examples of transition disks, as identified within previous mid-IR and sub-mm studies, and these targets and their corresponding references are identified in the notes of Table~\ref{tab:sample}.

At the age of Taurus, a spectral type of M6.25 is the demarcation between stars and brown dwarfs \citep[e.g.,][]{luhman05_ic348disks}. All spectral types for this sample were determined spectroscopically and have a typical uncertainty of $\pm$0.5 subclasses. Studies from the literature providing these spectral type values are the following, compiled by \citet{bulger14}: \citet{briceno02}; \citet{guieu06}; \citet{kenyonhartmann95}; \citet{luhman96, luhman06_spitzertaurus, luhman04_taurusbds, luhman09}; \citet{martin01}; \citet{slesnick06}; and \citet{white_basri03}. There are 14 M4-M5 stellar and 10 M6-M7 substellar objects in the sample. Previous single dish surveys \citep{andrewswilliams05, scholz06} have reported fewer M4-M5 sub-mm/mm detections than M6-M7 detections, and the sample is designed to characterize the transition from stellar to substellar disk properties with a sensitive ALMA survey.

\input{sample.txt}


\section{Observations and Data Reduction}
\label{sec:observations}

ALMA Band 7 observations were obtained for all targets in a series of tracks executed between November 2013 and July 2014 during the Cycle 1 Early Science campaign (program ID 2012.1.00743.S). Among the available ALMA Bands, Band 7 represented the best compromise between declining disk flux with wavelength and increasing ALMA sensitivity with wavelength. For example, ALMA sensitivity is 1.7 times deeper at 1.2mm than 850$\mu$m, but brown dwarfs with detections at both wavelengths are $\sim$2 - 4.5 times brighter at 850$\mu$m compared to 1.2mm \citep[e.g.,][]{bouy08}. The four spectral windows were centered on the following four frequencies: 331.8, 333.8, 343.8, and 345.7 GHz, providing a mean frequency of 338.8 GHz (885$\mu$m). Since the central goal of the continuum survey was the detection of faint sources, the correlator was configured to the widest available setting of 2 GHz for three of the four spectral windows; the fourth spectral window centered on the highest frequency was configured in the only slightly narrower 1.875 GHz mode to enable a search for $^{12}$CO(3-2) emission at a rest frequency of 345.70599 GHz. The aggregate sensitivity level across the full band pass was set to reach an RMS noise level of 0.15 mJy/beam to achieve an order of magnitude improvement over previous single dish surveys. The continuum observations are the subject of this paper, while a companion paper is focused on the spectral channel observations (van der Plas et al. 2017, \textit{in prep}).

\input{observations.txt}

The 24 targets were divided into three ALMA Scheduling Blocks (SBs) based on science goals and proximity on the sky to ensure target positions within a 10 degree radius. Two SBs were observed twice (``Taurus2a'' and ``Taurus2b''), consisting of targets of spectral type M5 and earlier) and one was observed three times (``Taurus1'', consisting of targets of spectral type M6 and later), as listed in Table~\ref{tab:obs}. The main observing sequence consisted of cycling through the Taurus sources and the gain/phase calibrators J0510+1800 and J0509+1806, depending on the observation. The phase calibrator J0509+1806 was fainter than expected based on extrapolating archive fluxes from the SMA Observer Center \footnote{\url{http://sma1.sma.hawaii.edu/callist/callist.html}}, but was still sufficient for the data analysis. In addition to the observations of the phase calibrators every $\sim$5-7 minutes, flux and bandpass calibrators were observed at the beginning of each track. Table~\ref{tab:obs} indicates which targets were allocated to each group, the observation dates, on-source time, the range of baselines, and environmental and system conditions. The time on-source ranged from 5 minutes to 10 minutes per target, and the precipitable water vapor (PWV) range of 0.36~mm--1.13~mm corresponds to 1st--3rd octile conditions for ALMA.

\input{positions.txt}



\section{Data Analysis}
\label{sec:data}

To convert raw ALMA observations into calibrated measurement sets, calibration and flagging tables derived from the ALMA Quality Assurance process \mbox{\citep{petry14}} were re-applied to the raw data in CASA 4.2.2 \citep[Common Astronomy Software Applications;][]{mcmullin07}. Minimal additional flagging was performed to remove data points that were identically zero and had been missed by the pipeline. 

We adopt a uniform approach to continuum imaging all of the targets within the three SBs in CASA. For each target, this included aligning the spectral windows between individual observations and concatenating the measurement sets, flagging all channels associated with CO emission as visually identified from plotting the amplitudes per channel, and averaging the remaining continuum channels after removing the CO-dominated channels\footnote{Example reduction scripts and auxiliary data are available at https://osf.io/9dyx4.}. Without flagging the CO channels, the median line flux for a target contributed $\sim$1\% additional emission over the full 7.875 GHz bandpass. Initial cleaned images were produced with natural weighting. From these images, 22/24 targets were detected, and the centers of continuum emission in the images were used to define new pointing centers, which were then applied to phase shift the measurement set of each target using the \emph{visstat} CASA task. These new target coordinates are provided in Table~\ref{tab:positions}, along with the offset from the 2MASS J2000 coordinates, and proper motion values from \mbox{\citet{zacharias15}}. The calibrated visibilities were then re-cleaned using natural, Briggs, and uniform weighting to compare the extracted flux values for each source. Average CLEAN beam sizes for the various weighting schemes were $0\farcs47 \times 0\farcs38$ (Natural), $0\farcs33 \times 0\farcs22$ (Uniform), and $0\farcs34 \times 0\farcs24$ (Briggs).

\begin{figure}
    \centering
    \includegraphics[scale=0.45]{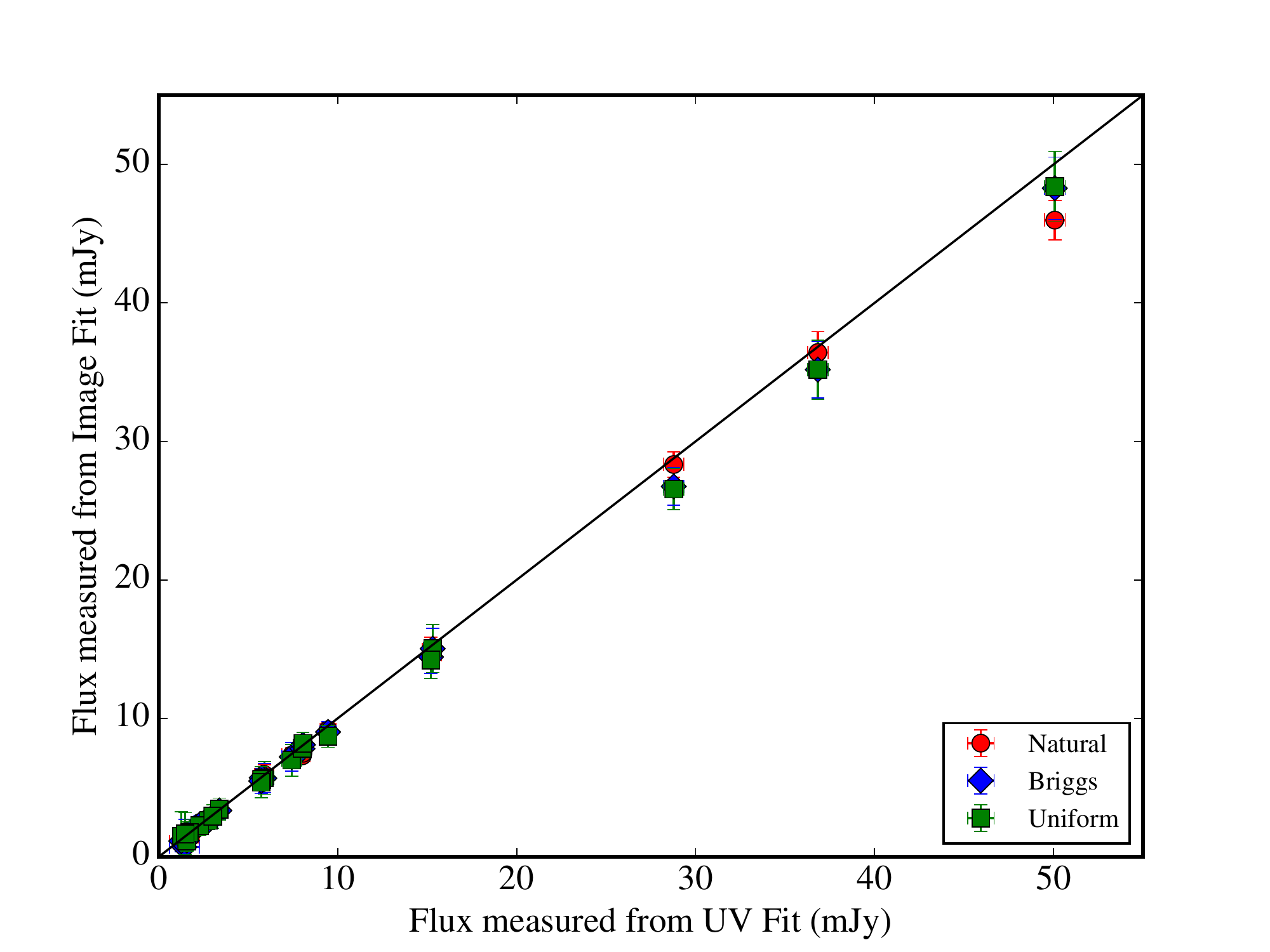}
    \caption{Flux density derived from CASA \textit{imfit} routine applied to the non--self-calibrated continuum maps generated with different weighting schemes (natural -- red, Briggs -- blue, uniform -- green) as a function of the flux density derived from CASA \textit{uvmodelfit} routine applied to visibilities. Errorbars shown are 3$\sigma$ uncertainties. The results are consistent, with an average difference of 7\%.}
    \label{fig:fluxcomparison}
\end{figure}

The \emph{imfit} task in CASA was used to fit the continuum emission in the image plane with 2D Gaussians for each of the 22 detections. The phase-shifted measurement sets were also used to fit the continuum emission in the \emph{uv}-plane using the CASA task \emph{uvmodelfit}, and the output source flux densities and uncertainties from the CASA tasks for each of the three weighting schemes in the image plane and \emph{uvmodelfit} results are provided in Table~\ref{tab:fluxes}. A comparison between the image plane fitting and \emph{uv}-fitting for the extracted fluxes is shown in Figure~\ref{fig:fluxcomparison}. The extracted fluxes agree within 7\% on average for all methods. 

For the 8 highest signal-to-noise ratio detections (SNR $>$ 40), we also performed self-calibration, consisting of 2 or 3 rounds of phase-only self-calibration. The number of iterations were determined by repeating self-calibration until the source residual emission matched the RMS noise level in the remainder of the field. For the self-calibrated sources, imaging was performed with Briggs weighting with \textit{``robust''}=0.5. For the remaining 16 sources with lower SNR, we adopt the fluxes obtained with natural weighting to maximize sensitivity in the image plane. The self calibration or natural weighting values from Table~\ref{tab:fluxes} are used for the subsequent analysis in the paper and an additional 10\% uncertainty was added to the uncertainties in Table~\ref{tab:fluxes} to account for the absolute flux scaling uncertainty; the $\pm$10\% absolute flux uncertainty dominates over the uncertainties from the measurements given in Table~\ref{tab:fluxes}.


\section{Results}
\label{sec:results}
Of the 24 Taurus low mass stars and brown dwarfs observed with ALMA, a total of 21 targets are detected at $>$8$\sigma$ levels above the background, a much higher detection rate than previous sub-mm/mm brown dwarf disk surveys with less sensitive instruments \citep[e.g.,][]{scholz06}. There is one marginal detection for J0414$+$2811 with SNR$\sim$3 in the cleaned image using Briggs weighting and SNR$\sim$5 in the cleaned image using natural weighting (this source was undetected with uniform weighting). Two sources -- J0419$+$2819 (V410 X-ray 6) and J0421$+$2701 -- are not detected. The flux densities of the detections range from 1.0 to 55.7 mJy. The non-detections have 3$\sigma$ upper limits of 0.27 mJy/beam for J04190110 and 0.29 mJy/beam for J04213459 based on the rms noise level in the map generated with natural weighting.

\input{fluxtable.txt}


\begin{figure}
    \centering
    \includegraphics[scale=0.45]{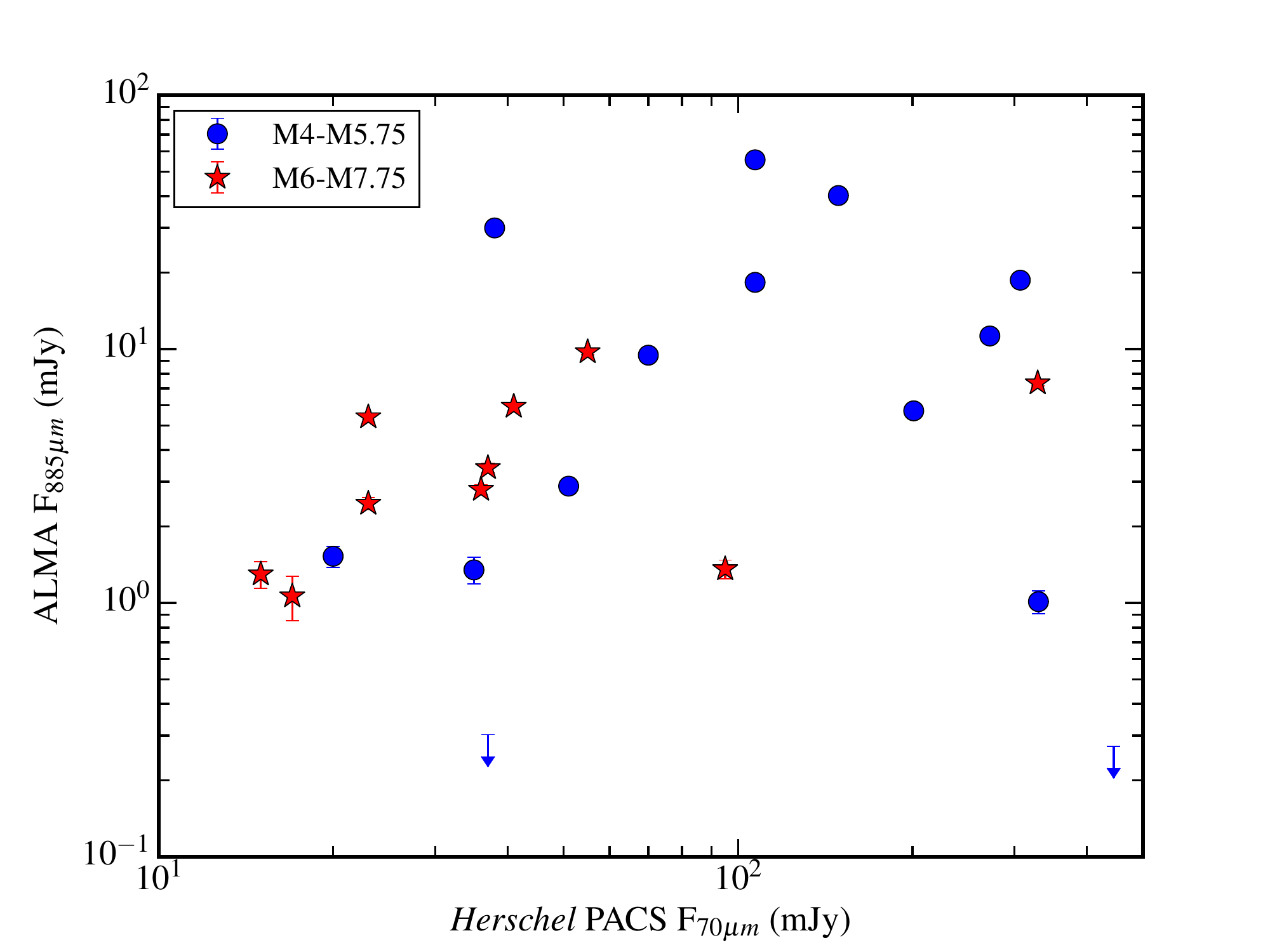}
    \caption{ Measurements and upper limits at 885$\mu$m from ALMA as a function of \textit{Herschel} measurements at 70$\mu$m for each source in the sample.  The M4-M5.75 subset is shown as blue circles and the M6-M7 subset is plotted as red stars.} 
    \label{fig:almaherschel}
\end{figure}

\begin{figure}
    \centering
    \includegraphics[scale=0.45]{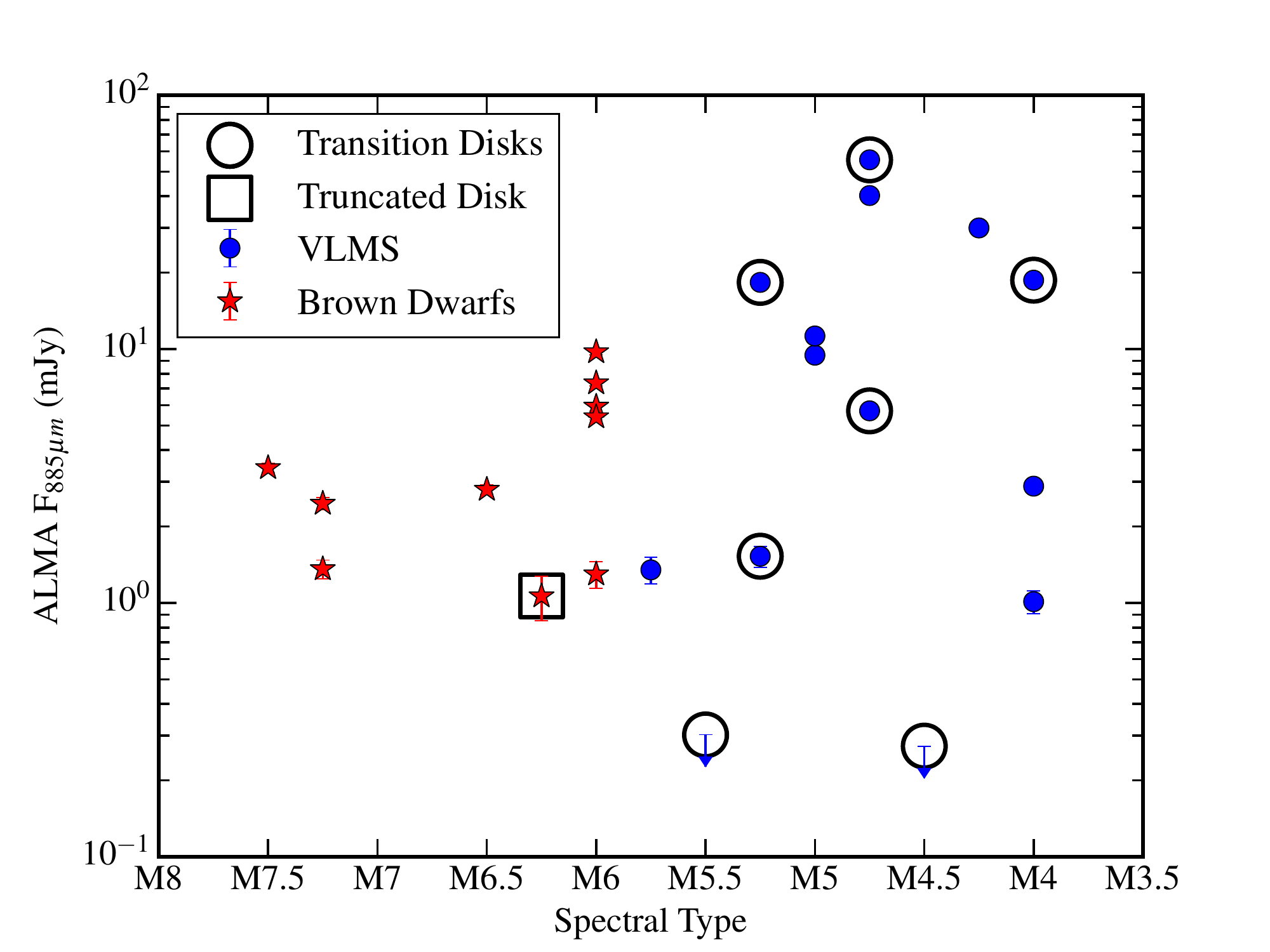}
    \caption{The new ALMA 885$\mu$m fluxes from the 24 targets in our study. All but two of the targets have continuum detections, and the two non-detections are both transition disks. However, additional transition disks (circled) are also found within the very low-mass star (VLMS) population within our sample, and a single truncated disk (square) was identified for one of the brown dwarfs in our sample.}
    \label{fig:flux_vs_spty}
\end{figure}

The ALMA 885$\mu$m flux densities are plotted against the selection criterion of the \textit{Herschel} 70$\mu$m flux densities in Figure~\ref{fig:almaherschel}. Although the detection of 70$\mu$m emission is well correlated with an ALMA 885$\mu$m detection, there is approximately an order of magnitude scatter in the 885$\mu$m flux density for a given 70$\mu$m level. The two 885$\mu$m upper limits are also not restricted to the faintest 70$\mu$m sources. There is no qualitative distinction in distributions of ALMA flux densities between the stellar M4-M5 and substellar M6-M7 populations. The transition disks identified by several studies \citep{currie_siciliaaguilar11, cieza12, bulger14} are labeled in Figure~\ref{fig:flux_vs_spty}. The transition disk flux densities from our ALMA study span the range of measured flux values for the full ALMA TBOSS sample, and they are not associated with lower 885$\mu$m emission. Previous disk surveys have noted that transition disks can have bright submm detections \citep[e.g.,][]{ansdell16, andrews13}.     

\begin{figure}
    \centering
    \includegraphics[scale=0.45]{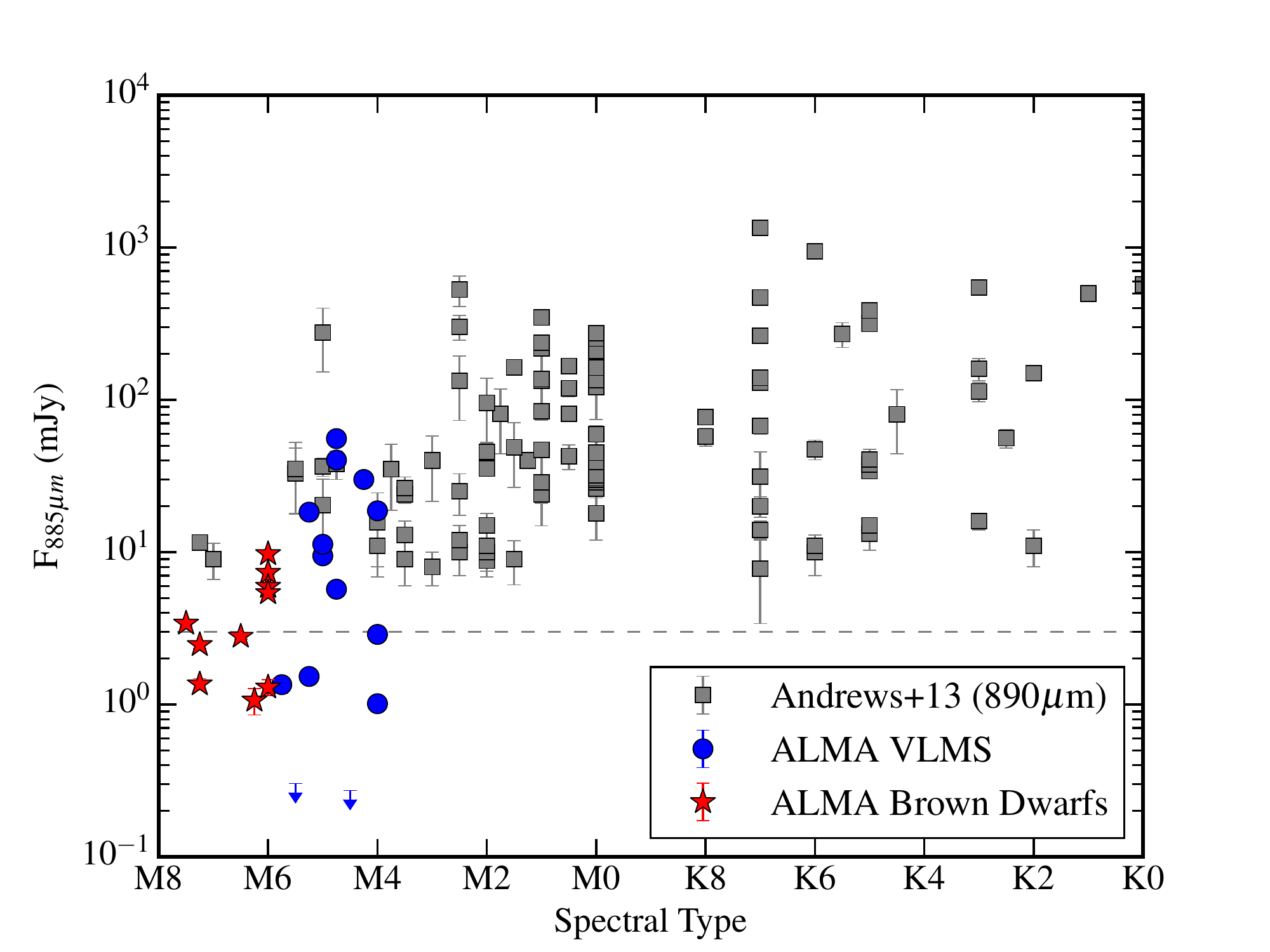}
    \caption{The new ALMA 885$\mu$m fluxes from the 24 targets in our study (red stars and blue circles), as a function of spectral type, shown with a previous compilation of measured or extrapolated 890$\mu$m fluxes for Class II Taurus members from \citet{andrews13} (gray squares), with the survey sensitivity limit shown for comparison (gray dashed line).}
    \label{fig:taurus_classII_fluxes}
\end{figure}

The ALMA results form one of the largest sets of sub-mm detections of low mass objects to-date and define the lower boundary of the detected flux densities as a function of spectral type for Taurus. Figure~\ref{fig:taurus_classII_fluxes} plots the Class II Taurus members with 850$\mu$m or 890$\mu$m detections. The faintest brown dwarf disks are a factor of $\sim$500 dimmer than the brightest disks around early K-stars. Despite the large difference in the typical level of emission, both the earlier and later spectral types exhibit a considerable dispersion of at least a factor of 10 about the average value. This large dispersion appears to be a universal characteristic of disk populations and is seen in surveys of a number of other regions such as Upper Sco \citep{barenfeld16}, Lupus \citep{ansdell16}, and Cha I \citep{pascucci16}.

\begin{figure*}
    \centering
    \includegraphics[width=0.2\textwidth]{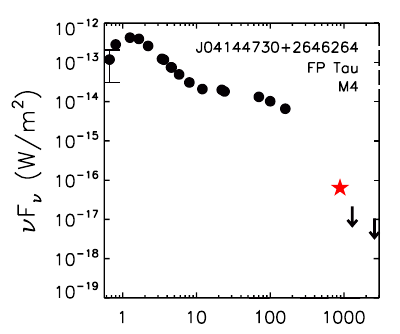}
    \includegraphics[width=0.2\textwidth]{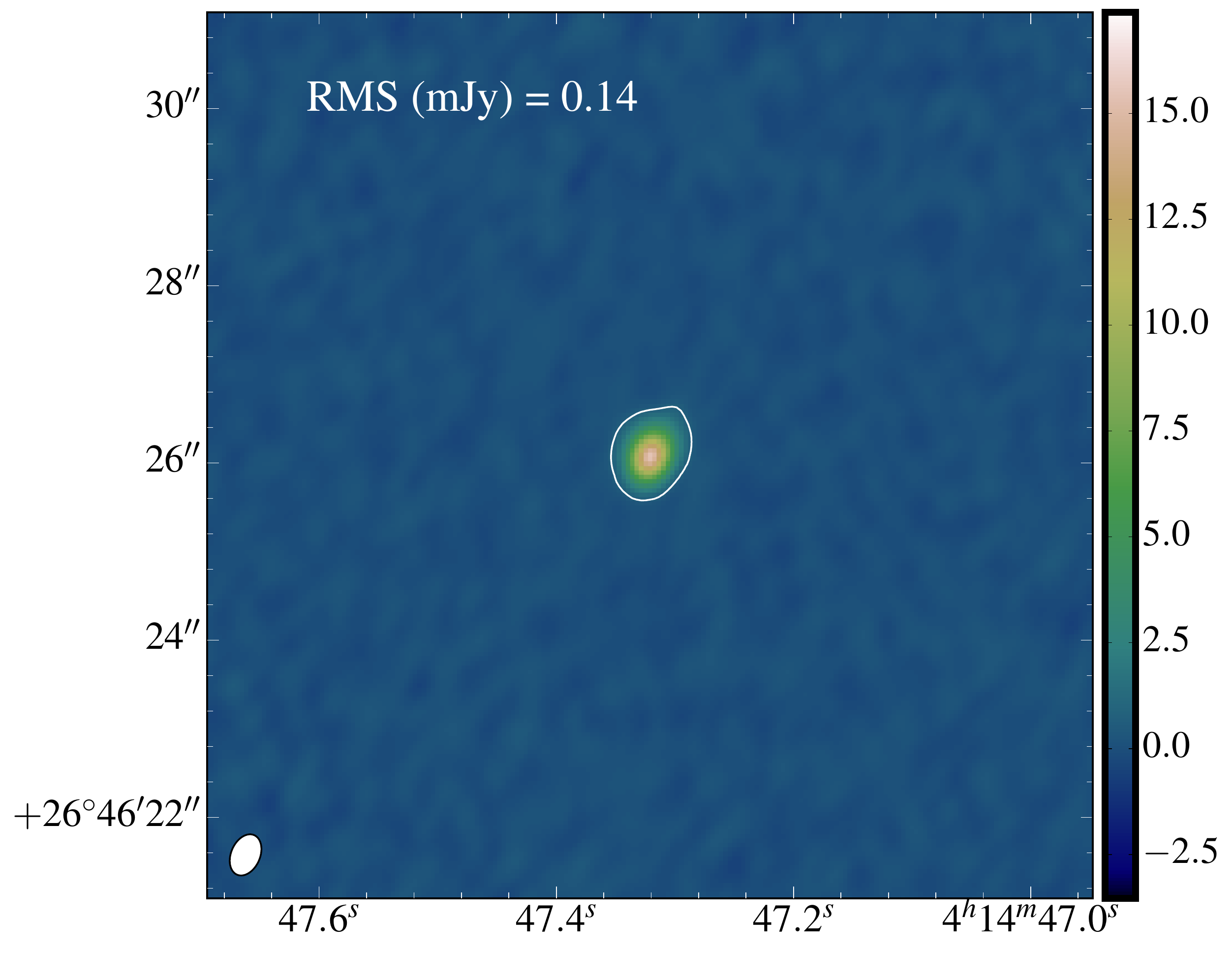}
    \includegraphics[width=0.2\textwidth]{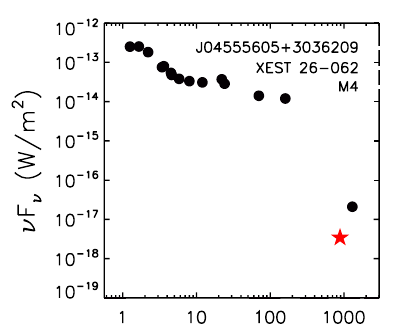}
    \includegraphics[width=0.2\textwidth]{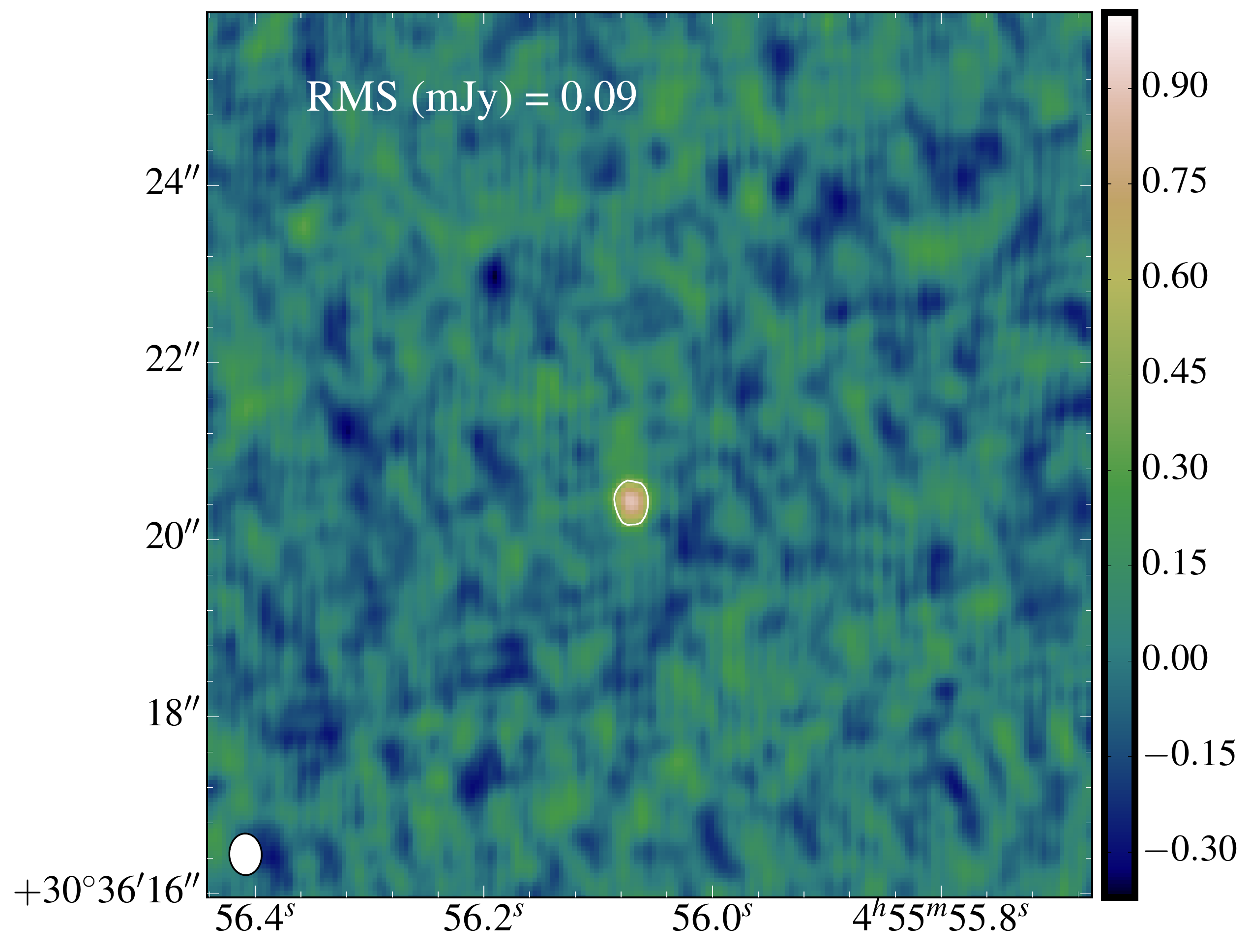}
    \\
    \includegraphics[width=0.2\textwidth]{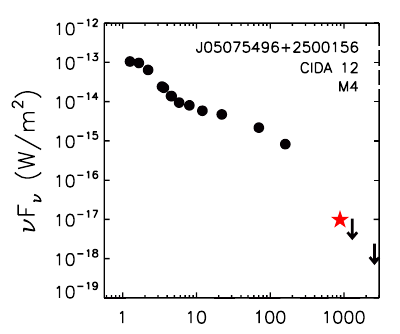}
    \includegraphics[width=0.2\textwidth]{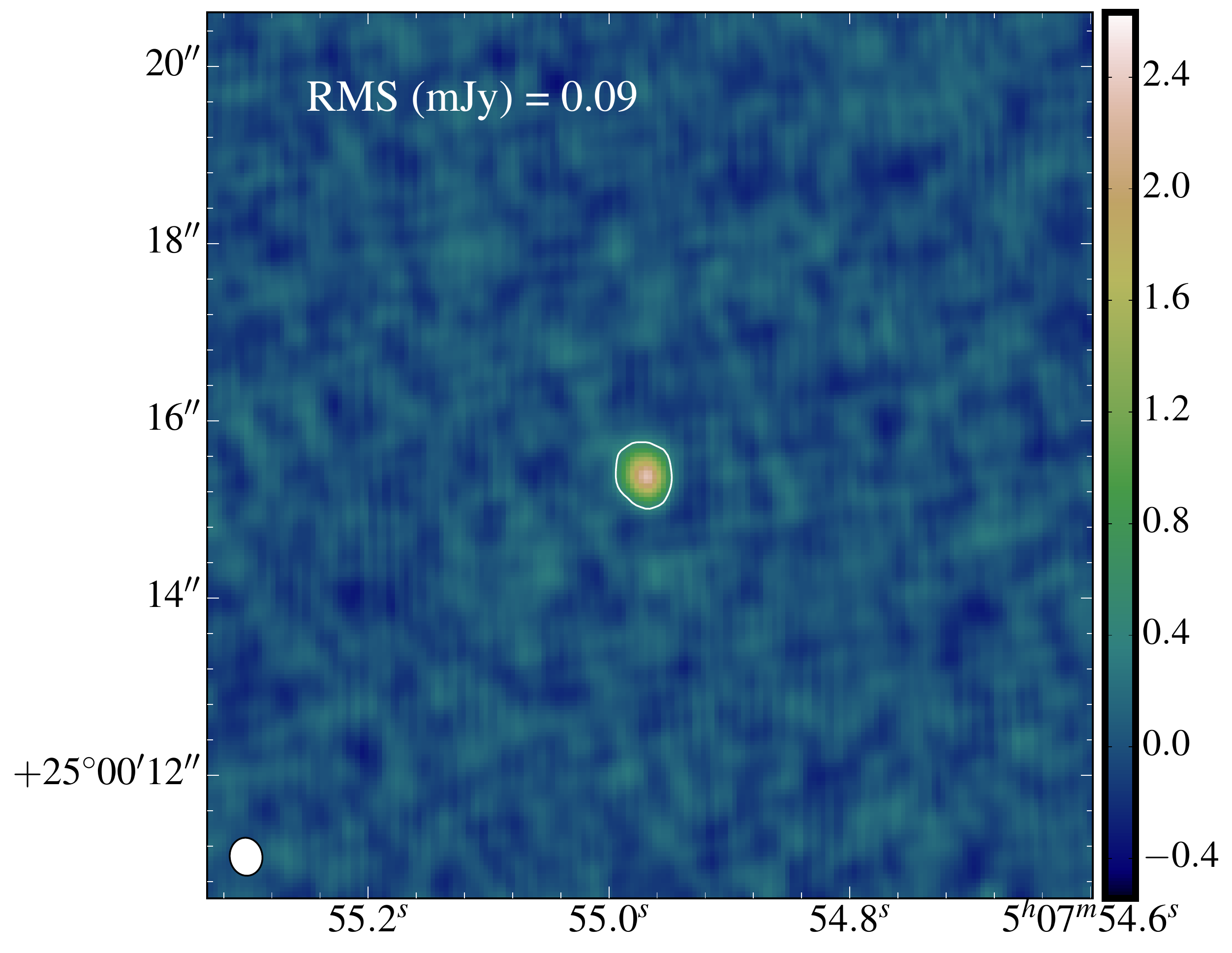}
    \includegraphics[width=0.2\textwidth]{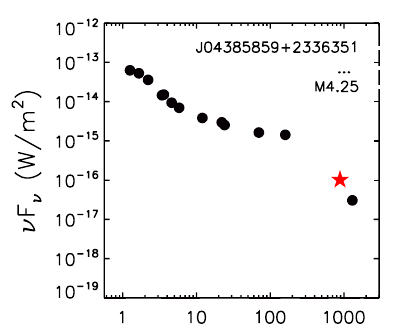}
    \includegraphics[width=0.2\textwidth]{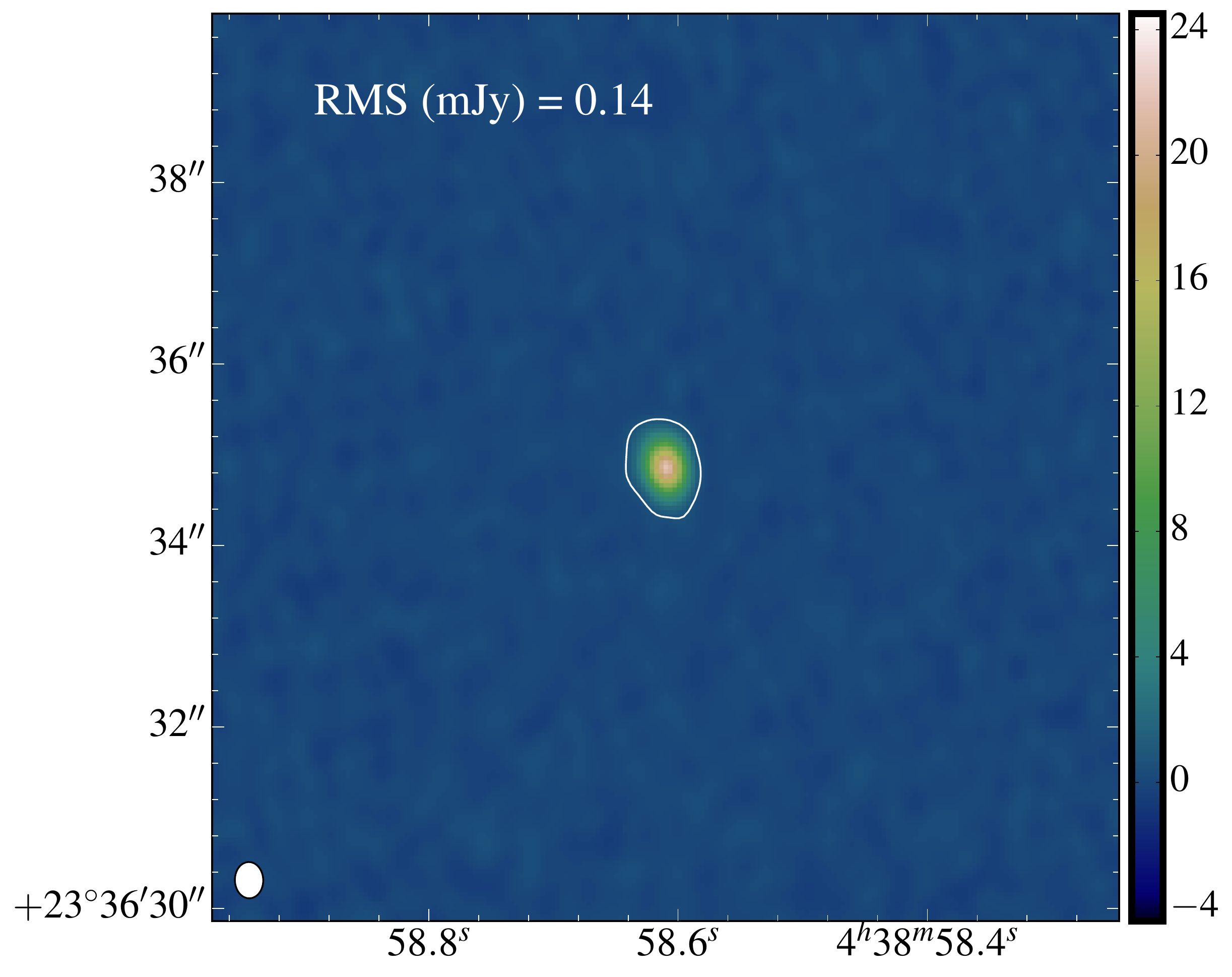}
    \\
    \includegraphics[width=0.2\textwidth]{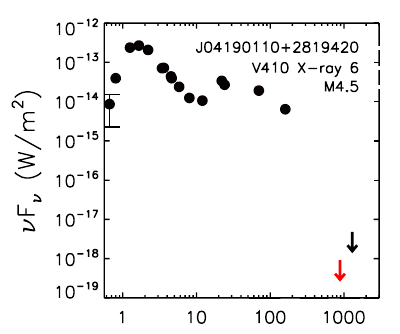}
    \includegraphics[width=0.2\textwidth]{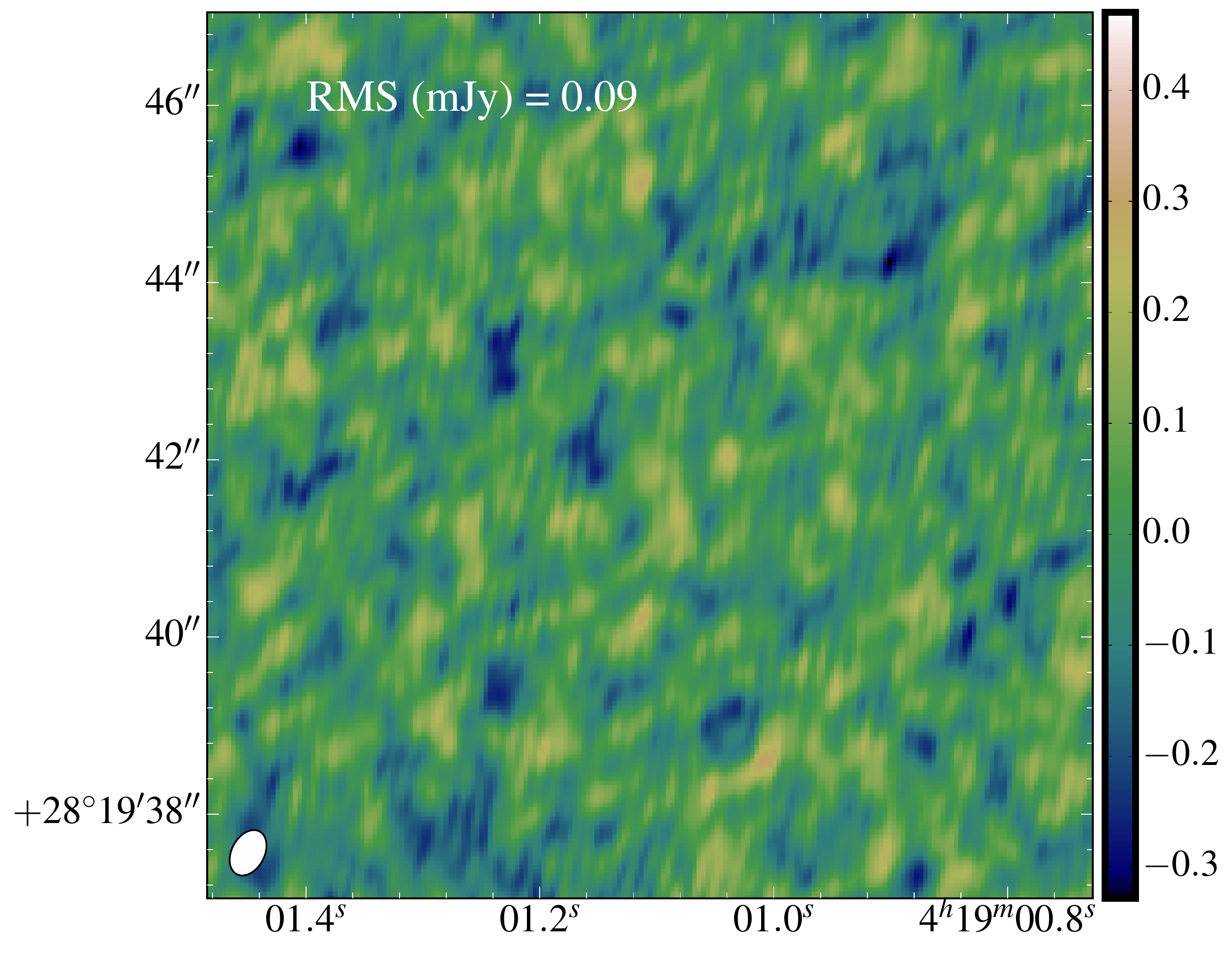}
    \includegraphics[width=0.2\textwidth]{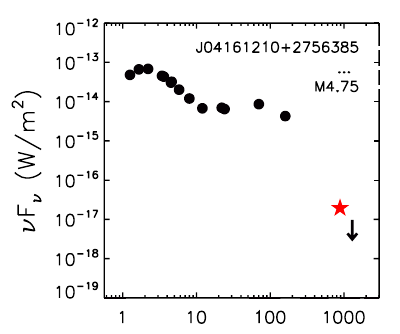}
    \includegraphics[width=0.2\textwidth]{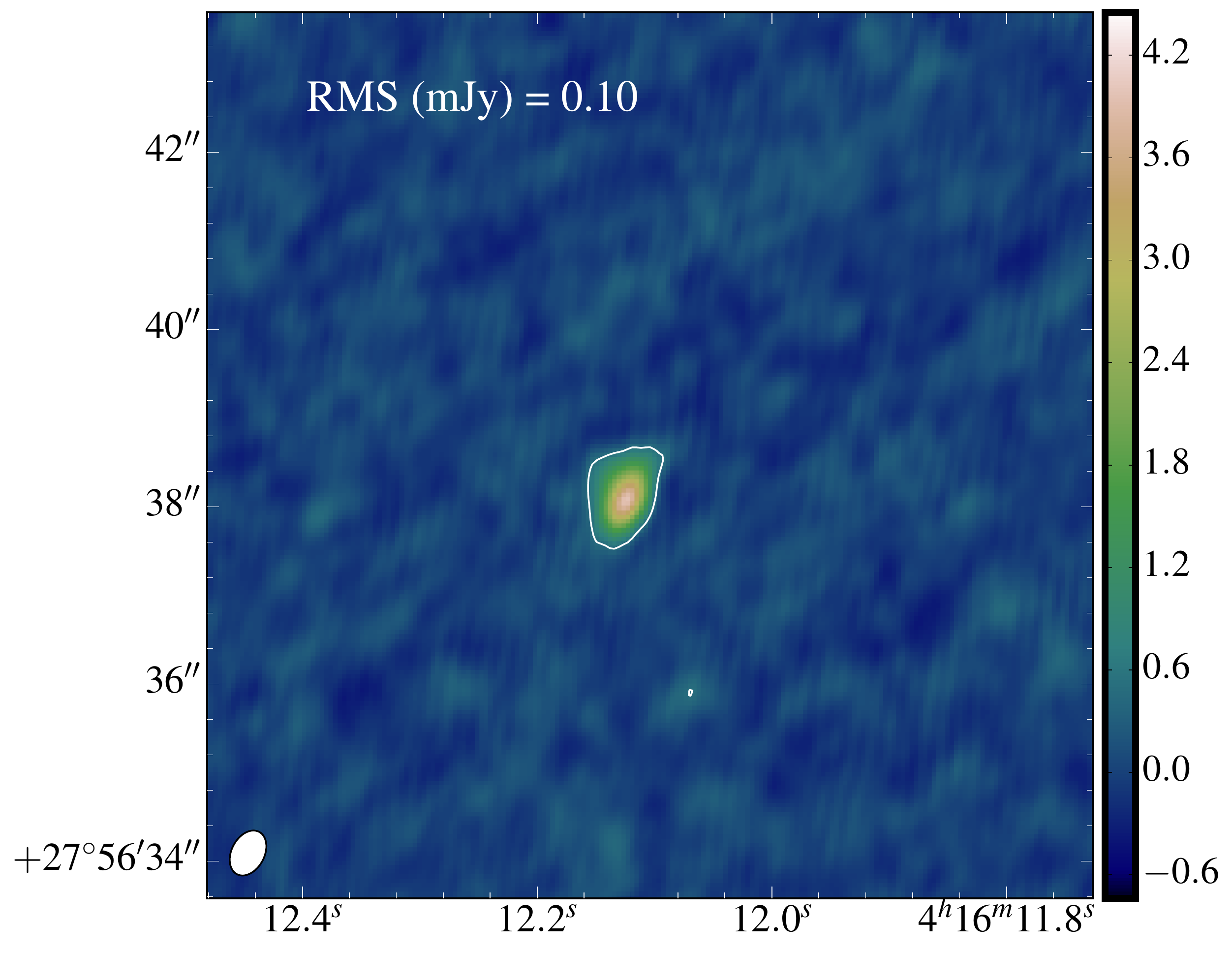}
    \\
    \includegraphics[width=0.2\textwidth]{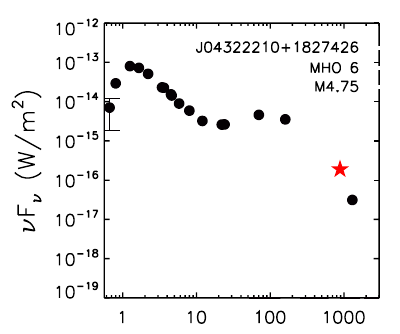}
    \includegraphics[width=0.2\textwidth]{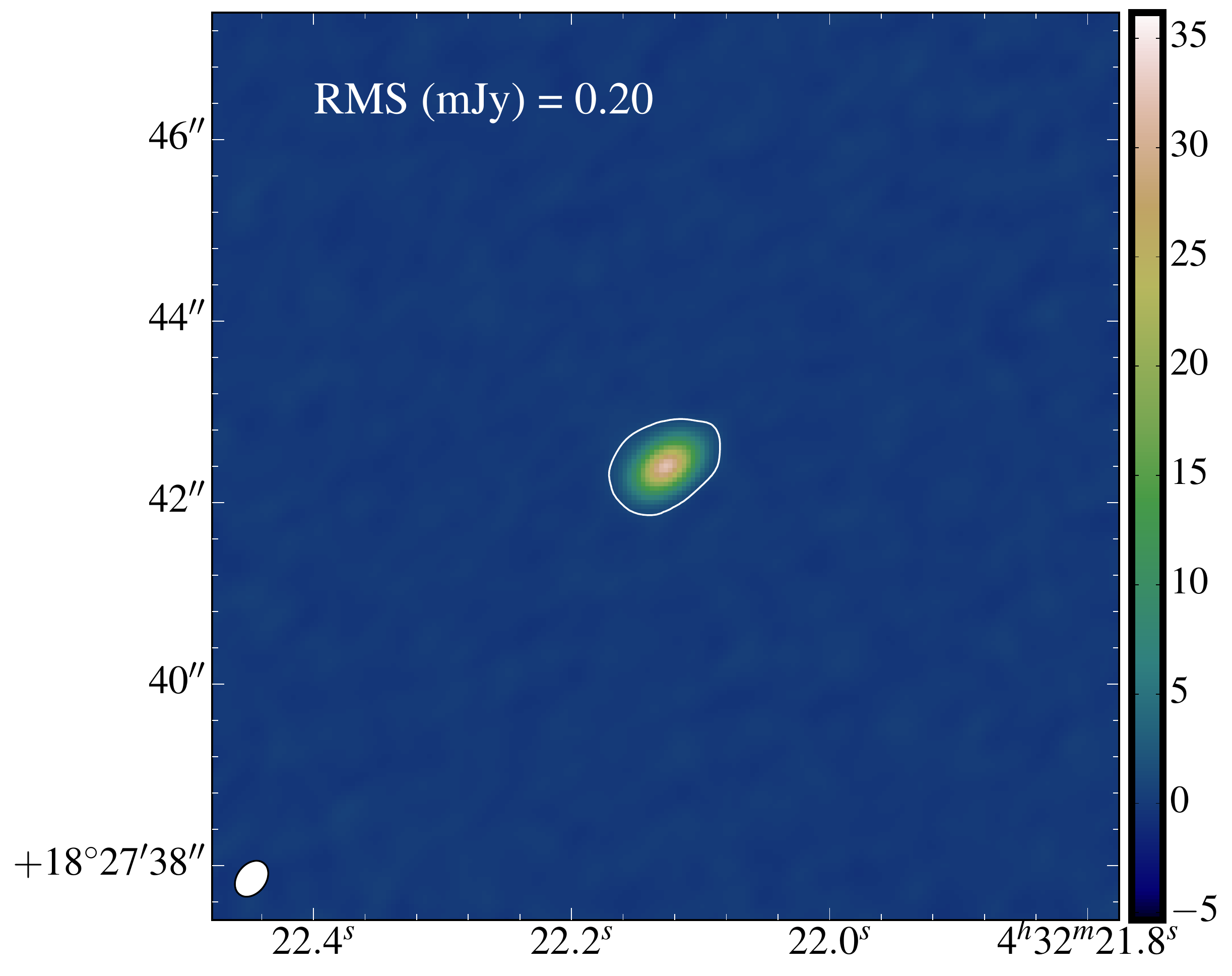}
    \includegraphics[width=0.2\textwidth]{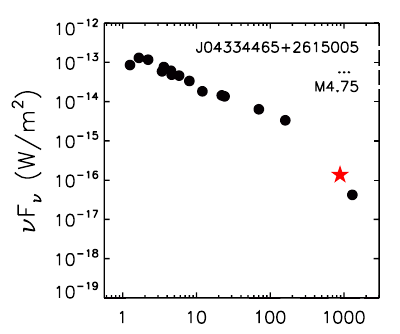}
    \includegraphics[width=0.2\textwidth]{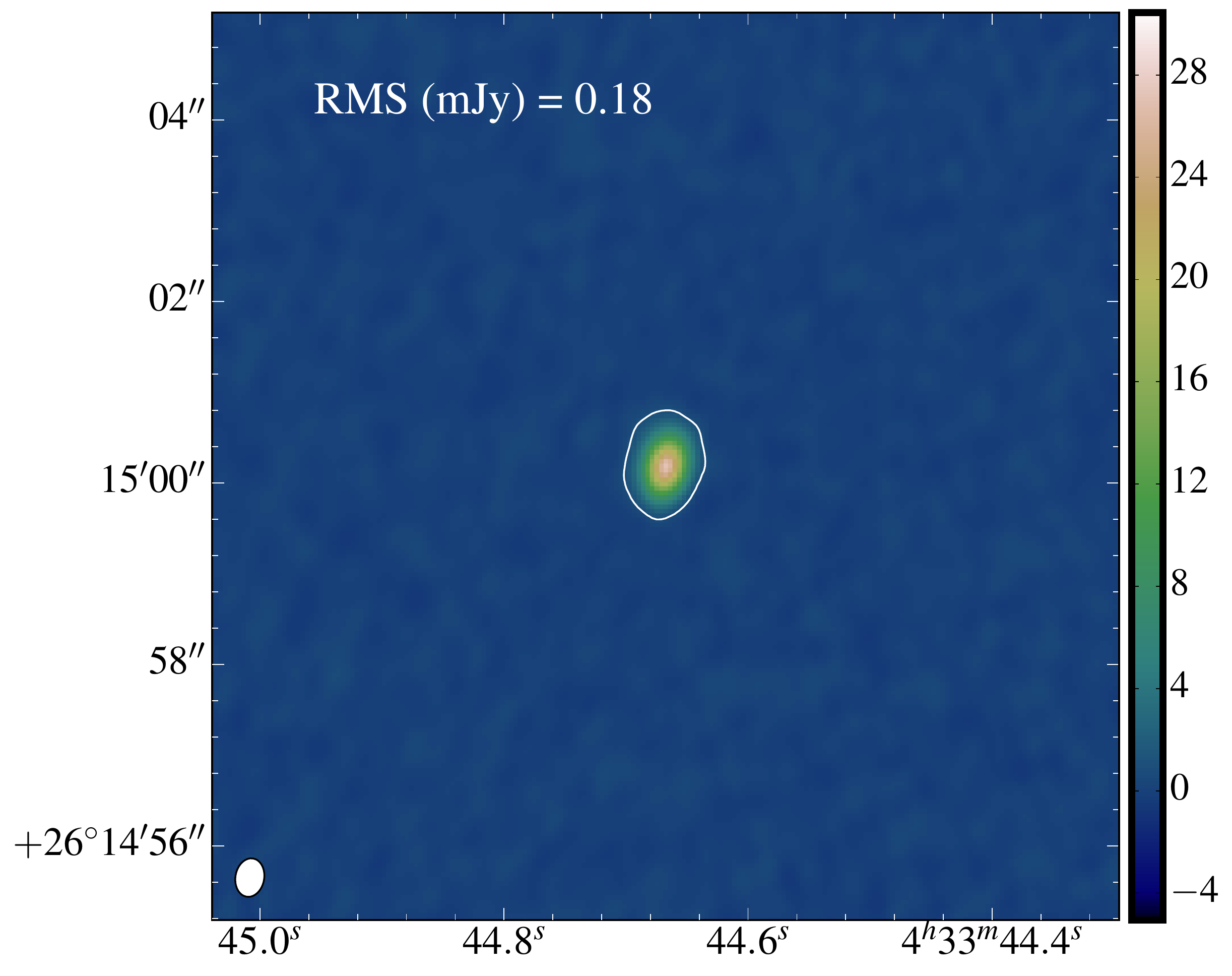}
    \\
    \includegraphics[width=0.2\textwidth]{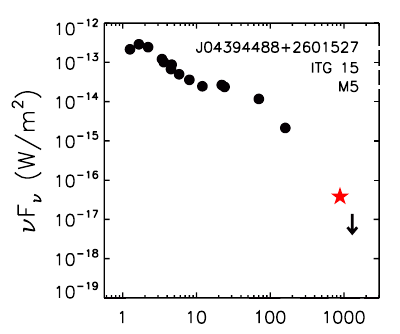}
    \includegraphics[width=0.2\textwidth]{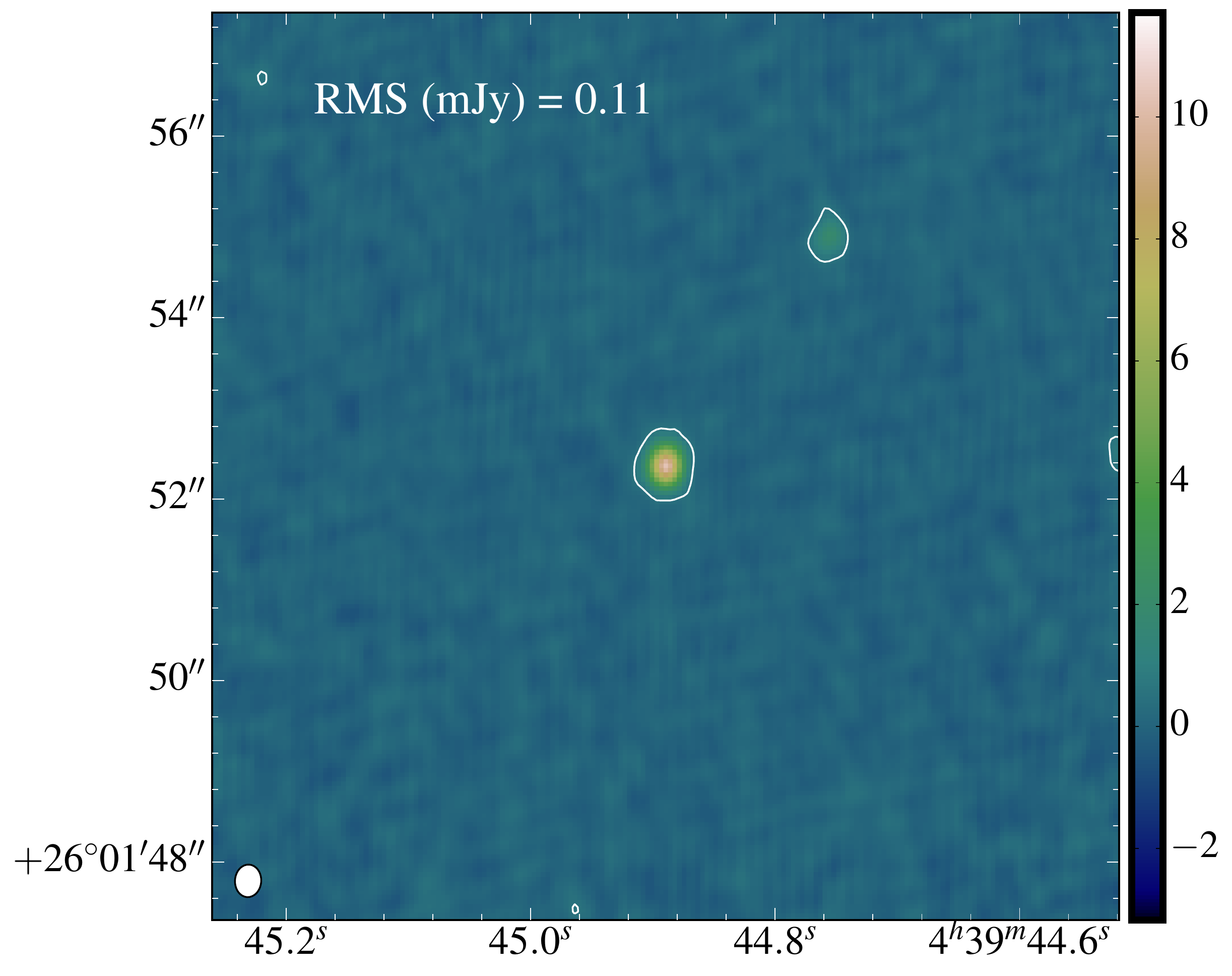}
    \includegraphics[width=0.2\textwidth]{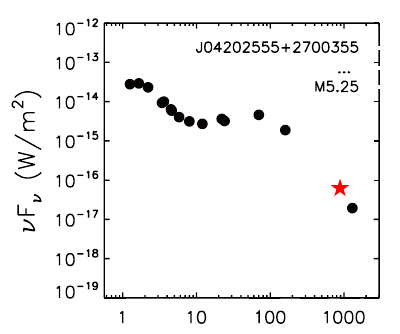}
    \includegraphics[width=0.2\textwidth]{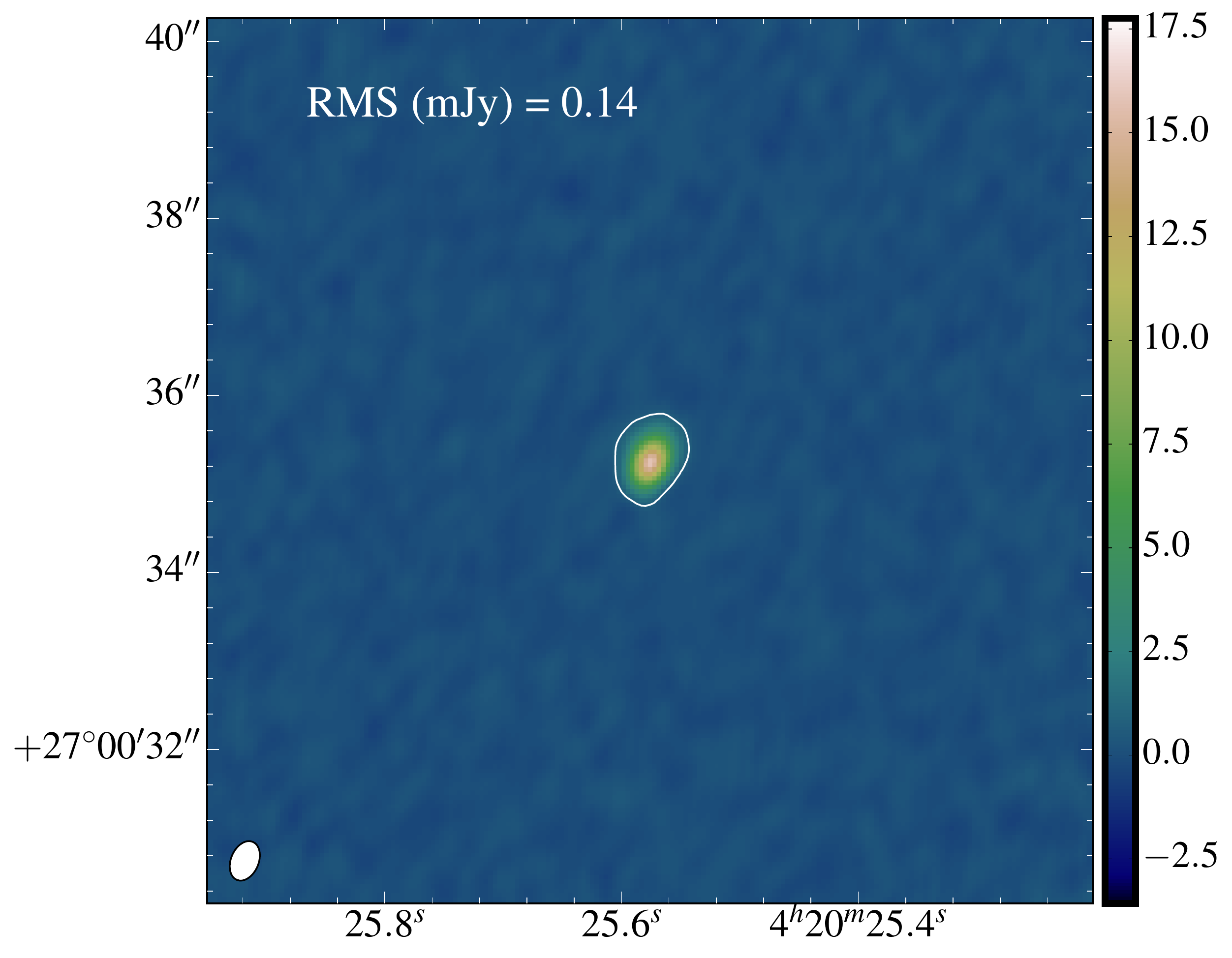}
    \\
    \includegraphics[width=0.2\textwidth]{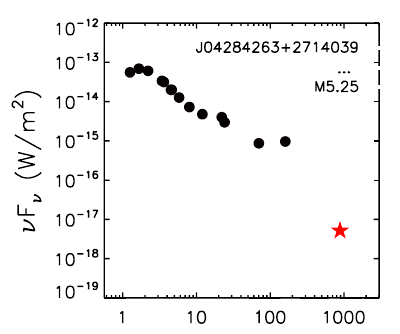}
    \includegraphics[width=0.2\textwidth]{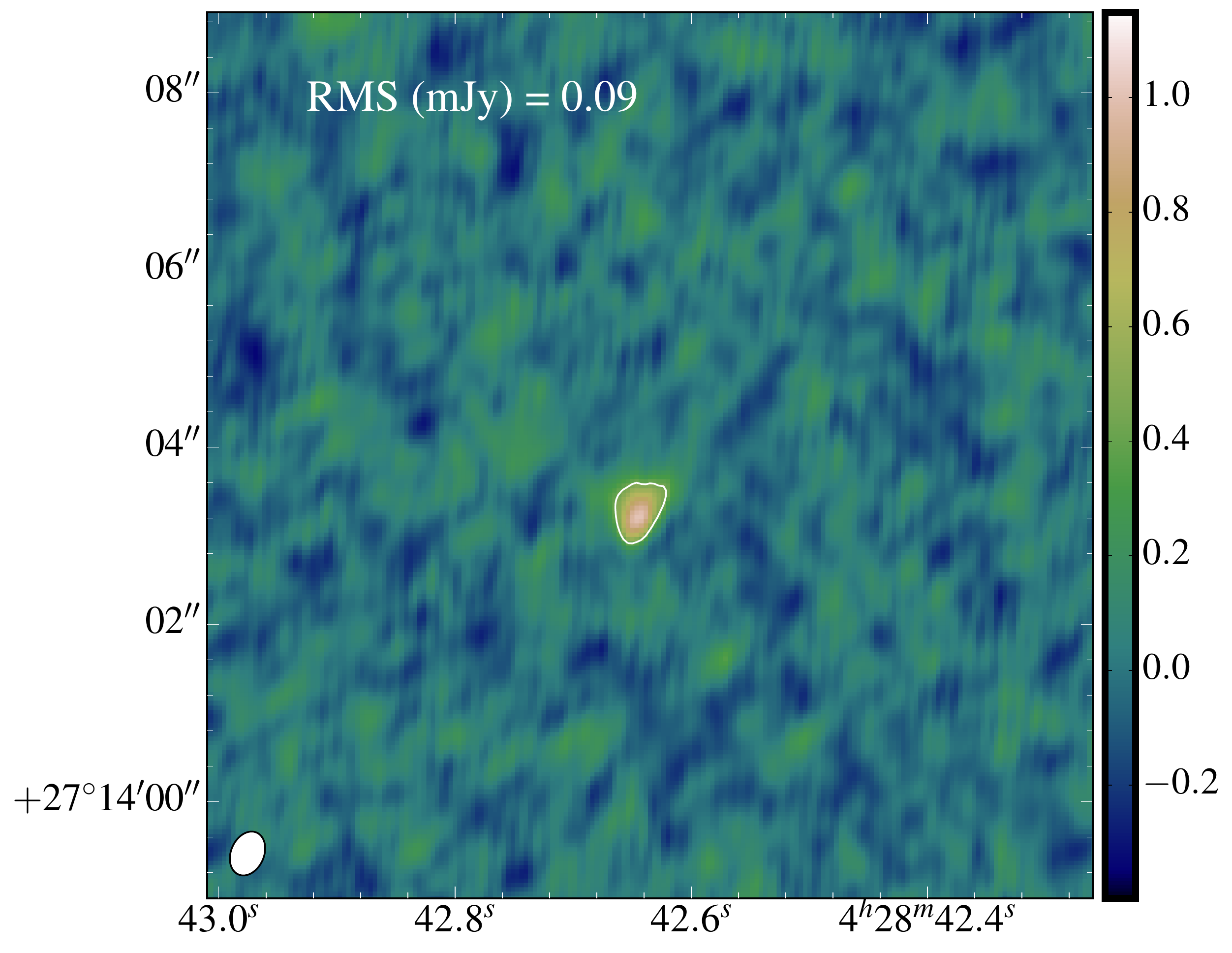}
    \includegraphics[width=0.2\textwidth]{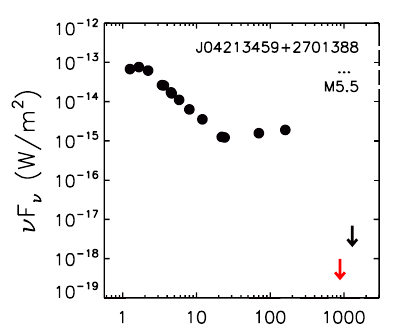}
    \includegraphics[width=0.2\textwidth]{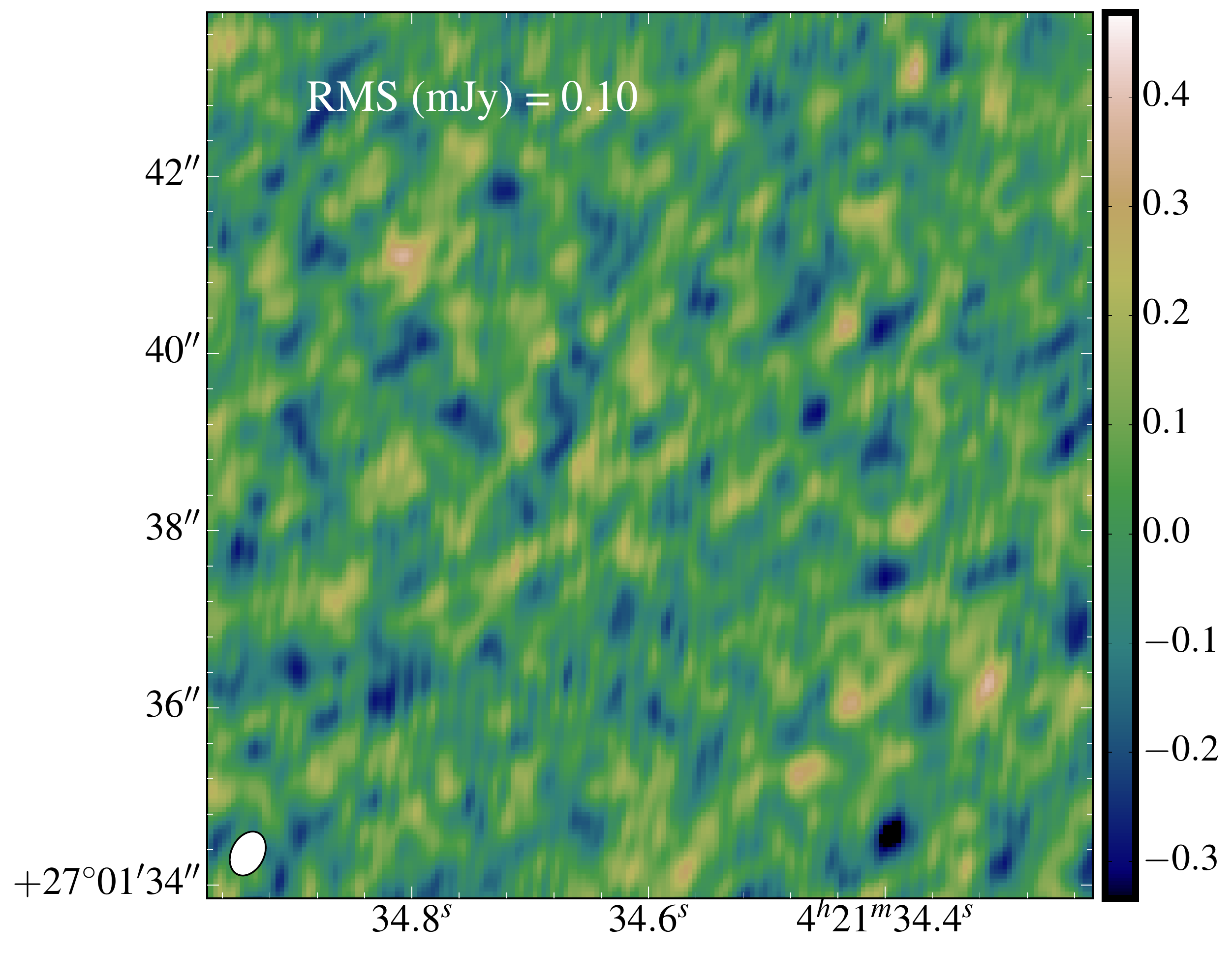}
    \\
    \includegraphics[width=0.2\textwidth]{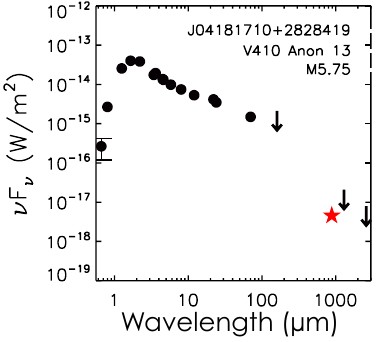}
    \includegraphics[width=0.205\textwidth]{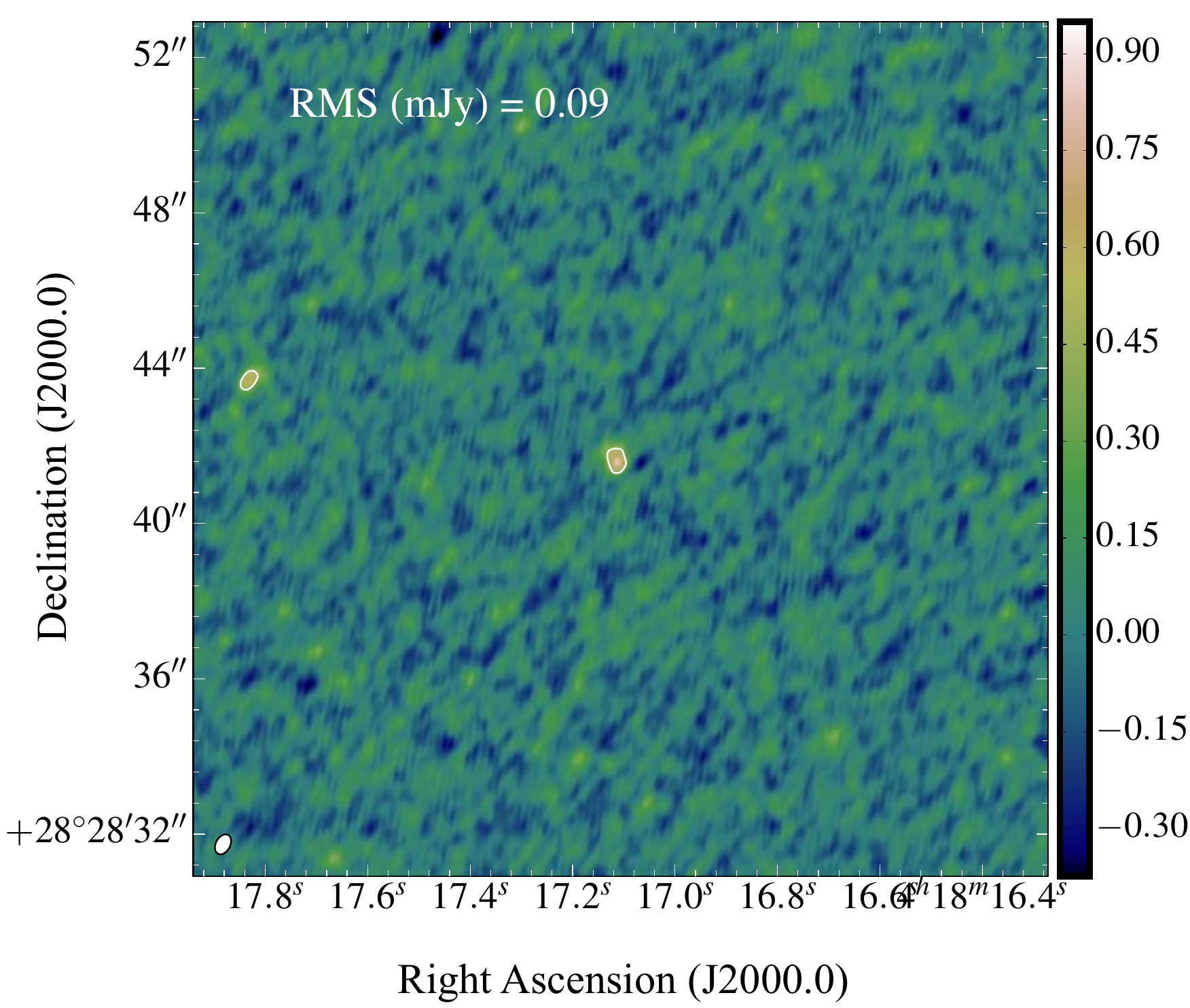}
    \includegraphics[width=0.2\textwidth]{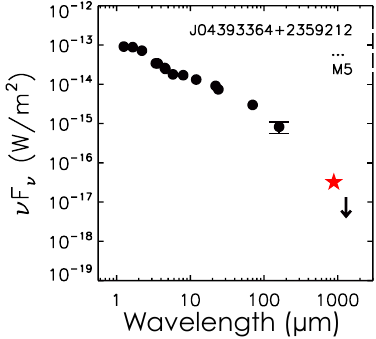}
    \includegraphics[width=0.205\textwidth]{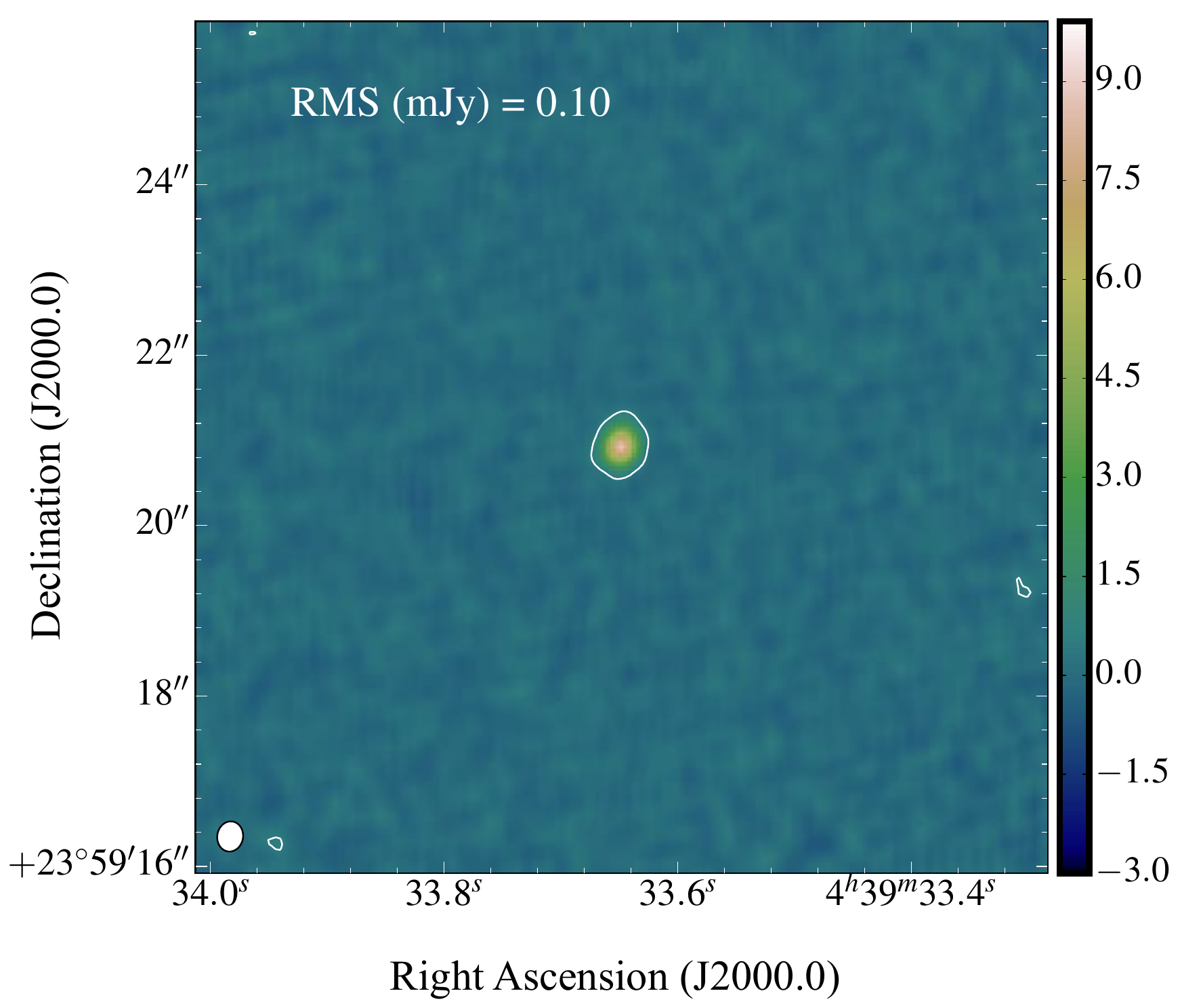}
    
    \caption{SEDs and continuum maps for targets with spectral types M4 -- M5.75. Map intensity corresponds to flux density in mJy. All contours shown are 5$\sigma$. For J04181710, the field of view has been increased to show a wide companion candidate detection. Beam sizes are indicated in the lower left corner with white ellipses, with typical sizes of $0\farcs47 \times 0\farcs38$.}
    \label{fig:gallery1}
\end{figure*}

Among the ALMA-observed TBOSS targets in this sample, three are known binaries \citep{itoh99, konopacky07, kraus12}, two are previously identified as binary candidates \mbox{\citep{kraus12}}, and a target within our sample also shows a 885$\mu m$ detection from a secondary source unassociated with any previously identified companions or candidates. Separations of the components are listed in Table~\ref{tab:binarytable}. For the binary with a separation less than the beam size -- J04292165 -- the continuum emission detection cannot be divided into primary and secondary disks, though the emission appears slightly extended and follow-up higher resolution mapping would determine the relative contributions from each component of the binary system. The total flux density is reported in Table~\ref{tab:fluxes} for this system. Two targets -- J04284263 and J04394488 -- are binaries with separations greater than the beam size. The subarcsecond pair J04284263 is not spatially resolved in the ALMA map in Figure~\ref{fig:gallery1}, while the $\sim$3$''$ pair J04394488 exhibits clear emission from both components. For the system J04181710, a secondary source 9$\farcs$6 in separation from the target was detected at 3$\sigma$; however, a corresponding source has not been previously reported in the literature for this target, making the background or associated nature of the source uncertain. For both the known binary and new candidate detections, the secondary disks are weaker in both cases, and the lower flux densities are reported in Table~\ref{tab:binarytable}. An additional two targets -- J04202555 and J04230607 -- were previously noted as binary candidates with separations $\leq4\farcs6$ \mbox{\citep{kraus12}}. Neither of these candidates are detected in the wider field maps in this study, and the 3$\sigma$ upper limits at the positions of the candidates are included in Table~\ref{tab:binarytable}.

\input{binarytable.txt}

\begin{figure*}
    \centering
    \includegraphics[width=0.2\textwidth]{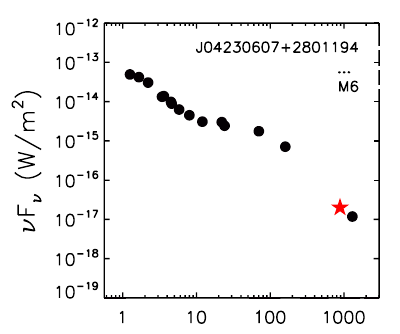}
    \includegraphics[width=0.2\textwidth]{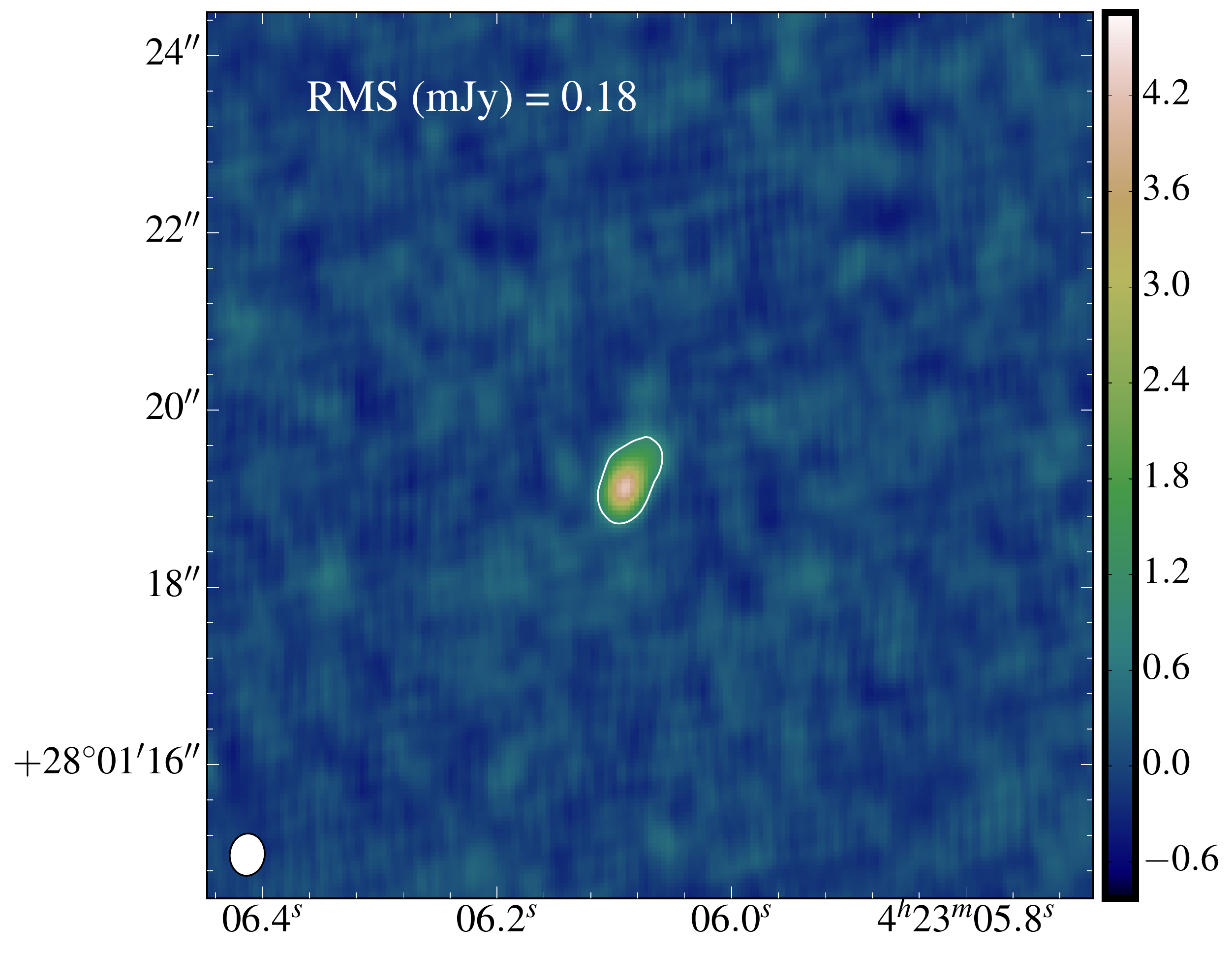}
    \includegraphics[width=0.2\textwidth]{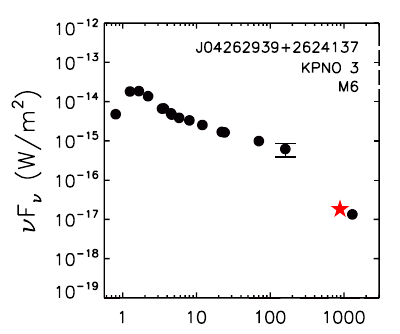}
    \includegraphics[width=0.2\textwidth]{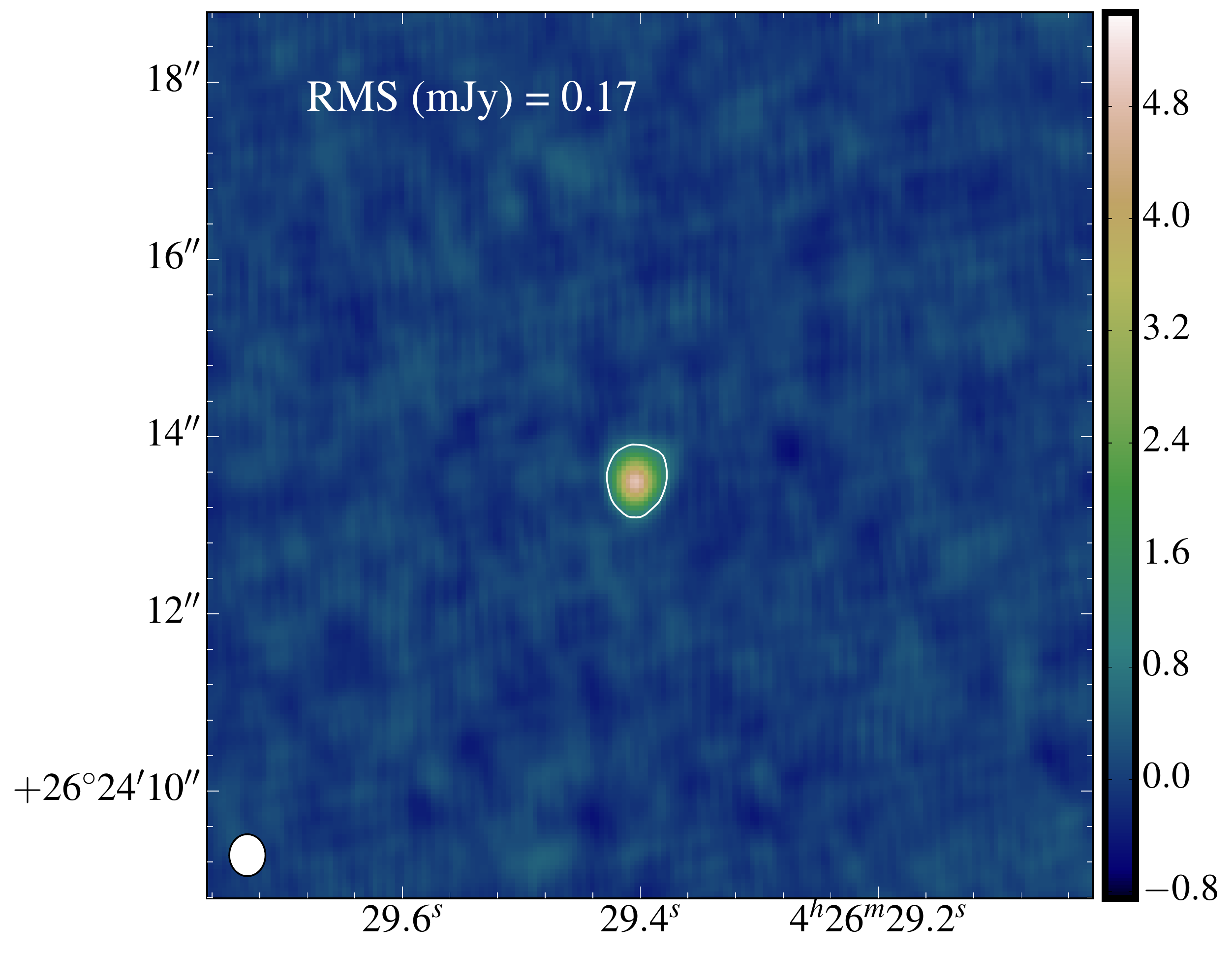}
    \\
    \includegraphics[width=0.2\textwidth]{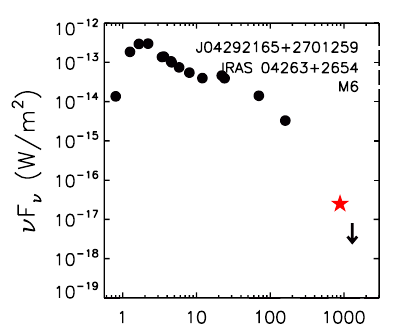}
    \includegraphics[width=0.2\textwidth]{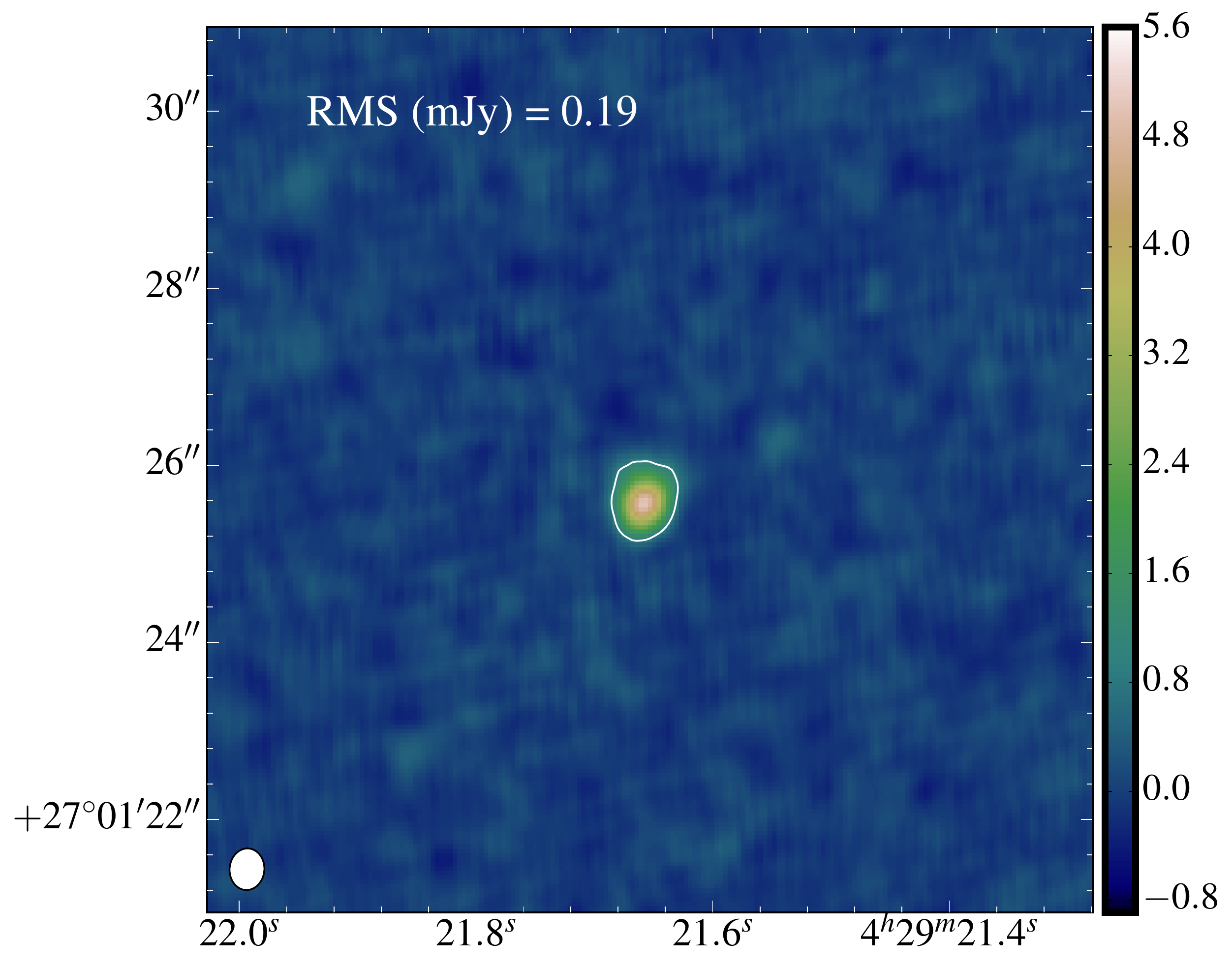}
    \includegraphics[width=0.2\textwidth]{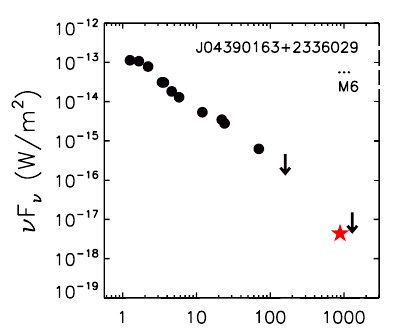}
    \includegraphics[width=0.2\textwidth]{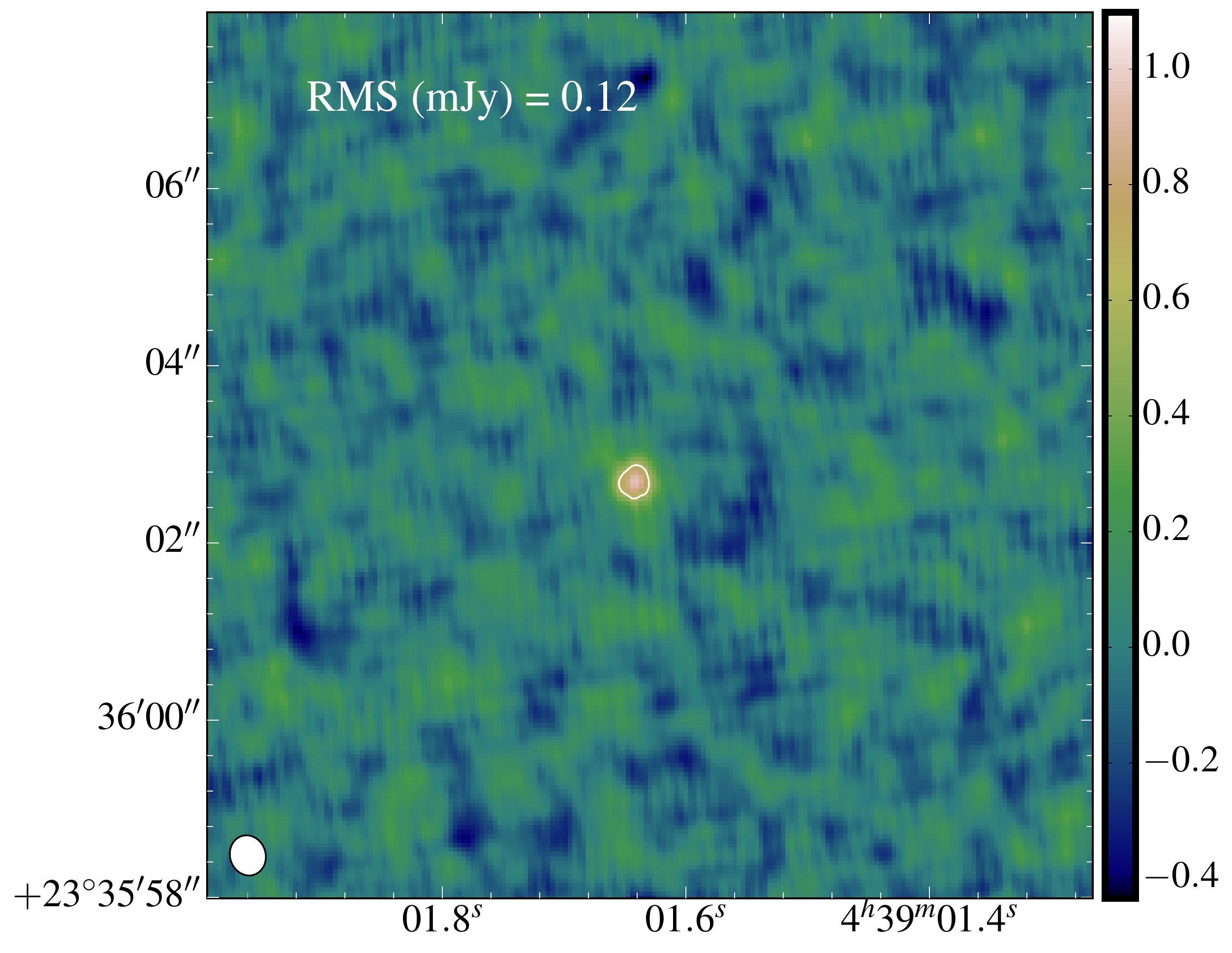}
    \\
    \includegraphics[width=0.2\textwidth]{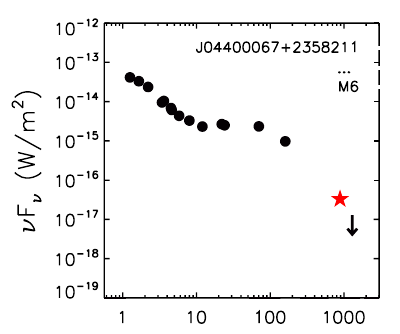}
    \includegraphics[width=0.2\textwidth]{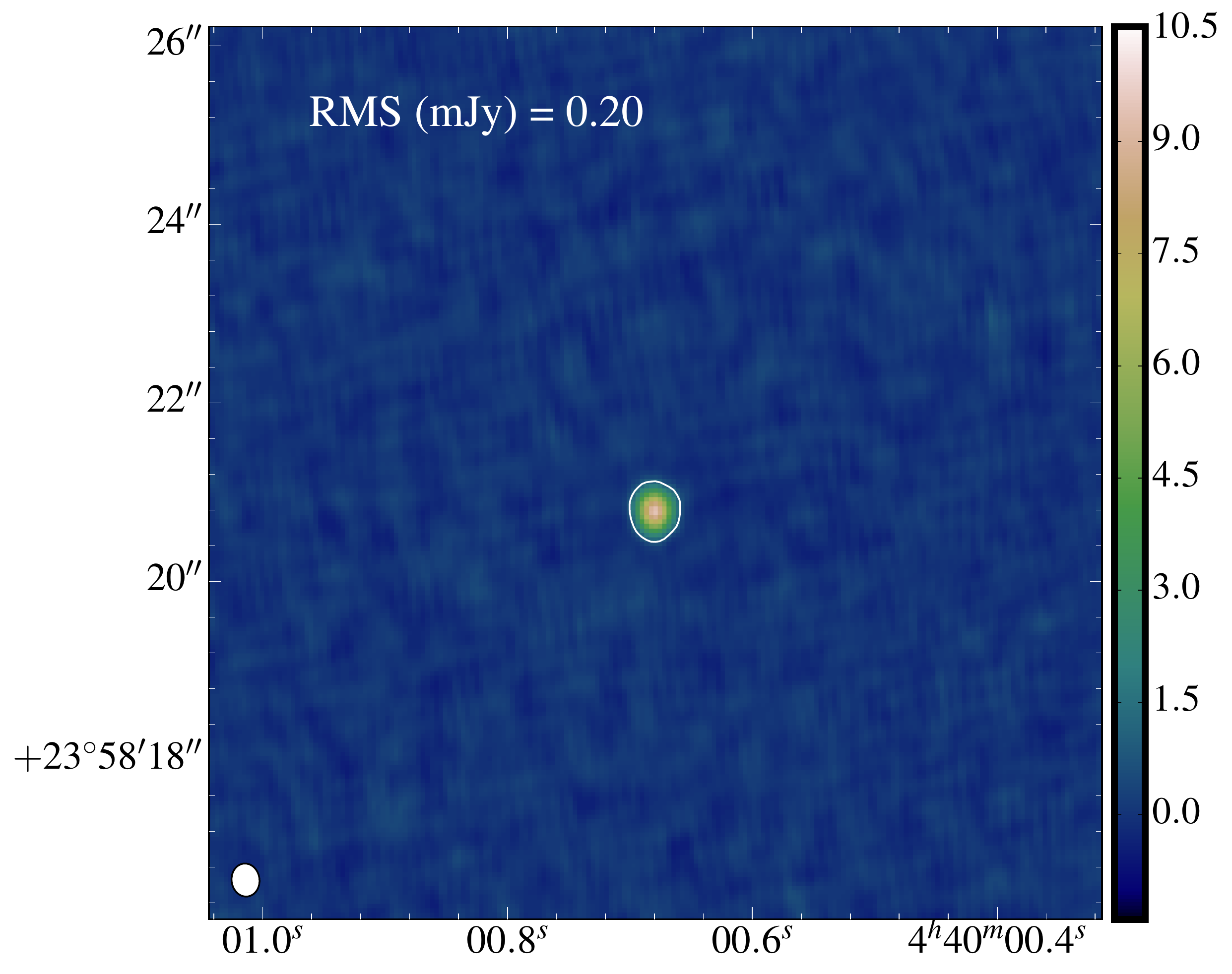}
    \includegraphics[width=0.2\textwidth]{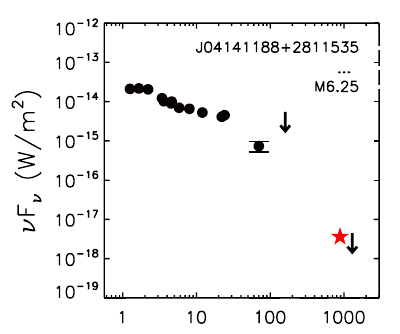}
    \includegraphics[width=0.2\textwidth]{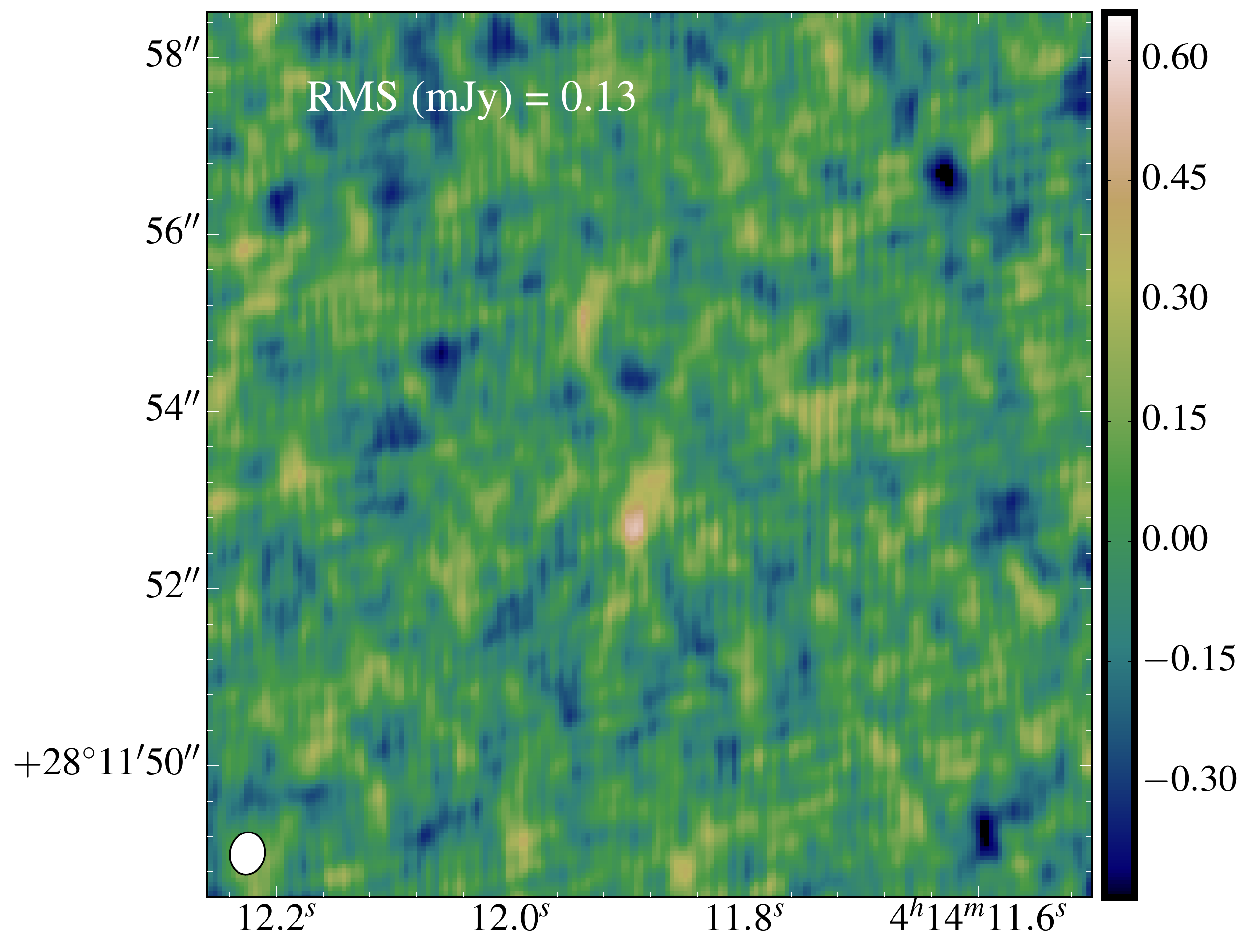}
    \\
    \includegraphics[width=0.2\textwidth]{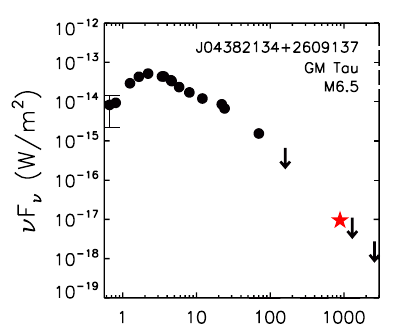}
    \includegraphics[width=0.2\textwidth]{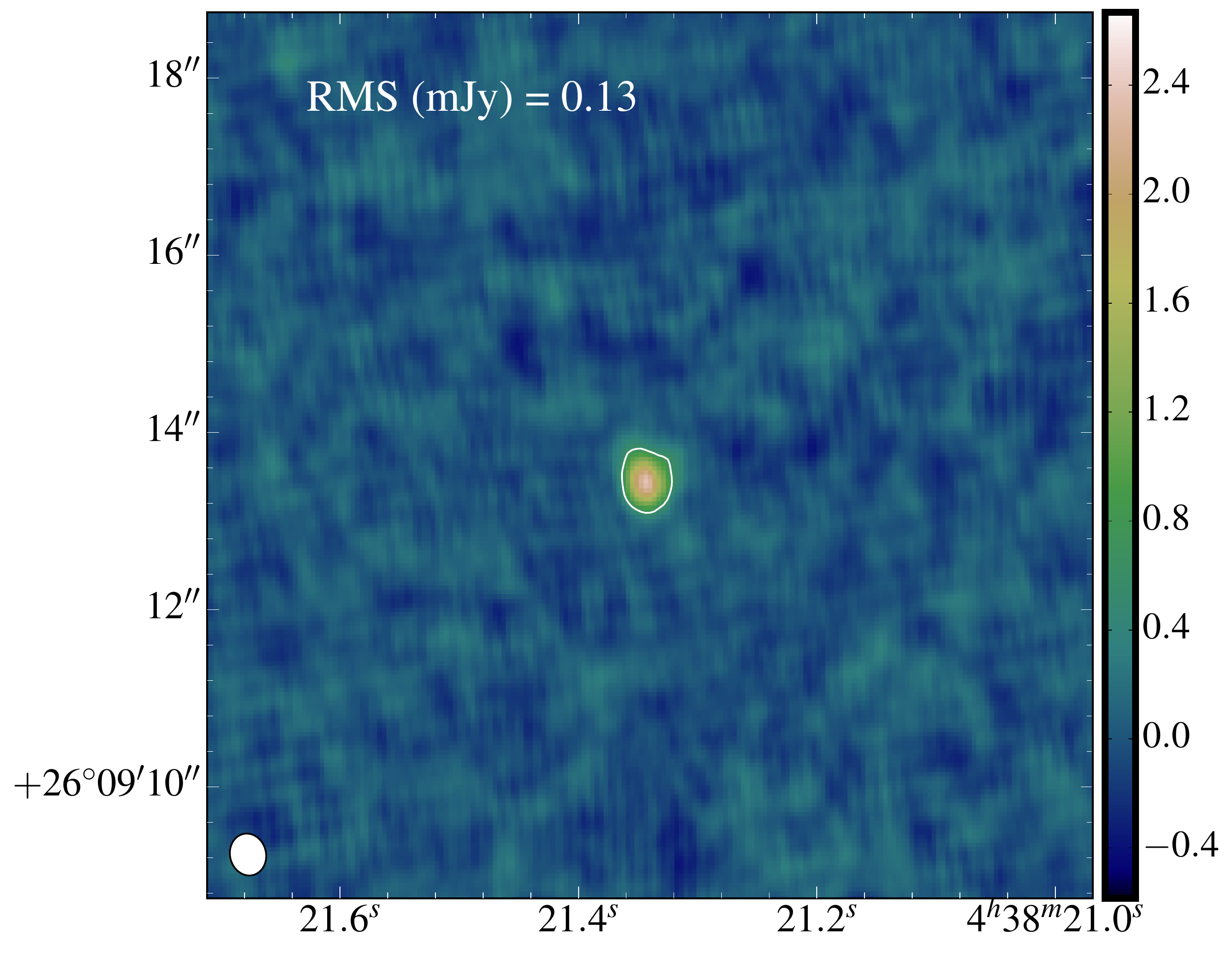}
    \includegraphics[width=0.2\textwidth]{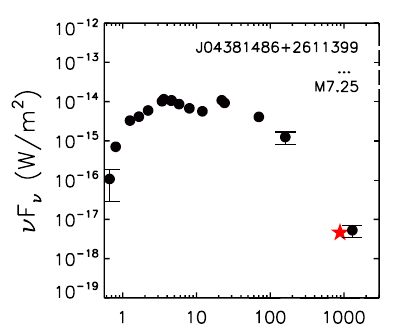}
    \includegraphics[width=0.2\textwidth]{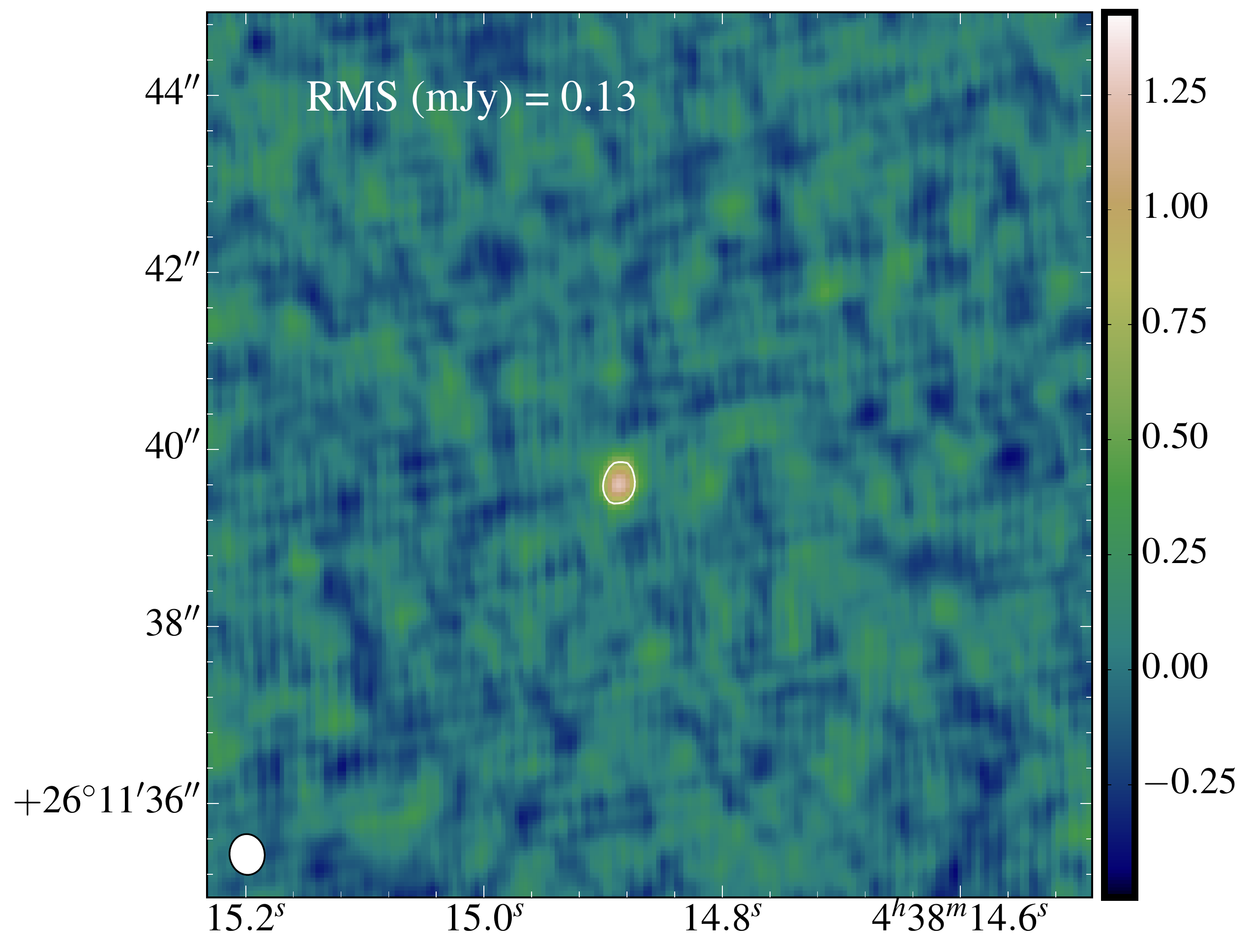}
    \\
    \includegraphics[width=0.2\textwidth]{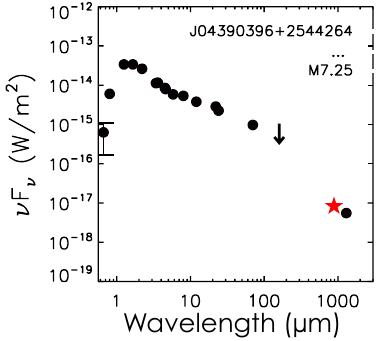}
    \includegraphics[width=0.205\textwidth]{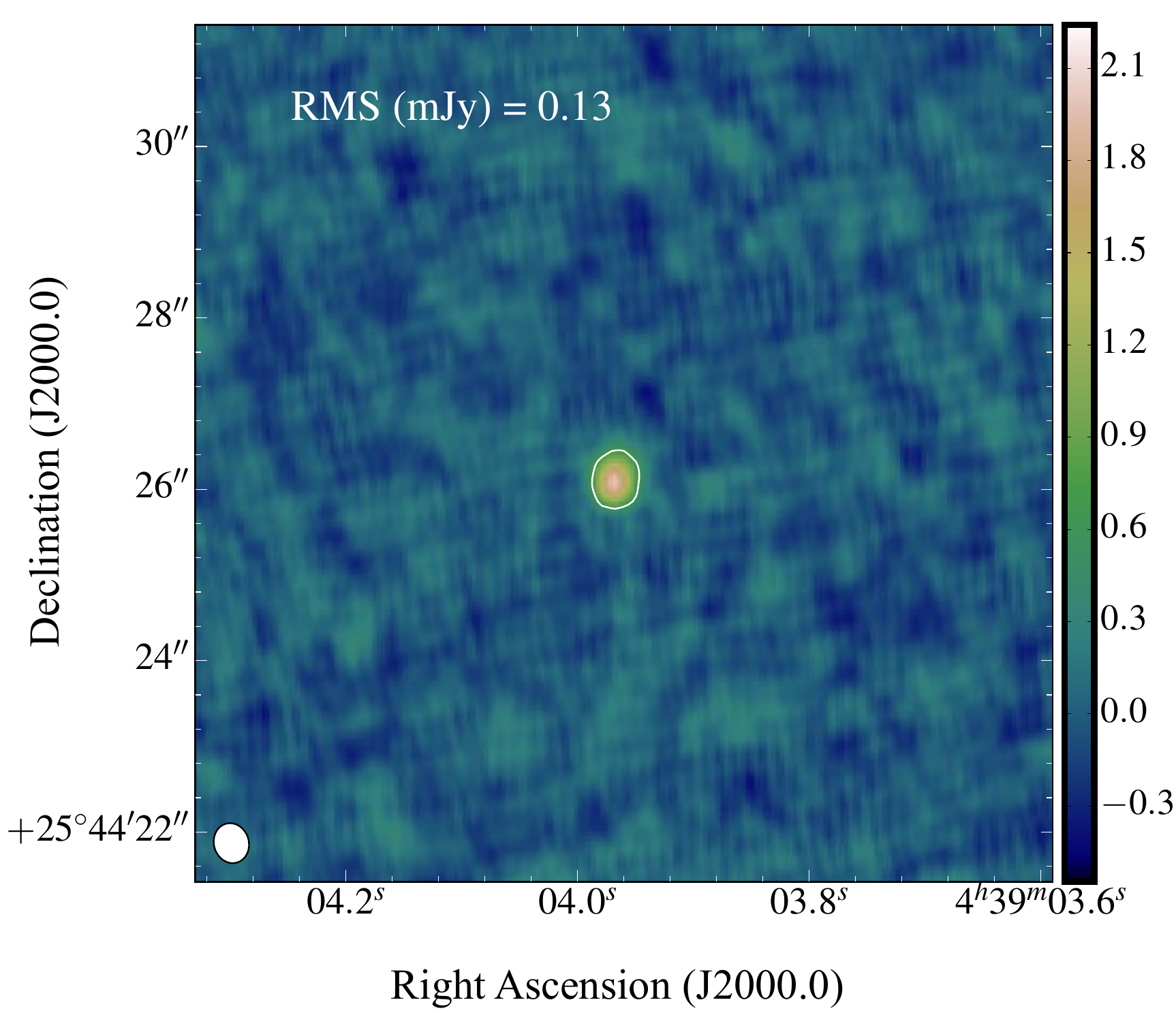}
    \includegraphics[width=0.2\textwidth]{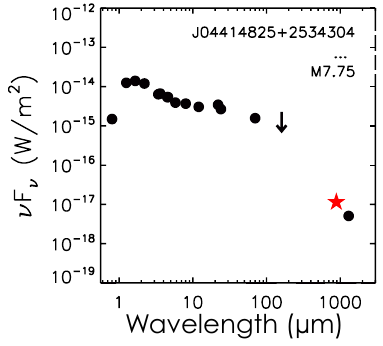}
    \includegraphics[width=0.205\textwidth]{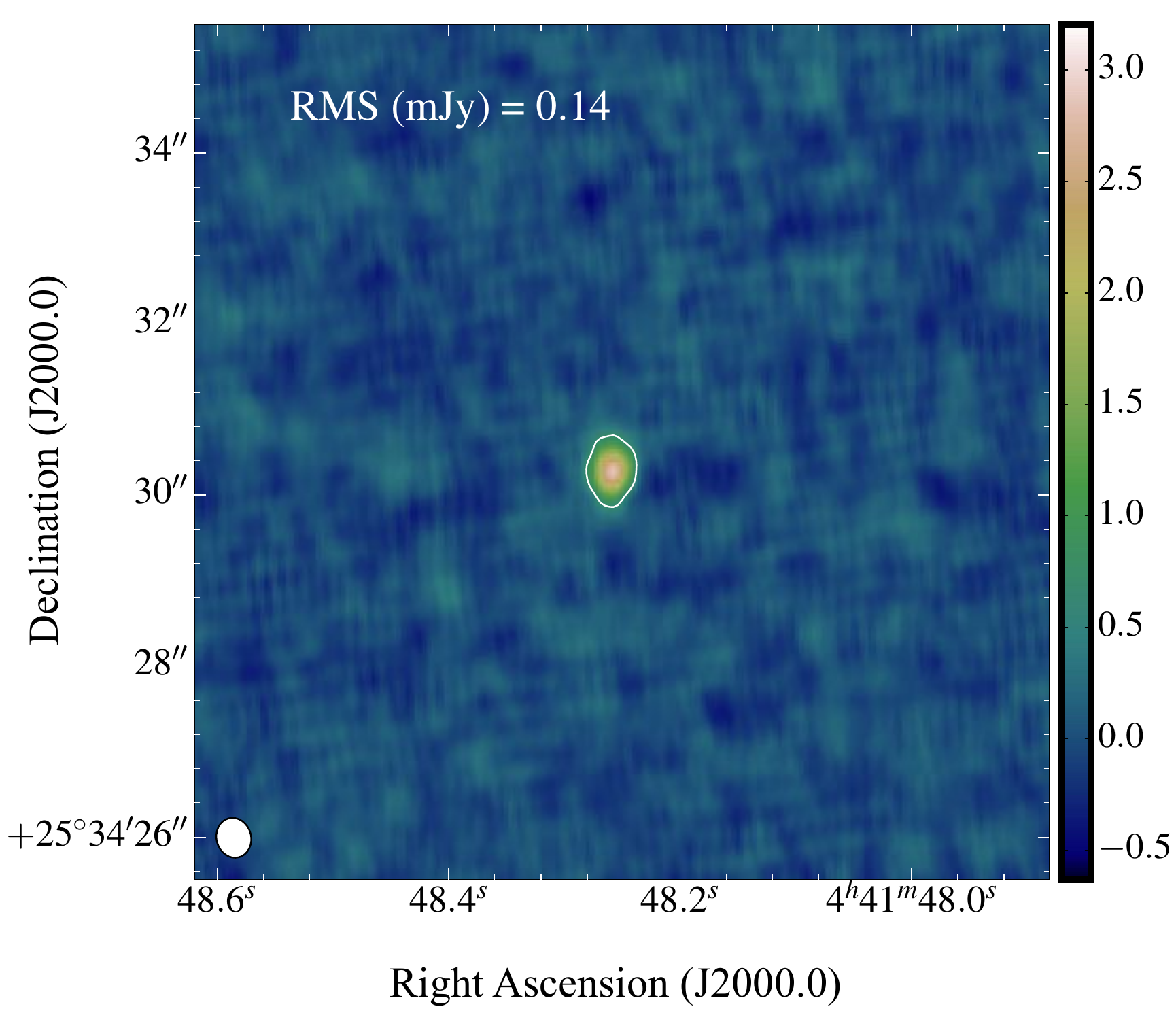}    
    
    \caption{SEDs and ALMA continuum maps for targets with spectral types M6 and later. Map intensity corresponds to flux density in mJy. All contours are 5$\sigma$. Beam sizes are indicated with white ellipses in the lower left corner, with typical sizes of $0\farcs47 \times 0\farcs38$.}
    \label{fig:gallery2}
\end{figure*}

By combining the new 885$\mu$m data with previously reported photometry from the literature \citep[compiled in mJy with original references in][]{bulger14}, the spectral energy distribution (SED) for each source was constructed. Each source SED is presented in Figures~\ref{fig:gallery1} and \ref{fig:gallery2}, along with the associated ALMA continuum map. For the majority of the targets, the ALMA flux density is the only detection in the submm/mm wavelength range critical for estimating disk masses.

\section{Discussion}
\label{sec:discussion}
\subsection{Calculations of Disk Masses from Analytic Relations}
\label{sec:analyticdust}
The Taurus target flux densities reported in Table~\ref{tab:fluxes} are converted into estimates of the disk dust mass through two approaches -- (1) applying flux-mass scaling relations and (2) fitting radiative transfer models to the SEDs including the new ALMA 885$\mu$m values. For this analysis, the natural weighting map fluxes are used for consistency, however the results are not dependent on the procedure applied to determine fluxes as shown in Figure~\ref{fig:fluxcomparison}. The analytic expression utilized to estimate disk masses is:

\begin{equation}
\label{eq:mdust}
\centering
\log M_{dust} = \log S_{\nu}+2 \log d-\log \kappa_{\nu}-\log B_{\nu}(\langle T_{dust} \rangle),
\end{equation}

where $S_{\nu}$ is the ALMA flux density, $d$ is the distance, $\kappa_{\nu}$ is the dust opacity, and $B_{\nu}(\langle T_{dust} \rangle)$ is the blackbody function at the dust temperature \citep{hildebrand83}. 

The first three terms of Eqn.~\ref{eq:mdust} are determined directly from measurements or standard assumptions. The ALMA flux $S_{\nu}$ for each source is given by the natural weighting or self-calibration value in Table 4. A distance to Taurus of 140pc \citep{kenyon94, bertout99, torres09} is used in the calculation. The opacity was scaled to the observation wavelength of 885$\mu$m from the assumptions of $\kappa_{1.3mm}$=2.3cm$^{2}$g$^{-1}$ and $\kappa \sim \nu^{0.4}$; this opacity normalization value and power law relation correspond to the opacity of a standard mixture of astronomical silicates with a maximum grain size $a_\textnormal{max}=1$mm and a grain size distribution following a power law  with slope=-3.5, similar to previous studies \citep{andrews13, carpenter14}.

Different approaches have been used in the literature to estimate the value of $T_{dust}$ needed for the final term of Eqn.~\ref{eq:mdust}. A fixed temperature, typically $\sim$20K, has been applied to early work on Taurus \citep{beckwith90} and recent ALMA surveys of Lupus and Cha I \citep{ansdell16,pascucci16}. A temperature scaling relation based on object luminosity was introduced and applied to surveys of more massive stars in Taurus and Ophiuchus \citep[e.g.,][]{andrews13}:

\begin{equation}
\label{eq:tdust}
\langle T_{dust} \rangle = 25 (L_{*}/L_{\odot})^{1/4} K .
\end{equation}

To estimate the luminosity required for Eqn.~\ref{eq:tdust}, measurements of the object photosphere such as a spectrum or photometric spectral energy distribution are compared with evolutionary models. For this study, we determine the target luminosities given in Table~\ref{tab:fluxes} from a scaled spectral type and effective temperature relation and evolutionary models assuming a fixed age for Taurus, and the procedure is described in further detail in Section~\ref{sec:diskmasses} and Appendix~\ref{sec:starmassestimation}. For low luminosity objects such as the targets in this study, the dust scaling given in Eqn.~\ref{eq:tdust} predicts very low $T_{dust}$ values, with average values of 12~K, comparable to the ambient molecular cloud. The values of $T_{dust}$ from Eqn.~\ref{eq:tdust} and the corresponding $M_{dust}$ are reported in Appendix~\ref{sec:staranddiskparams}.

To avoid the unphysically low temperatures implied by Eqn.~\ref{eq:tdust}, a different temperature-luminosity relation  more appropriate for samples extending to spectral types of $\sim$M5 and later was used, as explored in our previous paper \mbox{\citep{gvdp16}}:

\begin{equation}
\label{eq:newtdust}
\langle T_{dust} \rangle = A (L_{*}/L_{\odot})^{B} K
\end{equation}

Both the normalization factor $A$ and the power law index $B$ in Eqn.~\ref{eq:newtdust} vary depending on a number of factors, with the assumed outer radius of the disk being the dominant parameter; the coefficients $A$ and $B$ for different outer radii are reported in Table~\ref{tab:tdust}. For the subsequent analysis in the paper, the analytic estimate of the disk dust mass is based on Eqn.~\ref{eq:newtdust}, and we explore a range of radii from 10~au to 200~au. The full range of $T_{dust}$ and $M_{dust}$ for each target assuming different radii are given in Appendix~\ref{sec:staranddiskparams}, and a subset of values are listed in Table~\ref{tab:abridged_dustmasses}. As expected, the differences are most pronounced for the lowest luminosity objects, with variation in dust mass of $\sim$2.5$\times$ between the 40~au disks and 200~au disks. To account for a range of possible disk sizes, the $M_{dust}$ uncertainties incorporate both the $\pm$10\% flux scaling and sizes of $\pm$tens of au about a central disk size; we explore cases with central disk sizes of 100~au for all objects (used in previous studies), and cases with central disk size of 40~au or 20~au for the lower mass objects and 100~au for the higher mass objects.

\input{tdust_relations.txt}

\input{abridged_dustmass.txt}

\subsection{Calculations of Disk Masses from Radiative Transfer Models (MCFOST)}
\label{sec:mcfost}
The final approach to determining disk masses from the ALMA measurements involves a combination of the ALMA data with photometry at other wavelengths and a comparison with models generated with the Monte Carlo 3D continuum radiative transfer code MCFOST \mbox{\citep{pinte06, pinte09}} which produces synthetic SEDs. In the MCFOST routines, photons from the central object are propagated through the disk with a model incorporating a combination of scattering, absorption, and re-emission. The MCFOST parameters related to the central source are the central object effective temperature $T_\textnormal{eff}$, object radius $R_{∗}$, and luminosity $L_{*}$. These values are listed for each source in Table~\ref{tab:mcfost_stellar}, where the stellar radius and value of $A_\textnormal{v}$ for each source were derived with SED fitting in the previous \textit{Herschel} TBOSS study by \mbox{\citet{bulger14}}. The effective temperatures were estimated from the spectroscopically-determined spectral types reported in the literature (references in Table~\ref{tab:sample}) and the temperature scales from \mbox{\citet{luhman05_ic348disks}} and \mbox{\citet{kenyonhartmann95}}. A set of 9 parameters are used to define a disk structure and dust population and 5 are varied over ranges reported in Table~\ref{tab:mcfostparams}: dust mass $M_\textnormal{dust}$, inner radius $r_\textnormal{in}$, outer radius $r_\textnormal{out} = 100$AU, scale height $H_{0}$ at a reference radius $r_{o}$, flaring profile exponent $\beta$ for the disk height $H(r) \sim r^{\beta}$, surface density profile index $b$ where $\Sigma(r) \sim r^{b}$, minimum grain size $a_\textnormal{min} = 0.01\mu$m, maximum grain size $a_\textnormal{max} = 3$mm, and the grain size distribution $N(a) \sim a^{-3.5}$, with a corresponding continuum opacity $\kappa = 2.78\textnormal{cm}^{2}/\textnormal{g}$ at 870$\mu$m. The final parameters are the disk inclination $i$ and the reddening $A_\textnormal{v}$. Since none of the objects are in the more embedded Class I phase, a single continuous disk model was used, with no envelope component.

\input{mcfost_stellarparams_new.txt}

\begin{figure}
    \centering
    \includegraphics[scale=0.42]{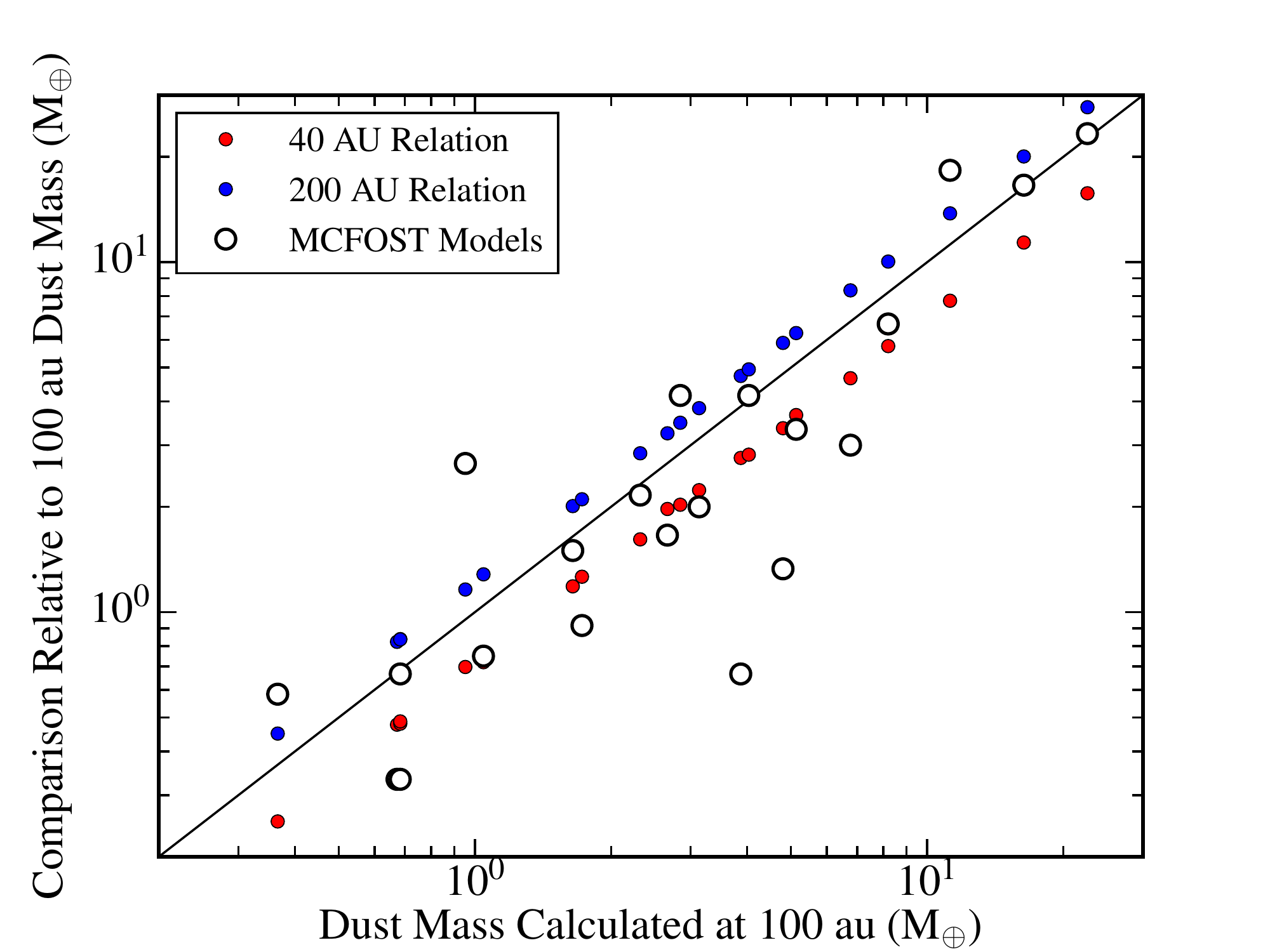}
    \caption{Comparison of the MCFOST model disk dust masses and the analytically-derived masses, calculated as described in Section~\ref{sec:analyticdust}. Estimated analytic masses assuming disk radii of 40 and 200~au (red and blue circles, respectively), and the masses derived from MCFOST radiative transfer modeling (open circles) are compared against the analytic result for a 100~au disk case on the x-axis. The black line represents the 1 to 1 relationship for the 100~au case plotted against itself. The MCFOST model results agree well within the ranges of masses inferred from the 40-200~au analytic estimates, and appear more consistent with the 40~au disk dust masses.}
    \label{fig:dustmass_modelanalytic}
\end{figure}

We apply a genetic algorithm approach, previously employed in \mbox{\citet{mathews13}}, to explore five free model parameters -- M$_\textnormal{dust}$, $H_{0}$, $r_\textnormal{in}$, $\beta$, and surface density index. These parameters are iteratively varied over a range of values to construct a minimal $\chi^{2}$ distribution. For each target, the genetic algorithm begins with an initial generation of models uniformly sampled over the free parameter minimum and maximum ranges given in Table~\ref{tab:mcfostparams}, and calculates $\chi^{2}$ values for each model. A successive generation of models is then generated by selecting from the previous generation of parent models, with parameters randomly sampled from the parent model parameters. Within the successive generation, a ``mutated'' subset of models is created by varying one-tenth of the parent parameter ranges for a fraction of models. The process is continued for following generations, with the range of parameter variation and mutation rate dependent upon the resulting $\chi^{2}$ values, optimizing to more densely sample the parameter space near the minimum of the distribution. The best-fit parameter values corresponding to the minimum $\chi^{2}$ for each SED fit are listed in Table~\ref{tab:mcfostresults}, and the dust masses are compared with the analytically-derived masses in Figure~\ref{fig:dustmass_modelanalytic}. SEDs with the resulting best-fit MCFOST models are provided in Appendix~\ref{sec:mcfostseds} for each of the stellar and brown dwarf targets (Figures~\ref{fig:mcfost_seds} and \ref{fig:mcfost_seds_BDs}, respectively).

\begin{table}
\centering
 \caption{MCFOST Model Parameter Ranges}
 \label{tab:mcfostparams}
 \begin{tabular}{lcc}
  \hline
  \hline
  Parameter & Minimum & Maximum\\
  \hline
  Disk Mass, M$_\textnormal{dust}$ & $10^{-8}$ & $10^{-4}$\\
  Scale Height, $H_{0}$ & 5 & 25\\
  Inner Radius, $r_\textnormal{in}$ & 0.01 & 1.0 \\
  Disk Flaring Index, $\beta$ & 1.0 & 1.3\\
  Surface Density Index & -1.5 & 0.0\\
  \hline
 \end{tabular}
\end{table}

\input{mcfost_results.txt}

\subsection{Disk Mass as a Function of Central Object Mass}
\label{sec:diskmasses}

The disk masses determined from the new ALMA data represent the lowest mass component of the Taurus population and can be placed in the context of the full spectrum of disks by combining with previous results on higher mass Taurus members. The results from an SMA snapshot survey combined with previous single dish measurements provide a catalog of measured or extrapolated 890$\mu$m flux densities for a sample of 179 Taurus systems \citep{andrews13}, to which the 24 ALMA results are added. The stellar mass of each Taurus member observed in either study is determined by relating the spectral type of the target to a corresponding effective temperature scaling from \citet{herczeg_hillenbrand14}, and a comparison of the evolutionary models of \citet{baraffe98} and \citet[hereafter BHAC15]{baraffe15}, and the MESA models for higher mass targets \citep{choi16}. Estimation of central object mass via spectral type has been performed in previous studies \citep[e.g.,][]{kraus07, pascucci16}, either alone or in tandem with other mass estimation approaches (e.g., model comparison with SED estimates of temperature and luminosity). In this study, we adopt a uniform mass estimation approach for all objects based on spectral type to avoid ambiguities in luminosity/age estimation due to the presence of edge-on disks. Further description of the mass and luminosity estimation method for the central stars/brown dwarfs is provided in greater detail in Appendix~\ref{sec:starmassestimation}.

\begin{figure*}
    \centering
    \includegraphics[width=0.49\textwidth]{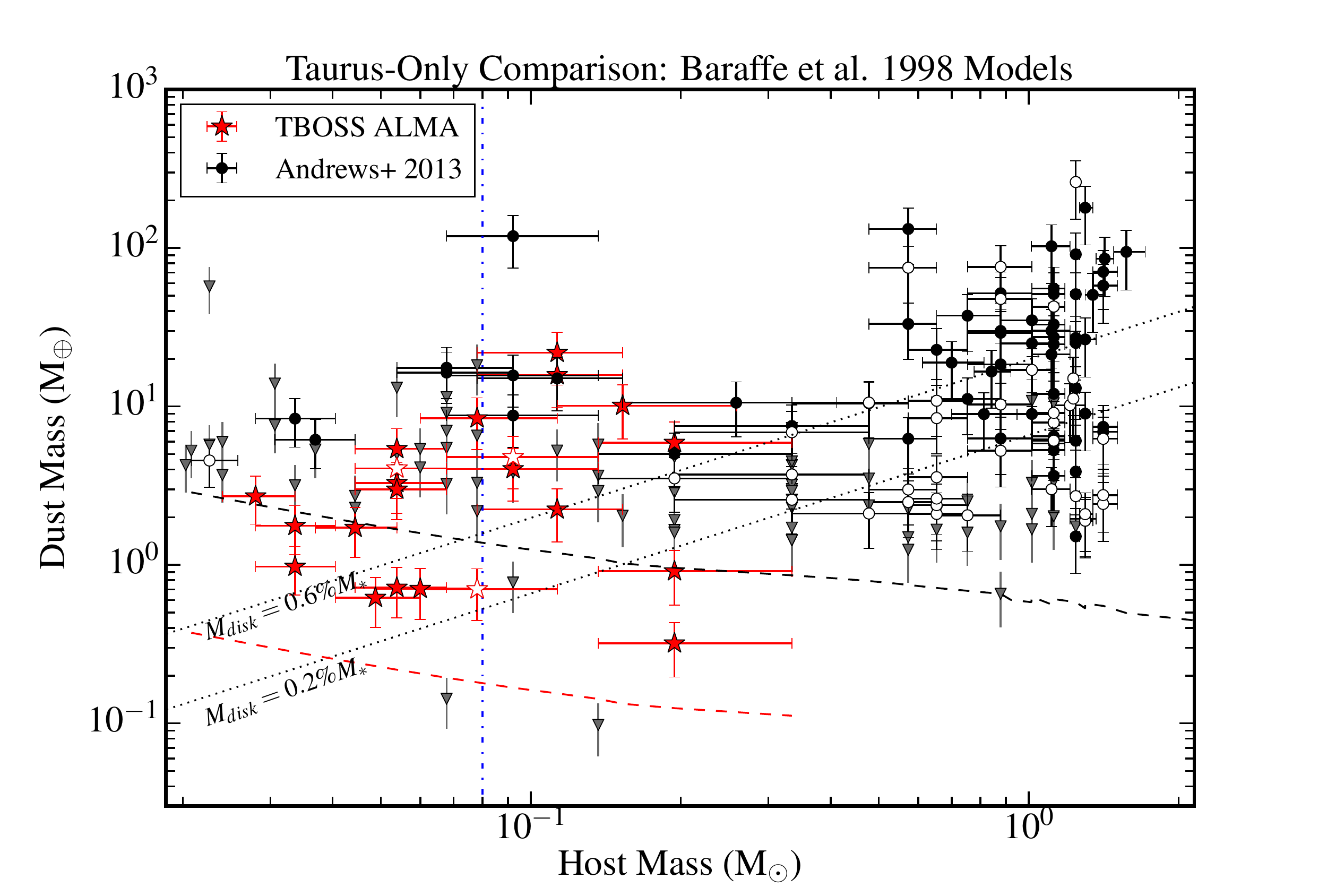}
    \includegraphics[width=0.49\textwidth]{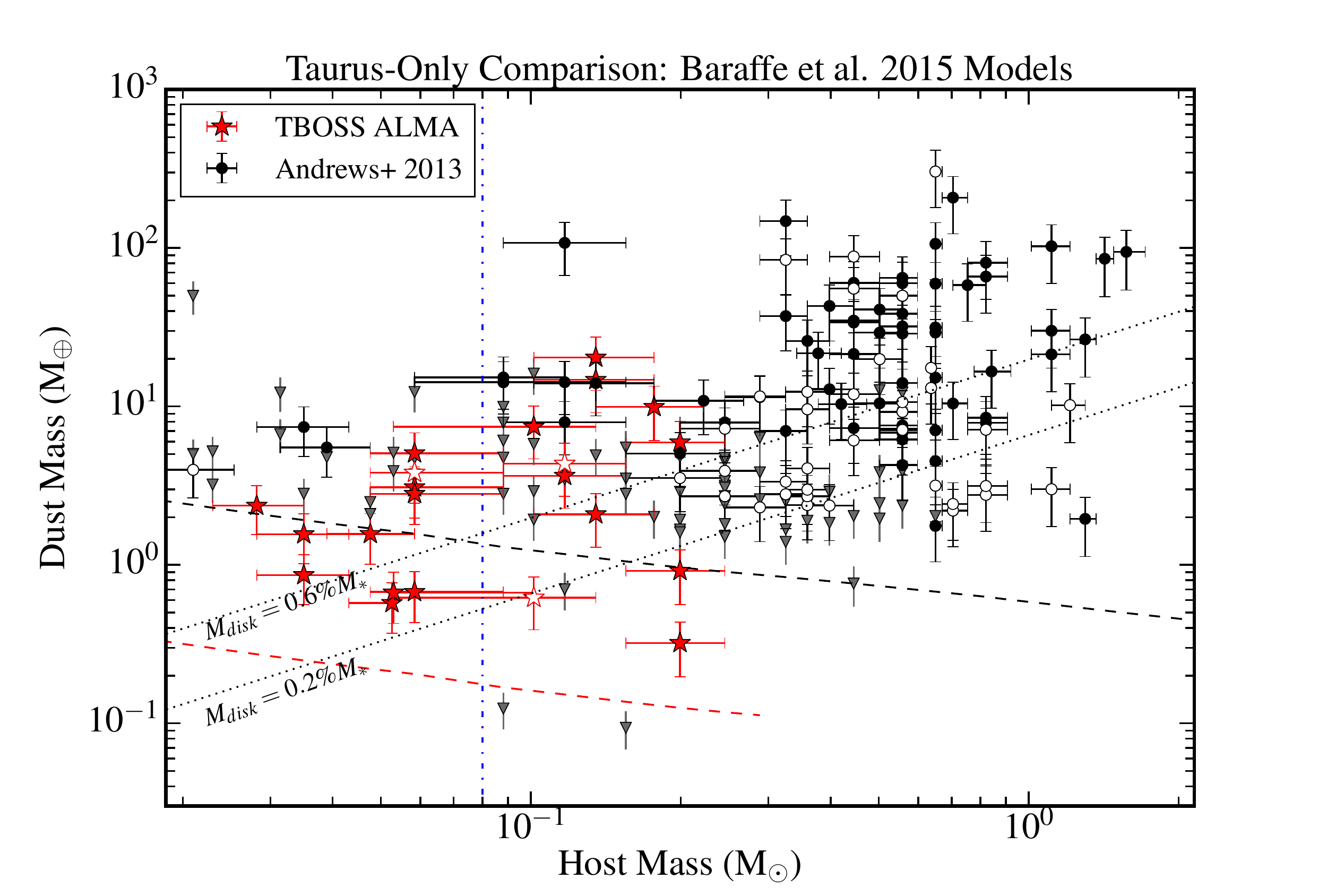}
    \caption{Taurus-only disk dust mass vs. object mass for detections within our sample (red stars) and the full Class II Taurus population with sub-mm detections from \citet{andrews13} (black points). Stellar parameters are derived from spectral types and the evolutionary models of \citet{baraffe98} (left figure) and \citet{baraffe15}  (right figure), assuming an age of 1~Myr. The x-axis errorbars correspond to the possible range of derived stellar masses assuming $\pm$0.5 subclass error on the spectral type. The y-axis errorbars correspond to the range of dust mass within the disks, assuming at minimum a disk radius of 40~au (lower limit) and maximum of 200~au (upper limit), and incorporate a 10~\% absolute flux calibration uncertainty. Open points correspond to identified binaries. Upper limits are provided as downward triangles, with the range denoting disk masses evaluated at disk radii of 40, 100, and 200~au. Overlaid in dashed lines are the 3$\sigma$ sensitivity limits for our survey (0.39 mJy; red line) and \citet{andrews13} (3 mJy; black line). Also shown are the lines of disk mass proportional to stellar mass (dotted black lines), and the stellar/substellar boundary at 0.08M$_{\odot}$ (blue vertical dot-dashed line).}
    \label{fig:taurusonlymasses}
\end{figure*}

The masses adopted from the new BHAC15 and MESA models are updated from those reported in the \citet{andrews13} compilation, which utilized an older suite of models \citep{dantona_mazzitelli97, baraffe98, siess00} that yield systematically lower masses at lower luminosities and higher masses at higher luminosities. The disk masses of the sources detected with the SMA or single dish surveys are estimated with Eqns.~\ref{eq:mdust} and \ref{eq:newtdust} and plotted on Figure~\ref{fig:taurusonlymasses} as a function of object mass, utilizing the dust temperature-luminosity scaling described in Section~\ref{sec:analyticdust}. In Figure~\ref{fig:taurusonlymasses}, the uncertainties in dust mass are derived from dust temperatures incorporating a range of disk sizes centered at 100~au disks, with the lower estimate of dust mass corresponding to 40~au disks and upper estimate corresponding to 200~au disks, and include the impact of a 10\% systematic uncertainty in flux.

Like the more massive host stars, the low mass ALMA-detected sources exhibit a large spread in disk mass for a given host mass, since the sensitivity limit is sufficient to detect most disks and not only the upper envelope of sources. To gauge the decline in disk mass as a function of central object mass, two comparison lines assuming a gas to dust ratio of 100:1 are also plotted, representing disks of 0.2\% and 0.6\% of the mass of the central object. The 0.2\%--0.6\% range, corresponding to the average scaling factor for the linear M$_\textnormal{disk}$ $\sim$ M$_\textnormal{star}$ range found by \citet{andrews13}, intercepts the median high-mass Taurus targets and the least massive disks for the lowest-mass hosts. With the large dispersion in dust mass at any given stellar mass, significant populations exist above and below the relations. 

Best-fit power laws to the detections and upper limits for the Taurus population are shown in Figure~\ref{fig:powerlaw1} (red points and lines), applying the Bayesian linear regression approach of \citet{kelly07} to incorporate both detections and upper limits. With greater numbers of targets at lower host masses, the Taurus best fit relation of $\textnormal{log}[M_\textnormal{dust} (M_{\oplus})] = (0.97\pm0.14) \textnormal{log}[M_\textnormal{star}(M_{\odot})] + (1.15\pm0.09)$ with an intrinsic scatter of $0.49$ dex in log$[M_\textnormal{dust} (M_{\oplus})]$ is consistent with a linear relation, similar to the relations reported for disks around Taurus stellar hosts in \citet{andrews13}, and the TBOSS data are consistent with the general trend of decreasing disk mass with declining central object mass, suggesting a common formation mechanism across the full mass spectrum.

\subsection{Disk Mass as a Function of Time and Environment}

\begin{figure*}
    \centering
    \includegraphics[width=0.49\textwidth]{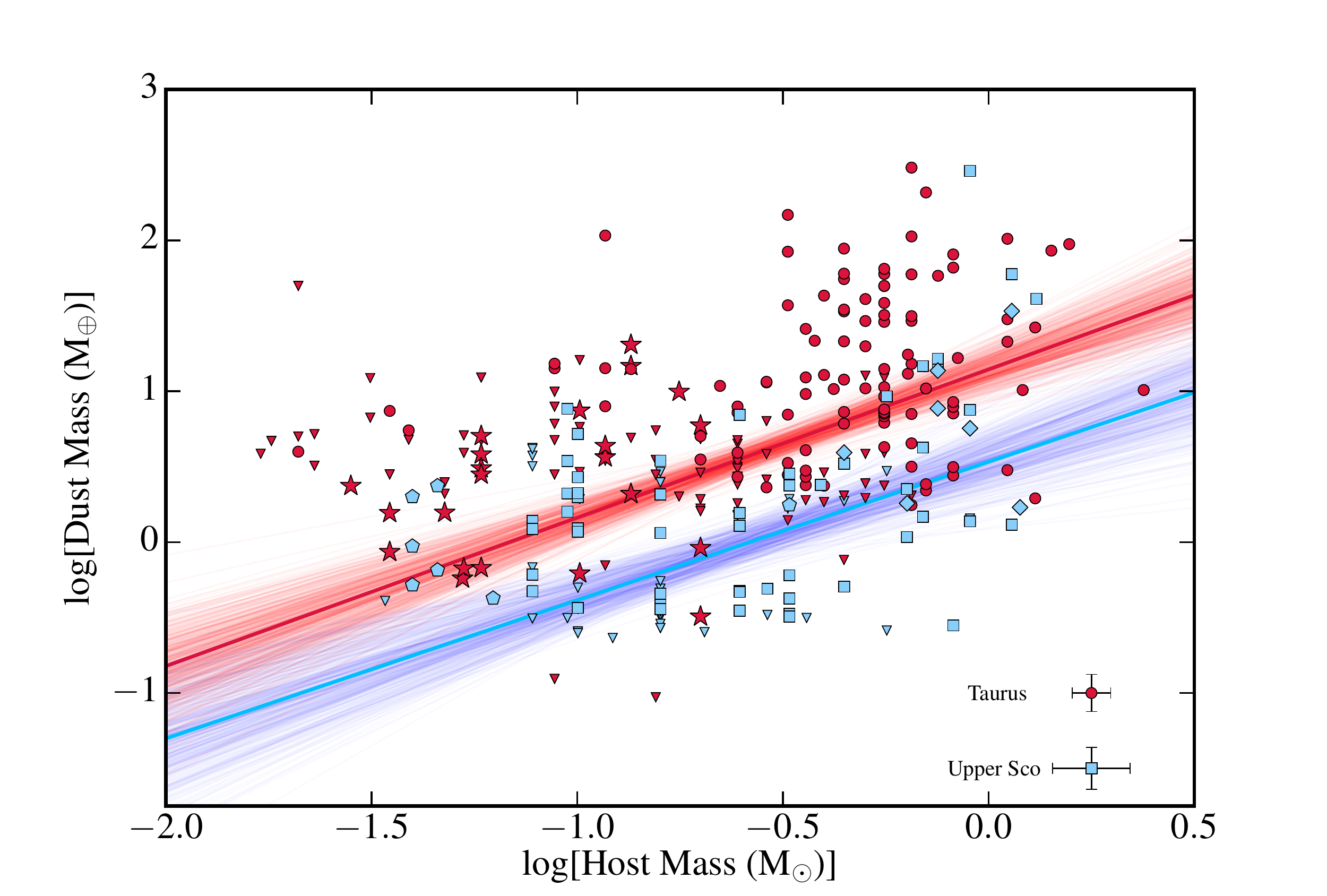}
    \includegraphics[width=0.49\textwidth]{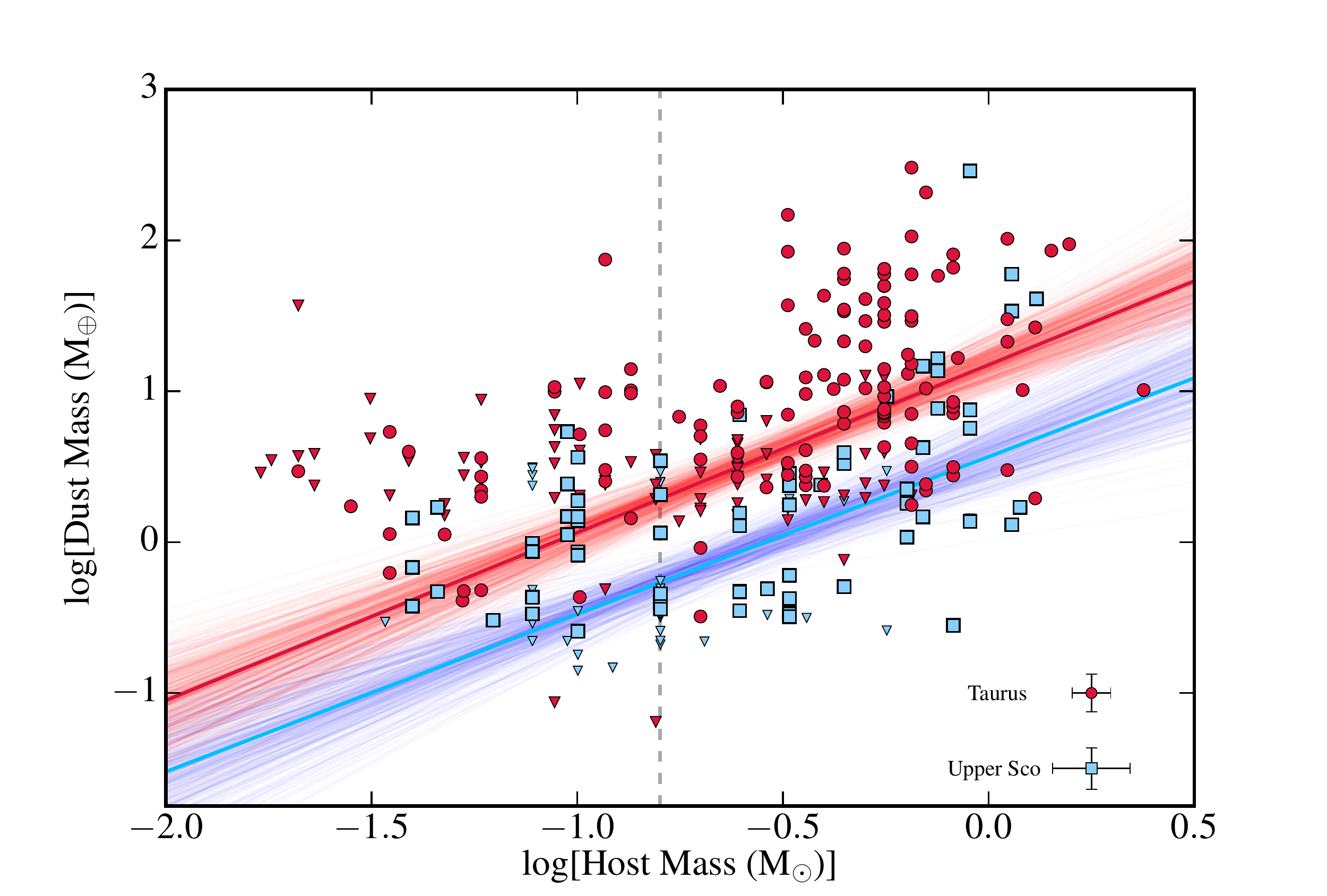}
    \caption{Disk dust mass as a function of stellar host mass for Taurus and Upper Sco with overlaid power-law fits to combined detections and upper limits.  (\textit{Left}) With a single disk size of 100~au for all objects and uncertainties (in the corner) incorporating disk sizes ranging from 40--200~au, the best-fit linear regression for Taurus is $\textnormal{log}[M_\textnormal{dust} (M_{\oplus})] = (0.97\pm0.14) \textnormal{log}[M_\textnormal{star}(M_{\odot})] + (1.15\pm0.09)$ with $0.49$ dex of intrinsic scatter (red lines), and for Upper Sco, $\textnormal{log}[M_\textnormal{dust} (M_{\oplus})] = (0.92\pm0.18) \textnormal{log}[M_\textnormal{star}(M_{\odot})] + (0.46\pm0.09)$ with $0.54$ dex of intrinsic scatter (blue lines). Symbols for combined studies include this work (stars) and \citet{andrews13} (circles) for Taurus in red, and \citet{gvdp16} (pentagons), \citet{barenfeld16} (squares), and \citet{mathews12} (diamonds) in blue for Upper Sco. The slopes between the Taurus and Upper Sco populations are similar within uncertainties. Dust mass and stellar mass estimations assume a population age of 10 Myr for Upper Sco vs. 1 Myr for Taurus. The three previous Upper Sco surveys cover a wide range of stellar masses and have significantly lower dust masses, corresponding to approximately 0.5 dex decrease between the two populations. (\textit{Right}) Assuming disk sizes of 40~au for targets M4 and later, and 100~au disks for $<$ M4, the slopes are slightly steeper ($1.11\pm0.14$ for Taurus; $1.05\pm0.18$ for USco), but agree with the 100~au case within uncertainties. The uncertainties (in the corner) include a range of disk sizes from 20--100~au.}
    \label{fig:powerlaw1}
\end{figure*}

To investigate the evolution of the disk dust mass, dust mass as a function of host mass is also plotted for the region of Upper Sco in Figure~\ref{fig:powerlaw1} (blue points and lines). The Taurus component is the same as in Figure~\ref{fig:taurusonlymasses}, described in Section~\ref{sec:analyticdust}. To explore the full range of stellar masses for targets in Upper Sco, a compilation of studies is used for comparison, with values drawn from a single dish IRAM survey of high-mass Upper Sco members \citep{mathews12} and a large recent ALMA study \citep{barenfeld16}. For the lowest-mass hosts, the results from the Taurus ALMA sample are compared with our ALMA pilot study of brown dwarf Upper Sco members \citep{gvdp16}. Both samples of brown dwarfs are too small in number and too biased toward detections to address the frequency of submm-detected disks over time, but the measured flux densities converted to disk masses can be used to study how the mass changes with age. Dust masses for all targets in Upper Sco were re-estimated with a self-consistent approach using Eqns.~\ref{eq:tdust} and \ref{eq:newtdust} (see Appendix~\ref{sec:DustMassComparisons}). While a considerable range of disk masses is present for any given object mass and the lowest mass systems in Taurus overlap with the highest mass examples in Upper Sco, there is a clear drop in the overall disk mass level with time. The ages of the two samples, with $\sim$1-2 Myr for Taurus \citep[e.g.,][]{kraus09} and $\sim$5-10 Myr for Upper Sco \citep{blaauw78, pecaut12}, cover important timescales in planet formation and disk evolution, including formation of giant planets by gravitational instability \citep[$<$1 Myr;][]{boss97} or core accretion \citep[$\sim$10 Myr; e.g.][]{pollack96}, the onset of terrestrial planet formation \citep[$\sim$3-10 Myr;][]{chambers_wetherill98}, and the dissipation of gas-rich primordial disks \citep[$\sim$3 Myr;][]{luhman2010}.

Applying the same linear regression analysis to the Upper Sco populations, the best-fit Upper Sco power law relation of $\textnormal{log}[M_\textnormal{dust} (M_{\oplus})] = (0.92\pm0.18) \textnormal{log}[M_\textnormal{star}(M_{\odot})] + (0.46\pm0.09)$ with an intrinsic scatter of $0.54$ dex in log$[M_\textnormal{dust} (M_{\oplus})]$ has a slope similar to that of the Taurus population fit in Section~\ref{sec:diskmasses} within uncertainties, and the combined populations are shown in Figure~\ref{fig:powerlaw1}. The comparison between intercepts of the fits to each of the two regions suggests a decline in disk mass by a factor of $\sim$4-5 over the critical $\sim$1-10 Myr time period between Taurus and Upper Sco, similar to the conclusion reached in previous studies \citep{ansdell16}. The total gas and dust disk mass decline is probably significantly larger than indicated by the drop in fit intercept values, as the gas to dust ratio likely evolves over time since Upper Sco targets typically only have upper limits \citep{gvdp16}. 

To measure the impact of adopting 100~au disk sizes for all of the objects, the Taurus and Upper Sco samples were broken into separate subsets at the M4 spectral type. Smaller disk radii of either 20~au or 40~au were then assumed for the M4 and later spectral types, with uncertainties corresponding to disk sizes from 10-100~au in the 20~au case, or ~20-100~au in the 40~au case. Figure~\ref{fig:powerlaw1} shows the fit to the populations with the 40~au disk size for lower mass objects. The slopes from the tests are listed in Table~\ref{tab:slopes}, showing that the results are within the uncertainty of the fit with the assumption of 100~au disks for all object masses. Regardless of the assumed disk size for the low mass component of the population, the Taurus and Upper Sco slopes are within 1$\sigma$ of each other. Finally, two separate power law fits were made to the Taurus population, splitting the sample at either M4 or M6 spectral types. The slopes for the high and low mass members are consistent within 2$\sigma$ of each other for a dividing spectral type of M4. The sample of substellar objects with spectral type M6 or later is too small and the fit to the brown dwarf population was unconstrained, ranging from positive to negative slopes. Within the limitations imposed by the current sample sizes, the brown dwarf disks do not appear to either dissipate more quickly than their counterpart disks above the substellar limit or to retain an elevated amount of disk dust material over time.

The fitted slope of $0.92\pm0.18$ for the combined Upper Sco population reported here is shallower than that of $1.67\pm 0.37$ reported in the large recent ALMA Upper Sco survey by \citet{barenfeld16}, and we investigate the source of the discrepancy. The additional detections and limits from \citet{gvdp16} and \citet{mathews12} do not change the slope at a significant level relative to including only the sample of \citet{barenfeld16}. Full details of the $M_{dust}$ and $M_{object}$ comparisons for Upper Sco are given in Appendix~\ref{sec:DustMassComparisons} and the results show that the key factor is the slope sensitivity to the choice of stellar evolutionary models --  \citet{siess00} models in the \citet{barenfeld16} analysis and the more recent \citet{baraffe15} models in this study. (Repeating our fitting technique for the Barenfeld et al. population with our re-calculated dust masses and their published stellar masses results in a slope of 1.87 $\pm$ 0.34, consistent with the \citet{barenfeld16} result.) Considering various treatments of dust temperature and stellar mass/luminosity, the range of slopes for both Taurus and Upper Sco reported within previous Taurus/Upper Sco surveys and recent ALMA surveys of regions such as Lupus III and Chamaeleon \citep[e.g.,][]{ansdell16, pascucci16}, have been consistent with both linear and steeper-than-linear relations. The choice of stellar evolutionary models and dust temperature relations are thus important factors in determining slope steepness and the fit parameters can only be compared if a uniform approach is adopted for all regions.

\begin{table}[]
\centering
\caption{Calculated slopes for Taurus and Upper Sco compilations.}
\label{tab:slopes}
\begin{tabular}{lcc}
\hline
\hline
Disk Size                                     & Taurus G8-M8.5 & U. Sco G7-M7.5 \\
\hline
Uniform 100 au                           & 0.98 $\pm$0.14    & 0.92 $\pm$ 0.18       \\
Uniform 100 au (det. only)              & 0.65 $\pm$ 0.11    & 0.42 $\pm$ 0.16       \\
40 au (M4+), 100 au ($<$M4) & 1.11 $\pm$ 0.14    & 1.05 $\pm$ 0.18       \\
20 au (M4+), 100 au ($<$M4) & 1.23 $\pm$ 0.14    & 1.16 $\pm$ 0.18    \\
\hline
\end{tabular}
\end{table}

\begin{figure}
    \centering
    \includegraphics[scale=0.45]{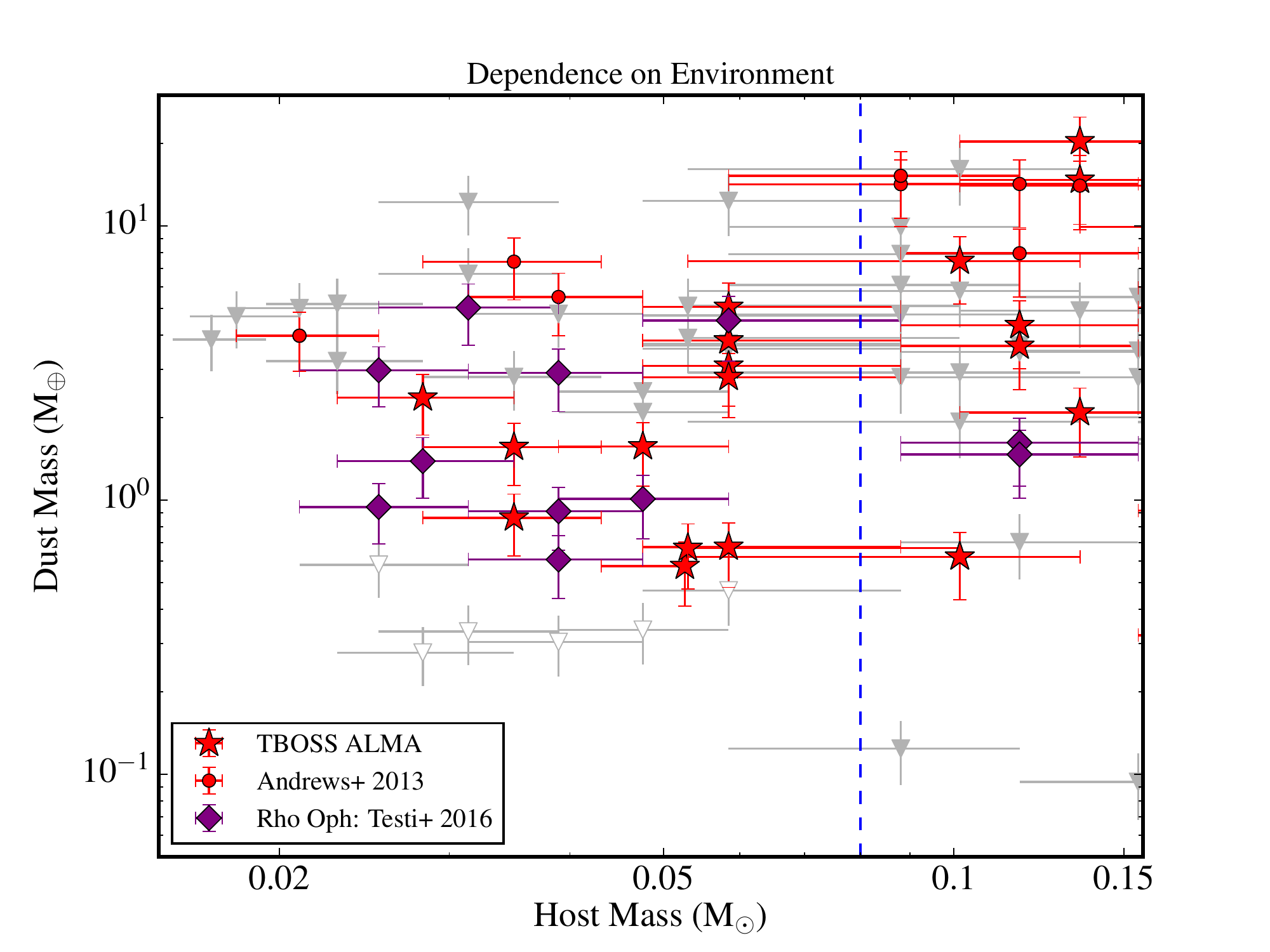}
    \caption{Comparison of the Taurus lowest-mass stars and brown dwarfs from \citet{andrews13} and our survey (red points and stars) and the Rho Ophiuchus population reported in \citet{testi16} (purple diamonds). Upper limits are shown as open downward triangles for Rho Oph and filled triangles for Taurus. While the age of the star forming regions are thought to be similar at $\sim$1~Myr, no statistically significant difference in dust mass is observed between the two regions, suggesting that any differing environmental effects may not be significant. The boundary between the stellar and substellar limit (0.08M$_{\odot}$) is shown with the vertical dashed line.}
    \label{fig:bds_only_dustmasses}
\end{figure}

To enable a comparison with a low-mass population at approximately the same age of Taurus, but in a different star-forming environment, the brown dwarf population of Rho Ophiuchus investigated by \citet{testi16} also with ALMA is shown for comparison with the Taurus population in Figure~\ref{fig:bds_only_dustmasses}. The Taurus and Rho Ophiuchus populations show similar mean and variance in dust masses for disk hosts with central object masses $< 0.08M_{\odot}$ (Taurus = 2.1 $\pm$ 1.4 M$_{\oplus}$, Rho Oph = 2.3 $\pm$ 1.6 M$_{\oplus}$). A two-sample Anderson-Darling (AD) test produced no statistically significant difference in dust mass with in brown dwarf and low-mass star disks between the TBOSS and Rho Oph (AD-statistic = 0.02, critical value for 5\% significance of 1.961, approximate $p$-value = 0.34).

\subsection{Implications for Planet Formation}

\begin{figure}
    \centering
    \includegraphics[scale=0.45]{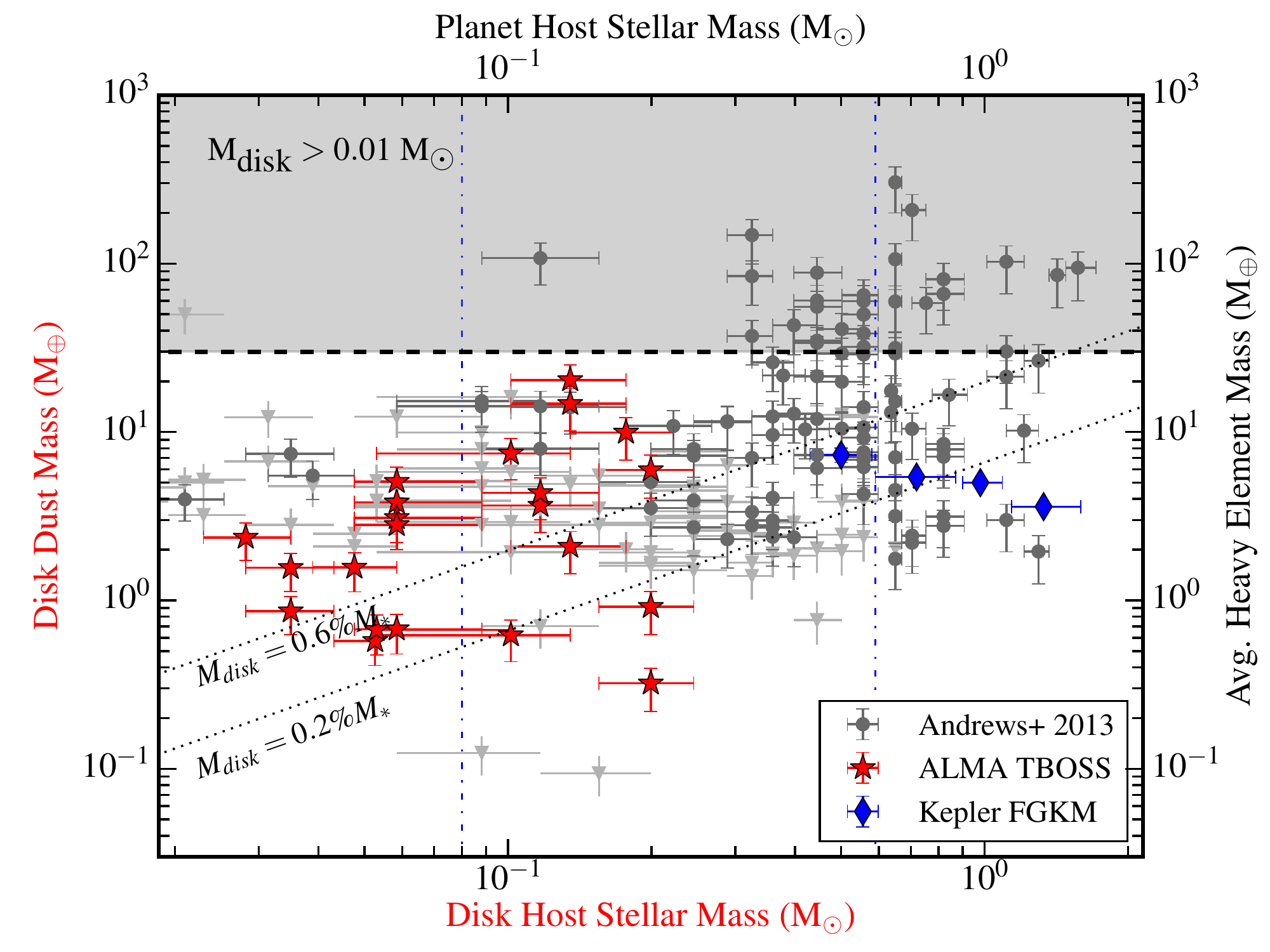}
    \caption{Comparison of our derived dust masses and the dust masses for higher-mass Taurus members from \citet{andrews13} with the heavy element distribution inferred from \textit{Kepler} FGKM stars (\citet{mulders15}; blue diamonds) and the giant-planet forming limit for the total mass of the disk (gas+dust) from the MMSN, assuming a gas:dust ratio of 100:1 (grey shaded region). Upper limits for the combined Taurus disk samples shown as downward triangles. and vertical blue dashed lines denote the range of Main Sequence M-dwarfs down to the 0.08M$_{\odot}$ limit.}
    \label{fig:kepcomp}
\end{figure}

The observed exoplanet population can provide insight into the amount of planet-forming material that must be available within primordial disks, enabling a comparison with the mass inventory in dust estimated from sub-mm flux densities of young Taurus objects. The average heavy element mass required to form the population of \textit{Kepler}-detected 2-50 day period planets was inferred by \citet{mulders15}. The \textit{Kepler}-inferred heavy element masses are plotted in Figure~\ref{fig:kepcomp} along with the Taurus ALMA results. Since the \textit{Kepler} results are confined to short period planets, corresponding to a limited radius within the disks, we also make a comparison with the Minimum Mass Solar Nebula \citep[MMSN, $\sim$35 Earth mass dust, $\sim$11 Jupiter mass gas+dust;][]{weidenschilling77}, since this covers the entire extent of the planetary system.  This is however a solar system-centric comparison, and it is not currently known how representative the MMSN is of a typical planetary system.  Indeed we know that many exoplanetary systems look very different from the solar system. In particular, it might well be expected that even if the MMSN is reasonably representative of G-type stars, it may not be applicable to other spectral types \citep[cf., a minimum-mass M-dwarf nebula of 53M$_{\oplus}$ of condensates for hosts of stellar mass 0.46M$_{\odot}$;][]{gaidos17}.  

\begin{figure}
    \centering
    \includegraphics[width=0.49\textwidth]{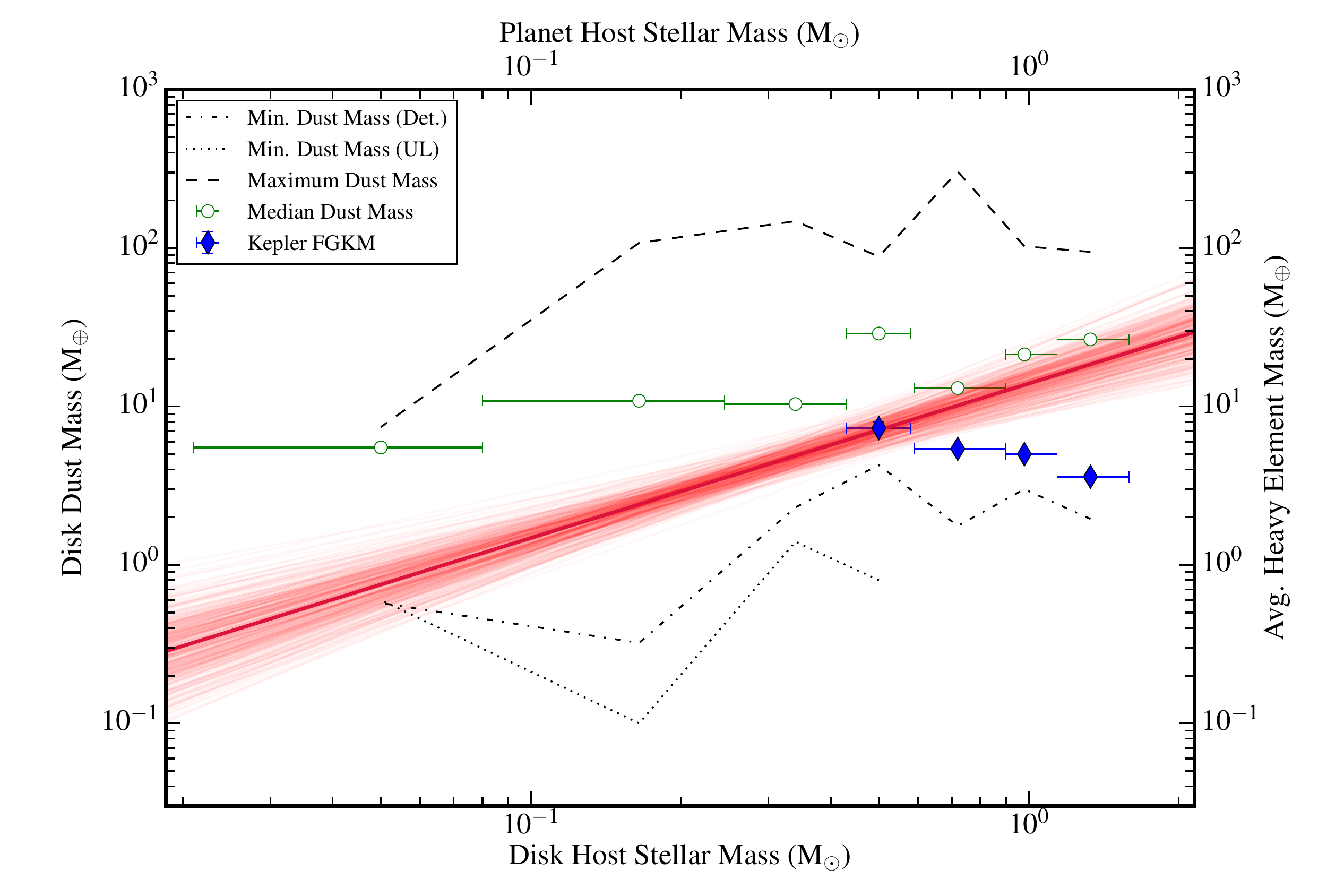}
    \caption{Comparison of the median, minimum (detections and upper limits), and maximum dust masses for Taurus in terms of disk dust mass (M$_{\oplus}$) as a function of the host stellar mass (M$_{\odot}$). As in Figure~\ref{fig:kepcomp}, the \textit{Kepler} FGKM heavy element masses estimate from Mulders et al. (2015) are shown as blue diamonds (with right y-axis and upper x-axis corresponding to the heavy element masses and \textit{Kepler} host star masses, respectively). The corresponding binned dust mass values are provided in Table~\ref{tab:binned_values}, and the overplotted linear regressions correspond to the Taurus best fit with 100~au disks in Figure~\ref{fig:powerlaw1}.}
    \label{fig:binneddust}
\end{figure}

The \textit{Kepler} planet host masses are determined from the stellar effective temperature and mass table given in \mbox{\citet{pecautmamajek}} and the \textit{Kepler} host star planets compiled in \mbox{\citet{mulders15}}.  Over 90\% of the M-star hosts analysed by \mbox{\citet{mulders15}} are M0-M3, and so the host mass range of the \textit{Kepler} results only extends down to $\sim$0.4M$_{\odot}$, as plotted in Figures~\ref{fig:kepcomp} and \ref{fig:binneddust}. The \textit{Kepler} and Taurus disk population results are summarized for comparison over common mass ranges in Table 12 which also quantifies the proportion of Class II disks that exceed the average heavy element mass estimated from \textit{Kepler} and the MMSN. Table 13 reports the minimum (both for detections and limits), maximum and median (including limits) disk dust mass values for the same mass ranges. The heavy element masses from \citet{mulders15} trend upward towards lower stellar masses for planetary systems with 2-50 day orbital periods. As shown in the dispersion of the points in Figure~\ref{fig:kepcomp} and the upper and lower envelopes in Figure~\ref{fig:binneddust}, the majority (57\%) of the Taurus sample has larger masses present in small particles than ultimately coalesce into planets with short periods measurable with \textit{Kepler}, and a smaller, but still significant fraction (24\%) contain more mass in dust than the MMSN. Considering only the best-fit relation for the full Taurus Class II population plotted in Figure~\ref{fig:binneddust}, the fit to disk dust mass exceeds the mass inventory in exoplanets around higher mass stars, and intercepts the expected exoplanet inventory for the lowest-mass hosts considered in the \emph{Kepler} study. From an ALMA survey of Cha I Class II members, \citet{pascucci16} similarly find that the best fit to the disk dust masses in Cha I is greater than the estimated material locked within the close-in exoplanet population for $\gtrsim$ 1~M$_{\odot}$ stars, but that the least massive (0.4 M$_{\odot}$) Cha~I hosts have median disk masses a factor of 2 lower than the average mass in exoplanets. Although the median Cha~I value for M-star hosts is lower than the inferred \emph{Kepler} value, the large dispersion in dust mass observed in Cha~I (similar to Taurus) is such that part of the M-star population retains disks with dust masses comparable to or larger than the \textit{Kepler} average heavy element mass.

\input{classIIIfrequency.txt}

\input{avgmedminmax_dustmasses.txt}

While our observations explore a range of grain sizes on the order of the observation wavelength, an outstanding question remains as to the fraction of mass in undetectable larger bodies by the age of Taurus. By the age of 1-2 Myr, the rate of dust detection in infrared and submm/cm surveys suggests that coagulation mechanisms in simulations, while efficient at growing grains up from sub-micron scales, are insufficient to maintain the small grain dust population on their own, which must be replenished. This could be achieved with an equilibrium reached between growth and collisional grinding and fragmentation processes \mbox{\citep{dullemond_dominik05}}. The model from \mbox{\citet{dullemond_dominik05}} incorporating coagulation with effects of grain settling and mixing as well as fragmentation, suggests that near $\sim$1~Myr, approximately 0.5 dex greater mass surface density of the disk is contained within cm-sized grains than submm grains, within a simulated vertical slice at 1~au. This factor of $\sim$3 in mass surface density can be compared with the observational results from longer wavelength studies of disks from the same or similar star-forming regions. For an M1 member of Taurus-Auriga, CY~Tau, \mbox{\citet{perez15}} analyzed spatially-resolved continuum measurements at 1.3, 2.8, and 7.1mm from the Disks$@$EVLA program. They find best fit model parameters on the disk structure which, at a radius of 1~au, correspond well with the surface density ratio of $\sim$3x more mass in larger grains inferred from \mbox{\citet{dullemond_dominik05}}, for the ratio of mass surface density from 1.3mm to 7.1mm. However, with resolved measurements, P\'{e}rez et al. find that the grain size distribution is strongly dependent on location within the disk, corresponding to a much larger population of small grains in the outer disk and providing strong evidence for radial drift effects. As the \mbox{\citet{dullemond_dominik05}} models present a simple case excluding factors such as radial drift and runaway growth, it is likely that simply scaling the submm-inferred dust mass by a factor of 3x presents a limiting case for mass in sub-mm to cm-sized objects.

\begin{figure}
    \centering
    \includegraphics[width=0.49\textwidth]{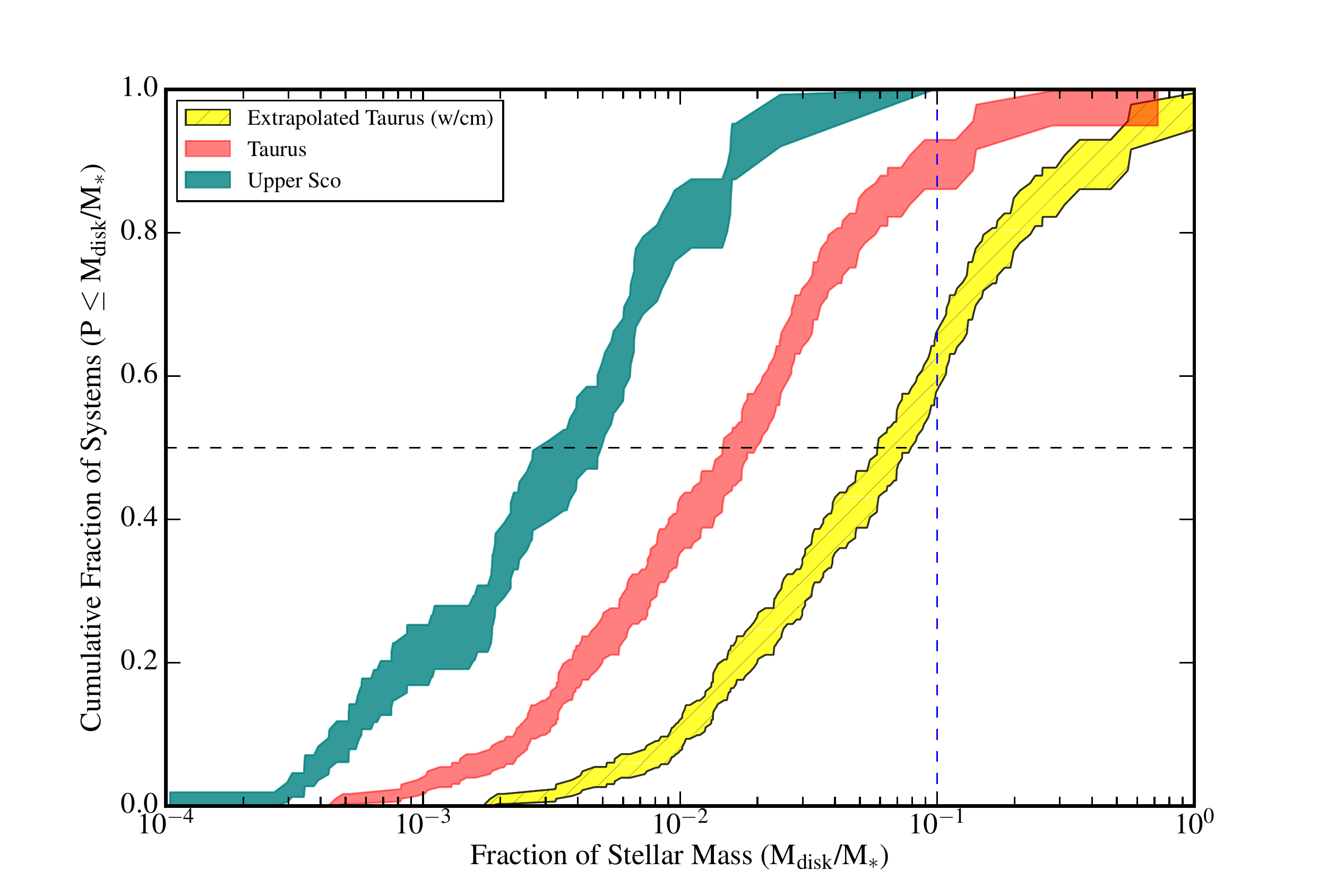}
    \caption{Cumulative distributions showing estimated total (gas + dust) disk masses as a fraction of total disk mass to stellar mass, assuming a gas:dust ratio of 100:1. The vertical blue dashed line indicates the gravitationally unstable limit (M$_\textnormal{disk}$ = 0.1M$_\textnormal{star}$), and the horizontal line indicates the median. Taurus populations are from this study and \citet{andrews13} (red curve), using analytically-derived masses assuming $r=100$~au disks. Upper limits are incorporated using Kaplan-Meier estimation, with distribution width indicating 1$\sigma$ confidence intervals. The Upper Sco population (green curve) is a combined distribution from \citet{mathews12}, \citet{barenfeld16} and \citet{gvdp16}, also incorporating upper limits. The yellow hatched distribution indicates a limiting case of extrapolating the Taurus mass in cm-sized grains as 3x the measured sub-mm dust masses, in which case $\sim35$\% of Taurus systems would be gravitationally unstable.}
    \label{fig:cumulativecomparison}
\end{figure}

To illustrate the distributions of disk masses derived from sub-mm observations and the potential impact of scaling up the Taurus disk masses to also include $\sim$cm-sized grains, we show the cumulative distributions of systems as a fraction of the gravitationally unstable disk mass limit in Figure~\ref{fig:cumulativecomparison}. The gas to dust ratio is assumed to be 100:1 as for the interstellar medium (ISM), and the limit for a gravitationally unstable disk is taken as M$_\textnormal{disk}$ = 0.1 M$_\textnormal{star}$. This places a representative upper limit on the possible mass of the disk and constrains the range of possible `unseen' mass in larger bodies within the disk.  Note that while it is possible that the gas to dust ratio at the age of Taurus is lower than 100:1, it would presumably have started at the ISM value and thus the gravitational stability limit we are comparing to would still have applied earlier in the disk evolution. As seen in Figure~\ref{fig:cumulativecomparison}, it is notable that the shape of the older Upper Sco distribution is very similar to that of the Taurus population, suggesting that the decrease in dust mass between the ages of Taurus and Upper Sco occurs uniformly across the distributions. For comparison, a scenario with three times the sub-mm dust mass in cm-sized grains is also shown for the Taurus samples (yellow hatched distribution). This leads to around 30-40\% of systems exceeding the gravitationally unstable mass, suggesting that the mass in larger objects not seen by our ALMA observations is not this large and that in many cases the dust we observe in the sub-mm constitutes the bulk of the mass of solid particles in the disk. As such, at the age of Taurus, planet formation may be in its very early stages.

To place these timescales within the context of our own solar system, isotopic studies have also placed limits upon the formation timescales of small grains and early parent bodies \citep{chambers10}, including: calcium aluminum-rich inclusions (CAIs, $\leq$ 0.2~Myr), iron meteorites ($\leq$ 1~Myr), chrondrules (1-3.5 Myr), and the cores of Mars and Vesta (ranging from 1-10 Myr, although earlier ages of 1.8 Myr for Mars have been posited; \cite{dauphas_pourmand11}). Given the relative size scales of CAIs and chondrules in meteorites, on the order of sub-mm and cm-sized grains, these timescales correspond well to the significant abundance of similar-sized grains detected in sub-mm/mm surveys of protoplanetary disks. Furthermore, the depletion when comparing with Upper Sco suggests that the majority of planet formation may be taking place between these age ranges, which would also be in agreement with the formation timescales of larger planetesimals in the Solar System. 

Theoretical models of giant planet formation \citep[e.g.,][]{alibert05} suggest that the MMSN is also roughly the minimum mass required for the formation of giant planets. As shown in Figure~\ref{fig:kepcomp}, while the upper envelope of disk masses exceeds this for hosts with masses above the stellar limit, this is not true for hosts below the stellar/substellar boundary. This suggests that the disks of substellar objects are not massive enough to support giant planet formation within the disks, and that planetary mass companions identified around brown dwarf primaries such as 2M1207b and 2M J044144 \mbox{\citep{chauvin04, todorov10}} may form through a process more similar to that of binary stars rather than within a planet-forming disk. This suggestion is reinforced by examining the 193 Taurus Class II and Class III objects with masses in the 0.08-0.6M$_{\odot}$ range (equivalent to main sequence M-dwarfs).  Of these 193 objects summarized in Table 12, 32 (17\%) have disk masses larger than the MMSN and thus are theoretically amenable to giant planet formation; this frequency assumes no Class III members have $\>$MMSN disks although there is not a comparably deep submm survey of Class III members. By comparison, large-scale exoplanet surveys indicate that the occurrence rate of giant planets around M-dwarfs is $\sim$2\% out to orbits probed by radial velocity surveys ($\sim$5.5yrs) \mbox{\citep[e.g.,][]{cumming08, johnson11}} and deep AO imaging surveys for giant planet companions to M-stars have reported null detections over the $\sim$10--100~au range \citep[e.g.,][]{bowler16}. Comparison of the frequencies of $\>$MMSN disks and M-star giant planets suggests that the efficiency of forming giant planets from MMSN disks is close to $\sim$10\%, and most disks that are theoretically capable of forming giant planets, at least around low mass hosts, do not do so.


\section{Summary and Conclusions}
\label{sec:summary}

In summary, the detections from this initial ALMA Cycle 1 study of 24 M4--M7.75 Class II Taurus members (21 detections at $>$8$\sigma$, one marginal detection at 5$\sigma$, and two non-detections) show that the dramatic increase in sensitivity achieved with ALMA combined with a target selection based on \textit{Herschel} PACS 70$\mu$m fluxes \mbox{\citep{bulger14}} enable investigations of the disk properties of the full mass spectrum of young star-forming regions. The targets represent half of the Class II members in this spectral type range with \textit{Herschel} detections and span the full range of PACS 70$\mu$m fluxes rather than a subset of the brightest members. This pilot study includes 7 transition disks and 1 truncated disk, and the non-detections are both transition disks, though other objects in this class are among the brightest ALMA detections; the truncated disk is the most marginal detection.

The 885$\mu$m continuum flux densities that are the subject of this paper range from 1.0 to 55.7~mJy. The results from the spectral line observations covering the $^{12}$CO(3-2) emission will be reported in the next paper in the TBOSS (Taurus Boundary of Stellar/Substellar) series (van der Plas et al. 2017, \textit{in prep}). Applying different approaches to converting the flux densities to dust masses -- several scaling laws and radiative transfer modeling with MCFOST -- results in a factor of 2.5 range in mass estimates, with the radiative transfer model estimate typically at the lower part of the mass range inferred from scaling laws based on different disk radii \mbox{\citep{andrews13, gvdp16}}. By employing the relations in Eqn.~\ref{eq:mdust} and Eqn.~\ref{eq:newtdust} that can be applied to all Taurus members with submm detections, the dust masses for the TBOSS ALMA sample range from 0.3~M$_{\oplus}$ to 20~M$_{\oplus}$, comparable to several times the mass of Mars to enough Earth masses to form a giant planet core \mbox{\citep{pollack96}}.

Combining the new ALMA results with the disks around more massive Taurus members shows a trend of declining disk dust mass with central object mass with a large amount of scatter (at least one order of magnitude) at any given mass. Considering a range of outer disk radii for the low mass object disks, the slope of the power law fit to the $M_{dust}$ vs. $M_{object}$ relation is consistent with linear over the host mass range of $\sim$35~M$_\textnormal{Jup}$ -- 1M$_{\odot}$ which encompasses most of Taurus. The specific value of the slope is very dependent on the choice of evolutionary model to determine the object masses, and a steeper than linear slope is obtained with a different model set. The brown dwarf disk population appears as a continuous extension of the low mass stars rather than a distinct set.

Comparing the Taurus detected disks with results from low mass stars and brown dwarfs in the older Upper Sco region shows that the Upper Sco members have disk masses comparable to or lower than the lowest mass disks around similar mass host objects. In contrast to the larger dust masses in Taurus, the decline in mass of dust in small ($\lesssim$ 1mm) particles in Upper Sco may be an indication that planet formation has progressed to the stage in which most solids are in the form of planetesimals and planets and undetectable at sub-mm wavelengths. It has long been noted that giant planet formation must complete before the gas disk dissipates so that they can accrete their gaseous envelopes.  Modern theories for the growth of solid planetesimals, such as the streaming instability \citep[e.g.,][]{youdin05, johansen07, youdin07} and pebble accretion \citep[e.g.,][]{lambrechts12, levison15a, levison15b}, which apply to both terrestrial planets and giant planet cores, proceed rapidly once the processes are initiated and also rely on the presence of gas. Furthermore, isotopic analysis of solar system meteorites indicates that large bodies had formed within a few million years of the condensation of the first solids \citep[e.g.,][]{bouvier10, connelly08, connelly12}. As such, the decline in dust mass from Taurus to Upper Sco is aligned with theoretical expectations for planet formation.

The mass inventory of solids in small particles detected by submm emission typically exceeds the average heavy-element mass inferred from \textit{Kepler} short period planetary systems \mbox{\citep{mulders15}}. This comparison quantifies that a sufficient mass reservoir exists to form the Super Earth and mini Neptune planets that constitute the bulk of the \textit{Kepler} exoplanet discoveries and that the timescale for formation may exceed the $\sim$1-2 Myr age of Taurus. While the majority of disks appear to be sites conducive to small planet formation, a much lower proportion of disks have a total mass large enough for giant planet formation based on a standard 100:1 gas:dust ratio and a threshold disk mass of $\sim$0.01M$_{\odot}$ \mbox{\citep{alibert05}}. Under these assumptions, few low-mass stars have disk masses meeting or exceeding the MMSN limit, commensurate with the limited numbers of giant planets detected around these hosts to-date. Direct imaging searches for sub-Jovian M-dwarf exoplanets with upcoming facilities like the \textit{James Webb Space Telescope (JWST)} anticipate reaching expected mass limits of $\sim$2 times that of Neptune \citep[]{schlieder16}, and the disk dust mass results suggest that higher-mass M-dwarfs may be more amenable to hosting low-mass gas/ice giant exoplanets than the lowest-mass M-dwarf hosts. 
Applying Solar System proportions of dust and ice in solids \citep[rocky material $\sim$1/3 and ice $\sim$2/3;][]{lodders03} to the composition of Neptune \citep[$\sim$13-15 M$_{\oplus}$ in heavy elements;][]{helled11} suggests that $\sim$4-5M$_{\oplus}$ in dust is required to form a Neptune-like planet. In a rough analogy to the MMSN estimate of the disk required to form a Jupiter-like planet, the minimum mass dust disk required to form a Neptune would contain $\sim$5M$_{\oplus}$ in rocky material, or $\sim$10M$_{\oplus}$ for the expected 2$\times$~Neptune \textit{JWST} imaging detection limit. As seen in Figure~\ref{fig:kepcomp}, few late-M Taurus disks contain $\sim$10M$_{\oplus}$ in dust particles measurable with ALMA.

Among Taurus members with masses in the range of Main Sequence M-stars (0.08-0.6 M$_{\odot}$), the frequency of observed candidate giant planet-forming disks is 17\%. This value exceeds the $\sim$2-3\% frequency of M-dwarf giant planets for periods $< 10{^4}$ days derived from the synthesis of radial velocity and microlensing surveys \citep[e.g.,][]{clantongaudi14b}, and with the null detection of wider orbit planets in M-dwarf direct imaging surveys \citep[e.g.,][]{bowler15}, suggests a relatively low efficiency for giant planet formation. By contrast, none of the brown dwarf Taurus members have total disk mass estimates above the giant planet formation threshold, suggesting that imaged planetary mass companions to brown dwarfs did not originate in disks.


\acknowledgments

The authors wish to thank the anonymous referee for providing a thorough review and helpful comments which improved this manuscript. We are grateful to Brian Mason, Sarah Wood, and the North American ALMA Science Center Staff for assistance with the data reduction for this work. We thank Steve Desch, Nat Butler, Maitrayee Bose, Prajkta Mane, Brian Svoboda, Anusha Kalyaan, Travis Gabriel, and Wanda Feng for helpful discussions. KWD was supported by the NSF Graduate Research Fellowship under Grant No. DGE-1311230 and support for this work was provided by the NSF through Award SOSPA3-007 from the NRAO (Student Observing Support Program). This work was also supported by an NSF Graduate Research Opportunities Worldwide supplemental award (Proposal 13074525) in partnership with CONICYT. The results reported herein benefitted from collaborations and/or information exchange within NASA's Nexus for Exoplanet System Science (NExSS) research coordination network at Arizona State University sponsored by NASA's Science Mission Directorate (Grant NNX15AD53G). GB, JB, JP, NJT, and KWD would like to acknowledge support from the Jet Propulsion Laboratory's Strategic University Research Partnerships (SURP) program. GvdP acknowledges support from the Millennium Science Initiative (Chilean Ministry of Economy) through grant RC130007 and from FONDECYT, grant 3140393. FMe, GvdP, and CP acknowledge funding from ANR of France under contract number ANR-16-CE31-0013 (``Planet-Forming-Disks''). APJ gratefully acknowledges funding through NASA grant NNX16AI31G (``Stop hitting yourself''). R.J.D.R has been supported by NSF grant AST-1518332, National Aeronautics and Space Administration (NASA) Origins grant NNX15AC89G, and NASA NExSS grant NNX15AD95G. This paper makes use of the following ALMA data: ADS/JAO.ALMA\#2012.1.00743.S. ALMA is a partnership of ESO (representing its member states), NSF (USA) and NINS (Japan), together with NRC (Canada), NSC and ASIAA (Taiwan), and KASI (Republic of Korea), in cooperation with the Republic of Chile. The Joint ALMA Observatory is operated by ESO, AUI/NRAO and NAOJ. The National Radio Astronomy Observatory is a facility of the National Science Foundation operated under cooperative agreement by Associated Universities, Inc. This research has made use of the SIMBAD data base and VizieR catalogue access tools, operated at CDS, Strasbourg, France. This research made use of APLpy, an open-source plotting package for Python \citep{aplpy}. This research makes use of the data products from the 2MASS, which is a joint project of the University of Massachusetts and the Infrared Processing and Analysis Center/California Institute of Technology, funded by NASA and the NSF.

\bibliographystyle{yahapj}
\bibliography{continuum_alma.bib}


\appendix

\section{Central Object Spectral Type, Temperatures, and Masses}
\label{sec:starmassestimation}
A uniform procedure was applied to estimate the central object mass for all of the Taurus, Upper Sco, and Rho Oph members considered in this study of disk mass as a function of host mass (Figures~\ref{fig:taurusonlymasses}--\ref{fig:binneddust})\footnote{Example analysis scripts and auxiliary data for the methods described in Appendices A and B are available at https://osf.io/9dyx4.}. The observable measured for all objects is the spectral type, and the transformation to mass required two main steps: (1) converting spectral type to effective temperature (T$_\textnormal{eff}$) with an empirical relation and (2) converting T$_\textnormal{eff}$ into mass with theoretical evolutionary models. The impact due to the choice of age and evolutionary model was investigated and comparisons with previous results for Taurus members were explored. Overall, the adopted evolutionary model had the largest effect on mass estimation, more important than the specific age assumed for the region or the spectral type-T$_\textnormal{eff}$ relation.

\begin{figure}
    \centering
    \includegraphics[scale=0.45]{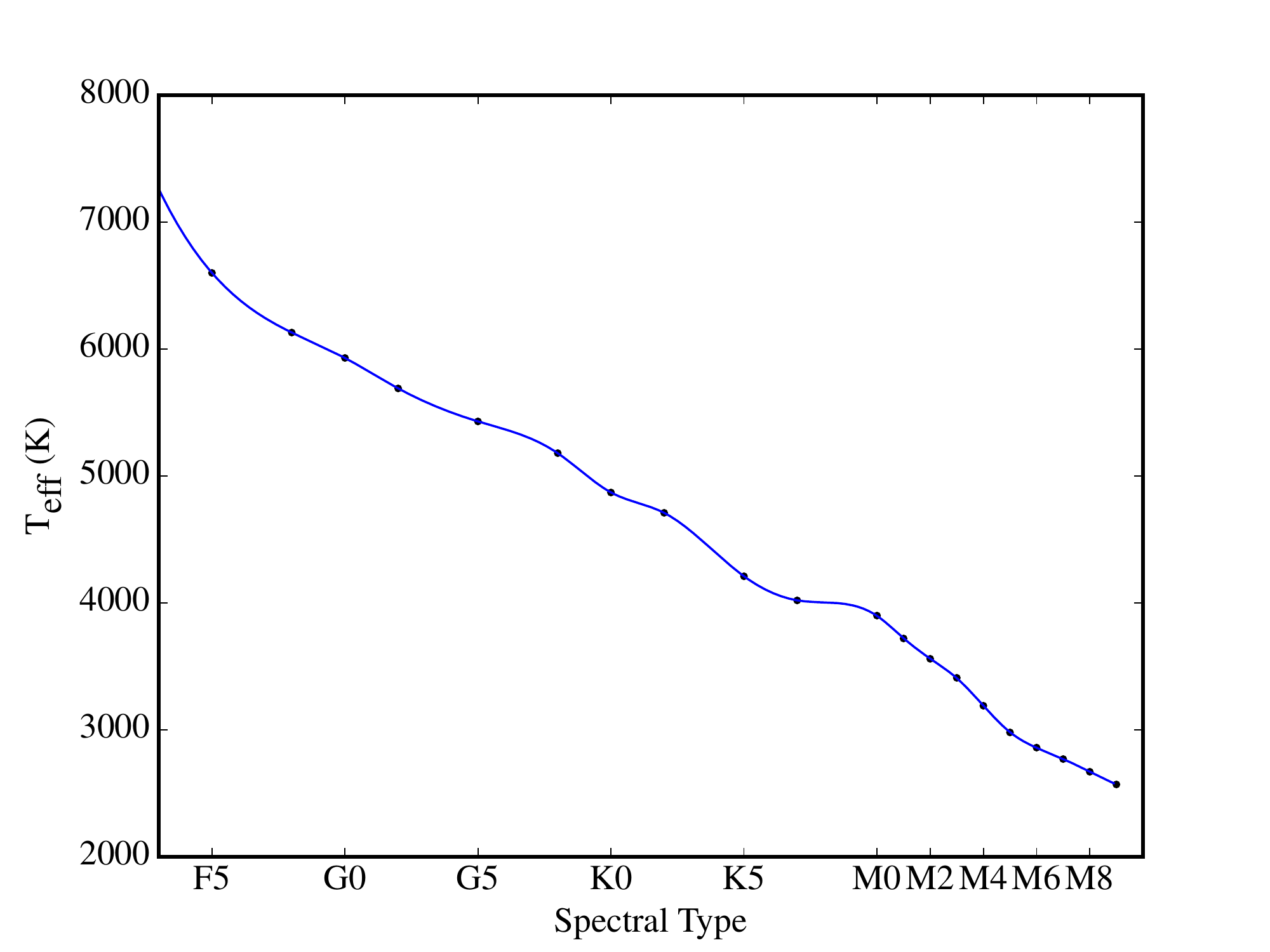}
    \caption{Correspondence between spectral type and effective temperature as provided in Herczeg \& Hillenbrand (2014). A least-squared univariate spline interpolation (blue line) is used to provide fractional subclasses of spectral types.}
    \label{fig:tempscale}
\end{figure}

The empirical relation developed to convert observed spectral types into effective temperatures is shown in Figure~\ref{fig:tempscale} \mbox{\citep{herczeg_hillenbrand14}}. This more recent transformation builds upon the relationship developed in \mbox{\citet{luhman03}} and covers a larger range of spectral types, and we applied a least-squared univariate spline interpolation to this relation to account for non-integer spectral types. Table~\ref{tab:recalc_dustmasses} reports the spectral types and uncertainties from the literature along with estimated T$_\textnormal{eff}$ for all objects considered in this study from Taurus \mbox{\citep{andrews13}}, Upper Sco \citep{mathews12, gvdp16, barenfeld16}, and Rho Oph \mbox{\citep{testi16}}. The most recent spectral type is used for sources with multiple values, and uncertainty of $\pm$0.5 subclasses is applied to any object without a reported uncertainty.

\begin{figure}
    \centering
    \includegraphics[scale=0.45]{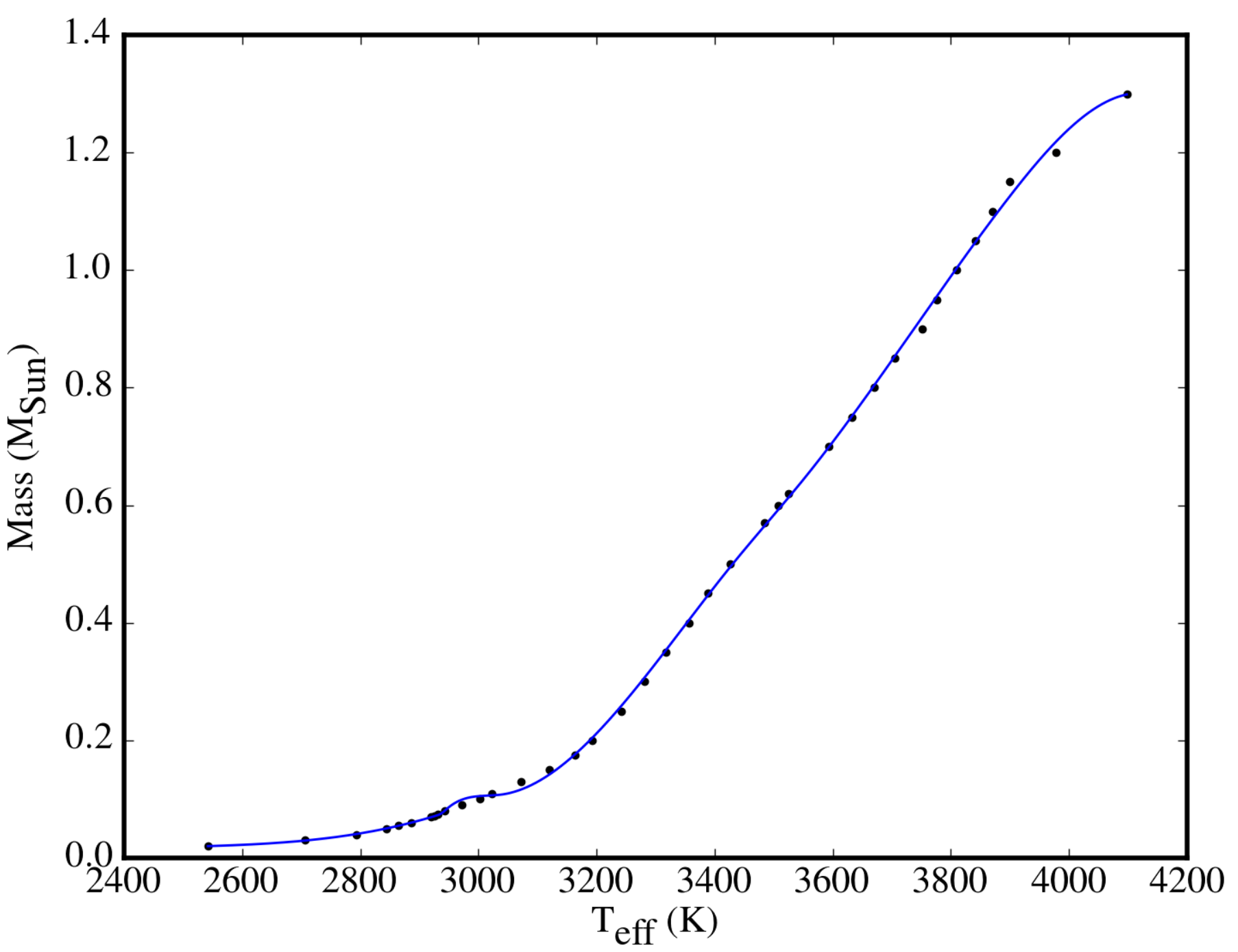}
    \caption{Interpolation over mass and effective temperature as provided in the 1 Myr \citet{baraffe98} models.}
    \label{fig:b98massteff}
\end{figure}

\begin{figure}
    \centering
    \includegraphics[scale=0.45]{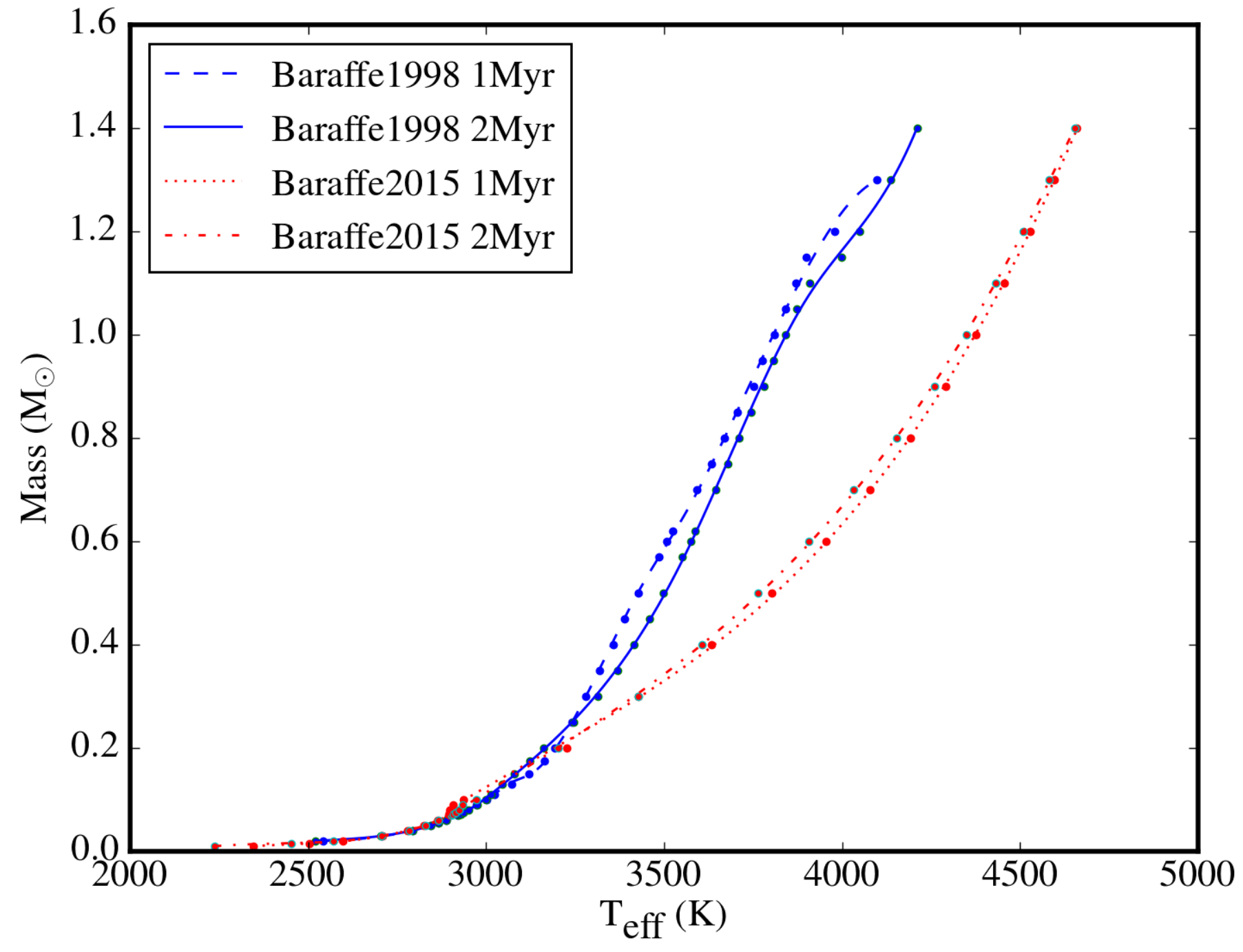}
    \caption{Interpolation over mass and effective temperature for each of the 1 and 2 Myr models provided by Baraffe et al. (1998, 2015).}
    \label{fig:mass_teff_vintage}
\end{figure}

\begin{figure}
    \centering
    \includegraphics[scale=0.45]{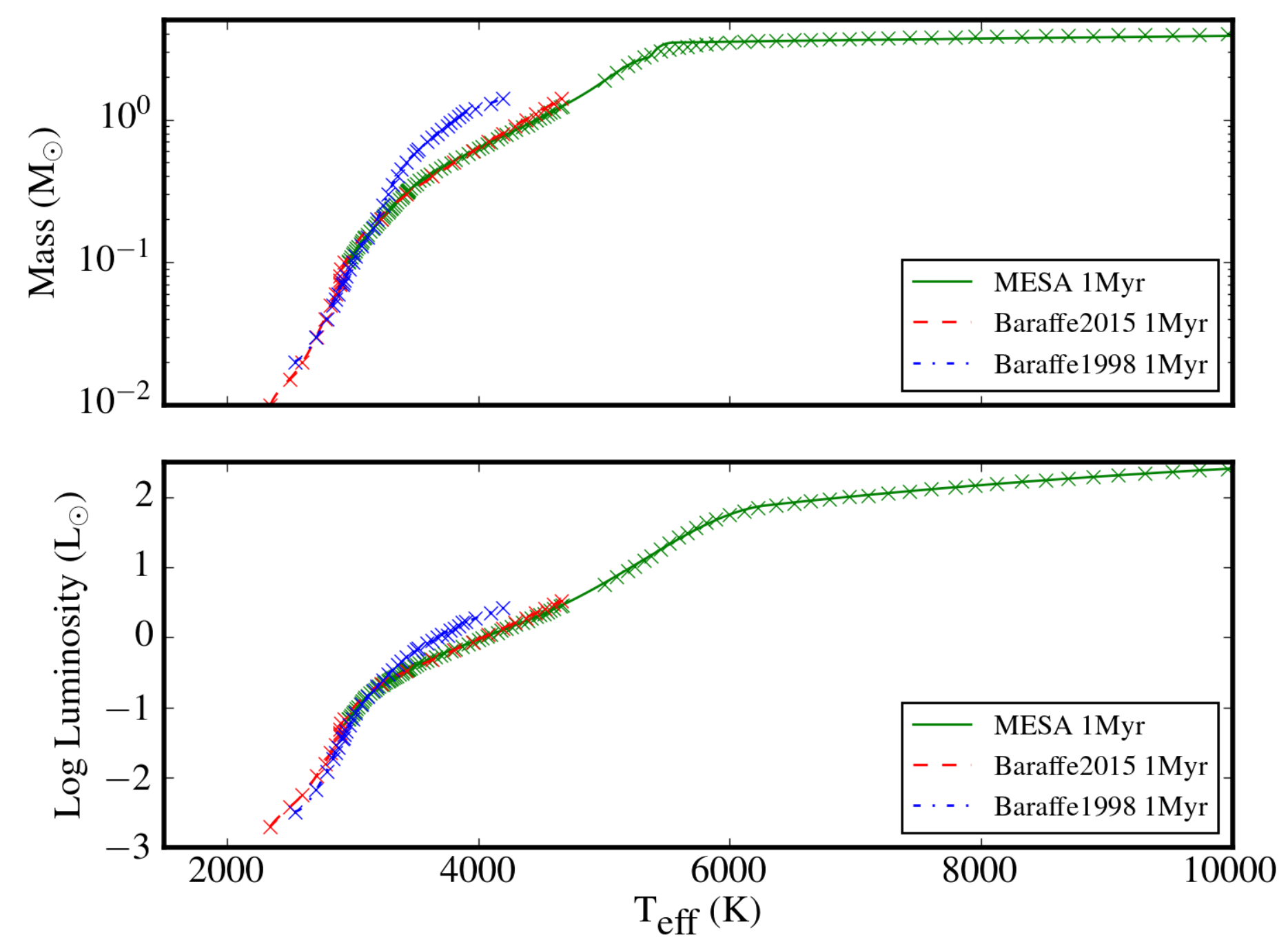}
    \caption{Interpolation over mass and effective temperature for the 1 Myr models provided by Baraffe et al. (1998, 2015), and from the MESA grid of models (Choi et al. 2016), adopted for the derivation of central object mass for stars with temperatures corresponding to masses greater than 1.4 M$_{\odot}$.}
    \label{fig:mass_teff_lum1}
\end{figure}

\begin{figure}
    \centering
    \includegraphics[scale=0.45]{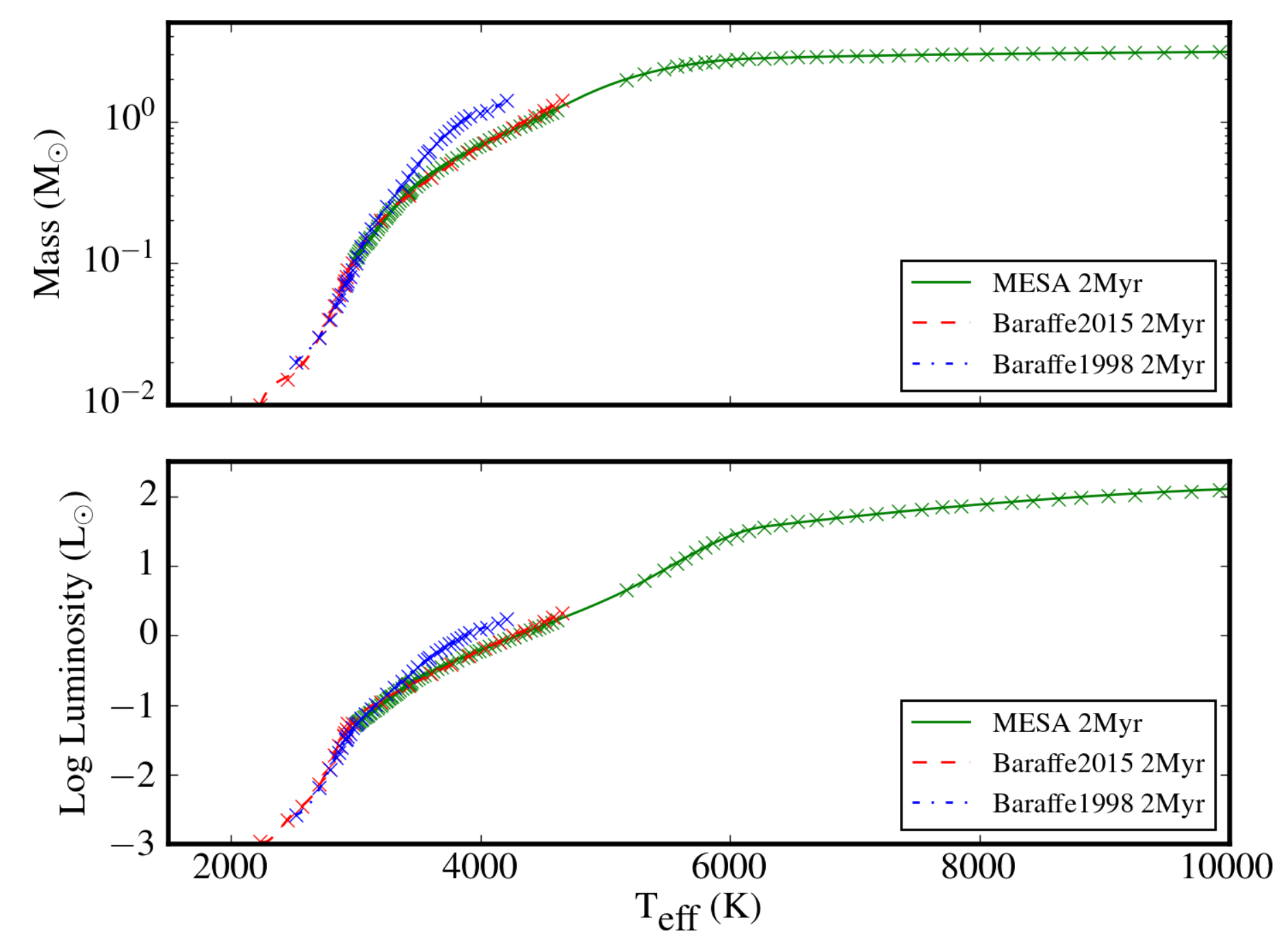}
    \caption{Interpolation over mass and effective temperature for the 2 Myr models provided by Baraffe et al. (1998, 2015), and from the MESA grid of models (Choi et al. 2016), adopted for the derivation of central object mass for stars with temperatures corresponding to masses greater than 1.4 M$_{\odot}$.}
    \label{fig:mass_teff_lum2}
\end{figure}

Combining the effective temperature with an age, the mass was estimated in conjunction with an evolutionary model. Typical ages reported for the Taurus region range from 0-5~Myr, with isochrone fitting to the cluster sequence tracing a canonical value of 1-2~Myr \mbox{\citep{kraus09}}. For objects cooler than $\sim$4500~K, the evolutionary models of \mbox{\citet{baraffe98}} have been widely adopted, although the updated grid from \mbox{\citet{baraffe15}} is now available. Figures~\ref{fig:b98massteff} and \ref{fig:mass_teff_vintage} show the impact of the choice of age and evolutionary model for the cooler Taurus members and highlight that the evolutionary model has the dominant systematic effect on the mass estimation, with the newer models yielding a lower mass for the same effective temperature. For this study, an age of 1~Myr and the \mbox{\citet{baraffe15}} model grid were used in the mass determination for objects with T$_\textnormal{eff} \leq$ 4211~K, corresponding to spectral types of approximately K7 or later. For earlier spectral types with higher effective temperatures, the MESA models \mbox{\citep{choi16}} were applied, and the full mass vs. T$_\textnormal{eff}$ sequences for 1 and 2~Myr models are shown in Figures~\ref{fig:mass_teff_lum1} and \ref{fig:mass_teff_lum2}. The newer \mbox{\citet{baraffe15}} models connect with the MESA models. Figures~\ref{fig:mass_teff_lum1} and \ref{fig:mass_teff_lum2} also plot the luminosity as a function of T$_\textnormal{eff}$ from the evolutionary models, since the central object luminosity is required to estimate the disk dust temperature, which is used for the calculation of disk dust mass.

The resulting masses for all Taurus members with submillimeter detections were compared with the masses reported for Class II members in the SMA dish survey which formed the higher mass comparison sample \mbox{\citep{andrews13}}. The approach to estimating masses in the SMA study applied Bayesian inference techniques from \mbox{\citet{jorgensen2005}} and \mbox{\citet{gennaro2012}}, first treating stellar luminosity and A$_{v}$ as free parameters in fits of template stellar photosphere models to optical/NIR SEDs, and then evaluating a corresponding conditional likelihood function to best-fit stellar masses and ages from three suites of pre-main sequence stellar evolution grids \mbox{\citep{dantona_mazzitelli97, baraffe98, siess00}}. For consistency, the values reported from the \mbox{\citet{baraffe98}} grid in \mbox{\citet{andrews13}} are used as comparison values for the results in this appendix.

\begin{figure}
    \centering
    \includegraphics[scale=0.45]{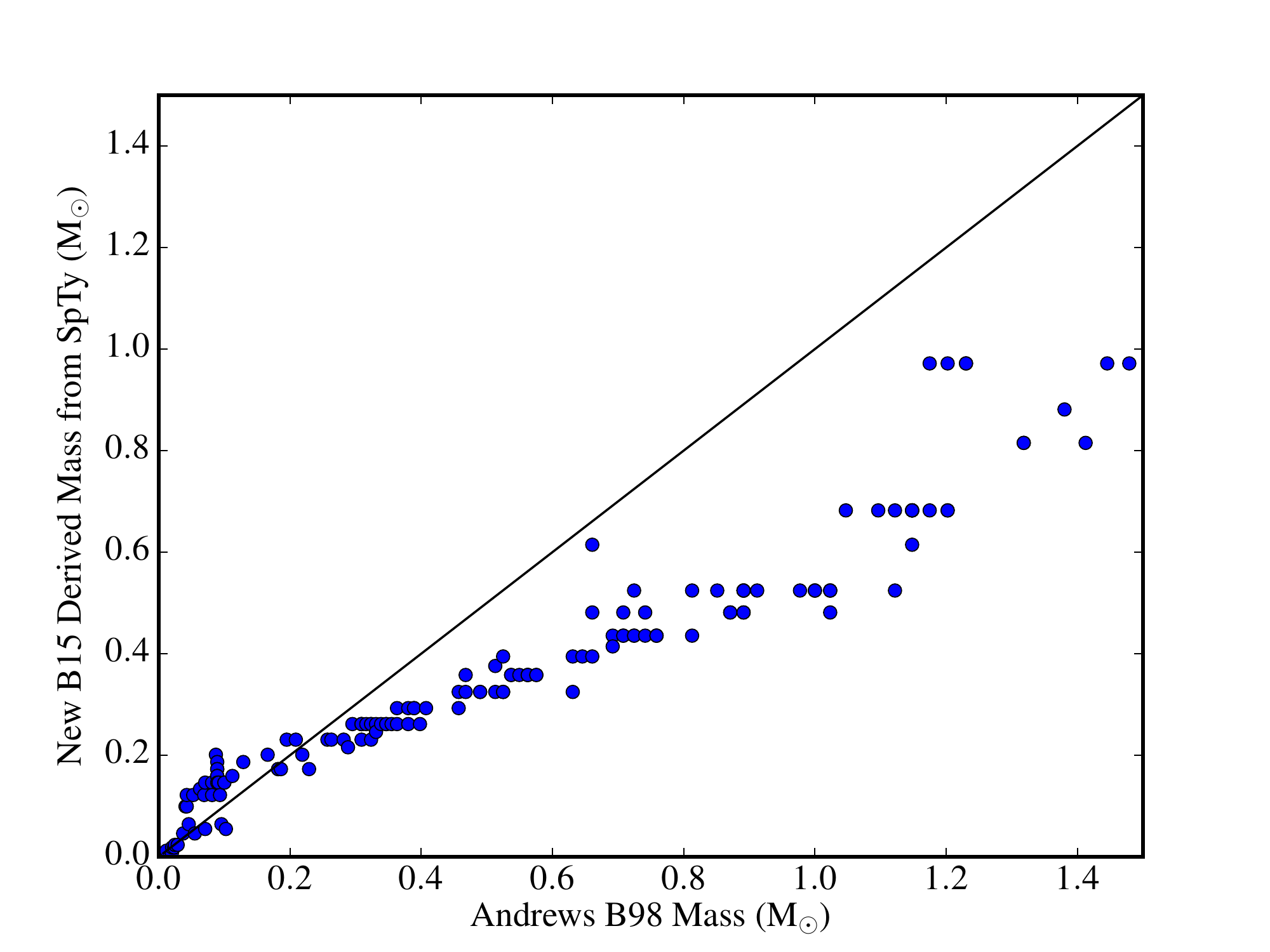}
    \caption{Comparison between the estimated stellar masses from \mbox{\citet{andrews13}}, using the \mbox{\citet{baraffe98}} model grid, and the newly derived masses using the approach described in this paper and the newer \mbox{\citet{baraffe15}} models.}
    \label{fig:masscomp1}
\end{figure}

\begin{figure}
    \centering
    \includegraphics[scale=0.45]{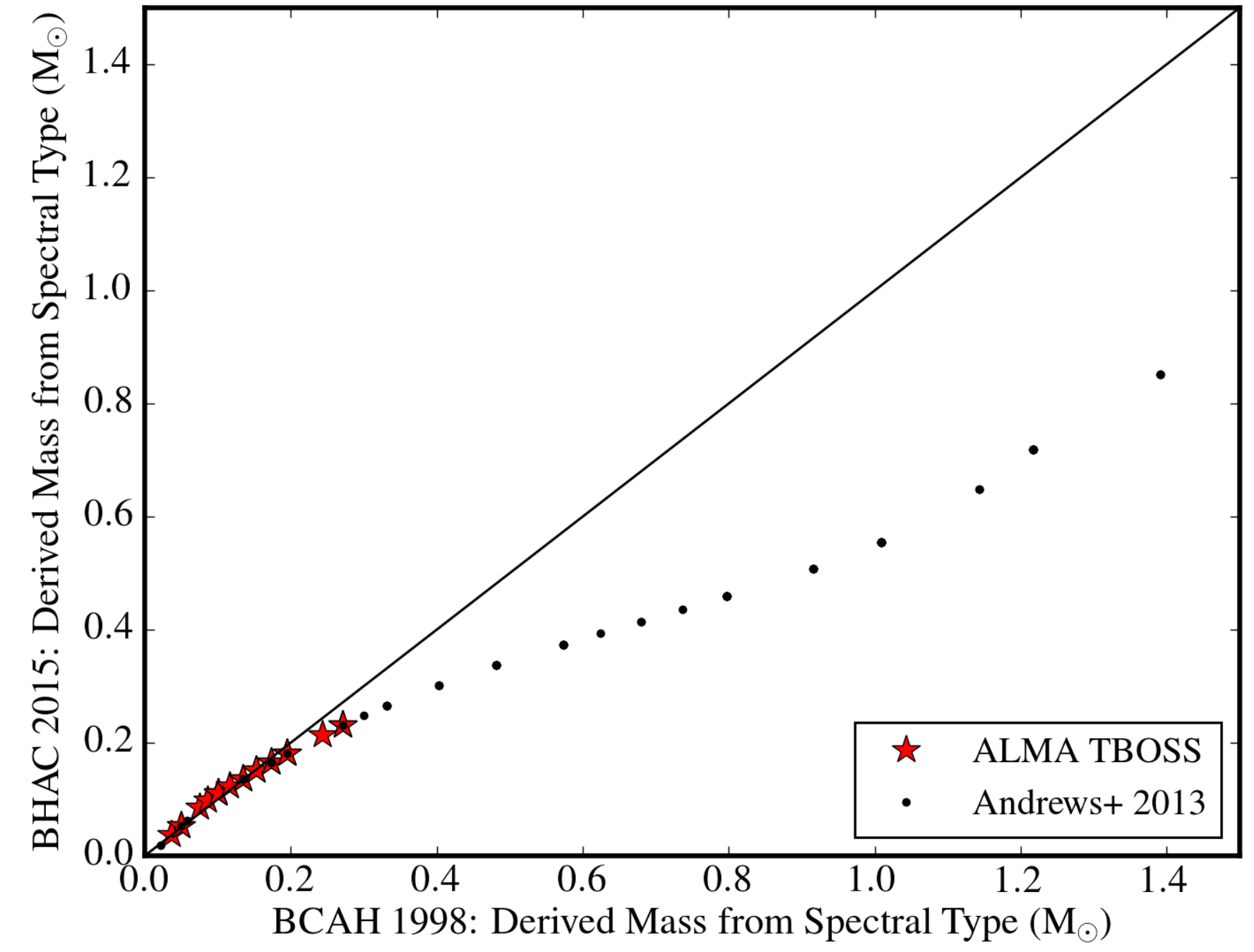}
    \caption{Comparison of stellar masses derived using the \citet{baraffe98} and \citet{baraffe15} models.}
    \label{fig:bcahmasses}
\end{figure}

\begin{figure}
    \centering
    \includegraphics[scale=0.45]{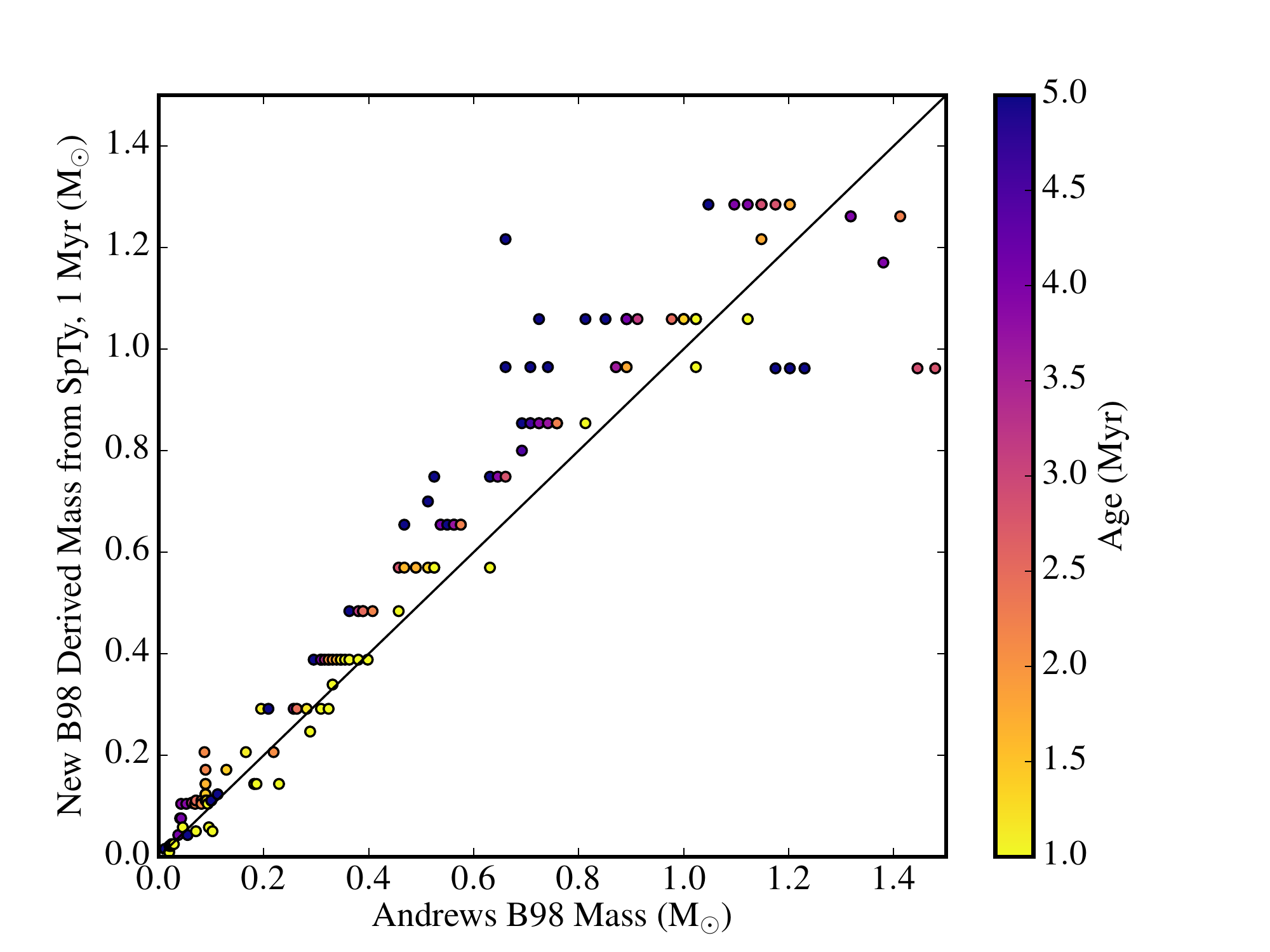}
    \caption{1 Myr comparison of the stellar masses derived using the approach described in this paper and the estimated stellar masses from \citet{andrews13}, both using the \citet{baraffe98} evolutionary model grid. The colorbar represents the derived age of the targets from the Bayesian analysis presented in \citet{andrews13}.}
    \label{fig:masscomp_age1}
\end{figure}

\begin{figure}
    \centering
    \includegraphics[scale=0.45]{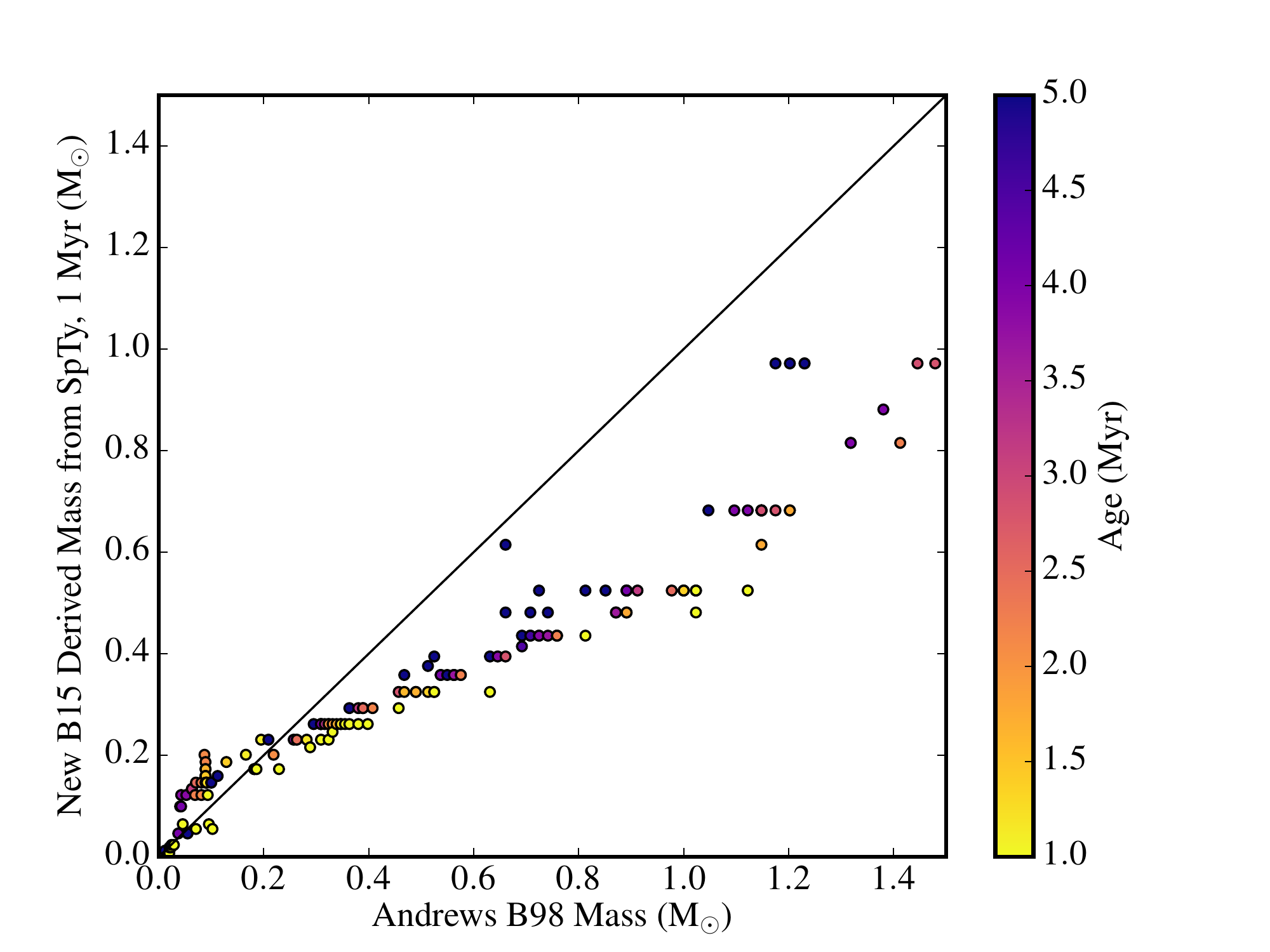}
    \caption{The same figure as in Figure~\ref{fig:masscomp1}, underscoring the difference between the older and newer evolutionary grids. The points are color-coded to represent the derived age of the targets from the Bayesian analysis presented in \citet{andrews13}.}
    \label{fig:masscomp_age2}
\end{figure}

\begin{figure}
    \centering
    \includegraphics[scale=0.45]{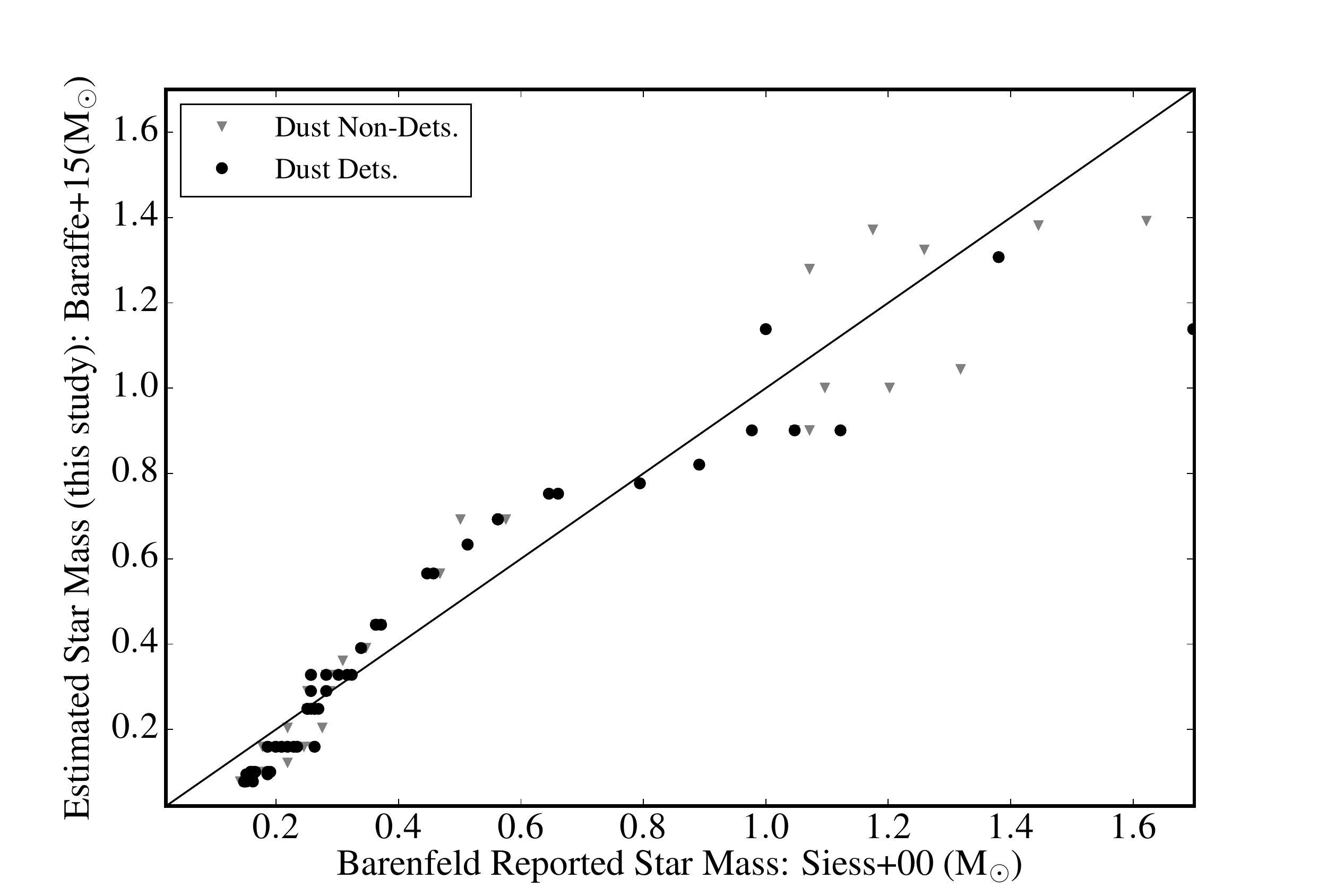}
    \caption{Comparison of the published stellar masses from \citep{barenfeld16} and the re-derived stellar masses for the same population used in this work, applying the uniform method described in Section~\ref{sec:diskmasses} and Appendix~\ref{sec:starmassestimation}.}
    \label{fig:masscomp_barenfeld}
\end{figure}

The comparison with the literature masses and mass estimated from the spectral type, age of 1~Myr and \mbox{\citet{baraffe98}} grid is given in Figure~\ref{fig:masscomp1} to consider the effect of using different approaches. The differences are negligible for objects fit with an age of 1~Myr, within $\pm$0.1 $M_{\odot}$ (comparable to the range from $\pm$0.5 uncertainty in spectral type) for masses $\leq$1$M_{\odot}$ and show a larger scatter of $\pm$0.3$M_{\odot}$ for the higher mass objects. Figure~\ref{fig:bcahmasses} shows the Taurus masses used in this study compared to the SMA survey, and the larger, systematic difference is dominated by the adoption of the newer model grid, following the trend shown in Figure~\ref{fig:mass_teff_vintage}. To understand whether the departure between older and newer vintages of the evolutionary model grids followed a systematic trend with the estimated ages of the targets from \mbox{\citet{andrews13}}, the estimated stellar mass comparisons are also shown color-coded with the ages from the Bayesian inference approach in Figures~\ref{fig:masscomp_age1} and \ref{fig:masscomp_age2}. While Taurus members with older assigned ages from \mbox{\citet{andrews13}} show a more significant departure from the 1:1 relation, as to be expected, the difference in model grids remains the dominant factor. A similar comparison was performed for the Upper Sco population from \citet{barenfeld16} in Figure~\ref{fig:masscomp_barenfeld}, demonstrating a similar significant impact of stellar model selection on estimated stellar mass, in this case between the \citet{baraffe15} models with the \citet{siess00} models.

\clearpage

\section{Comparison of estimated and literature dust masses}
\label{sec:DustMassComparisons}
As described in Section~\ref{sec:analyticdust}, the calculated disk dust mass is dependent upon the assumed dust temperature, $T_\textnormal{dust}$, which in turn can be related to the central object luminosity, $L_{\odot}$, by the following prescription:

\begin{equation}
\label{eq:generictdust}
\langle T_{dust} \rangle = A (L_{*}/L_{\odot})^{B} K .
\end{equation}

\begin{figure}
    \centering
    \includegraphics[scale=0.5]{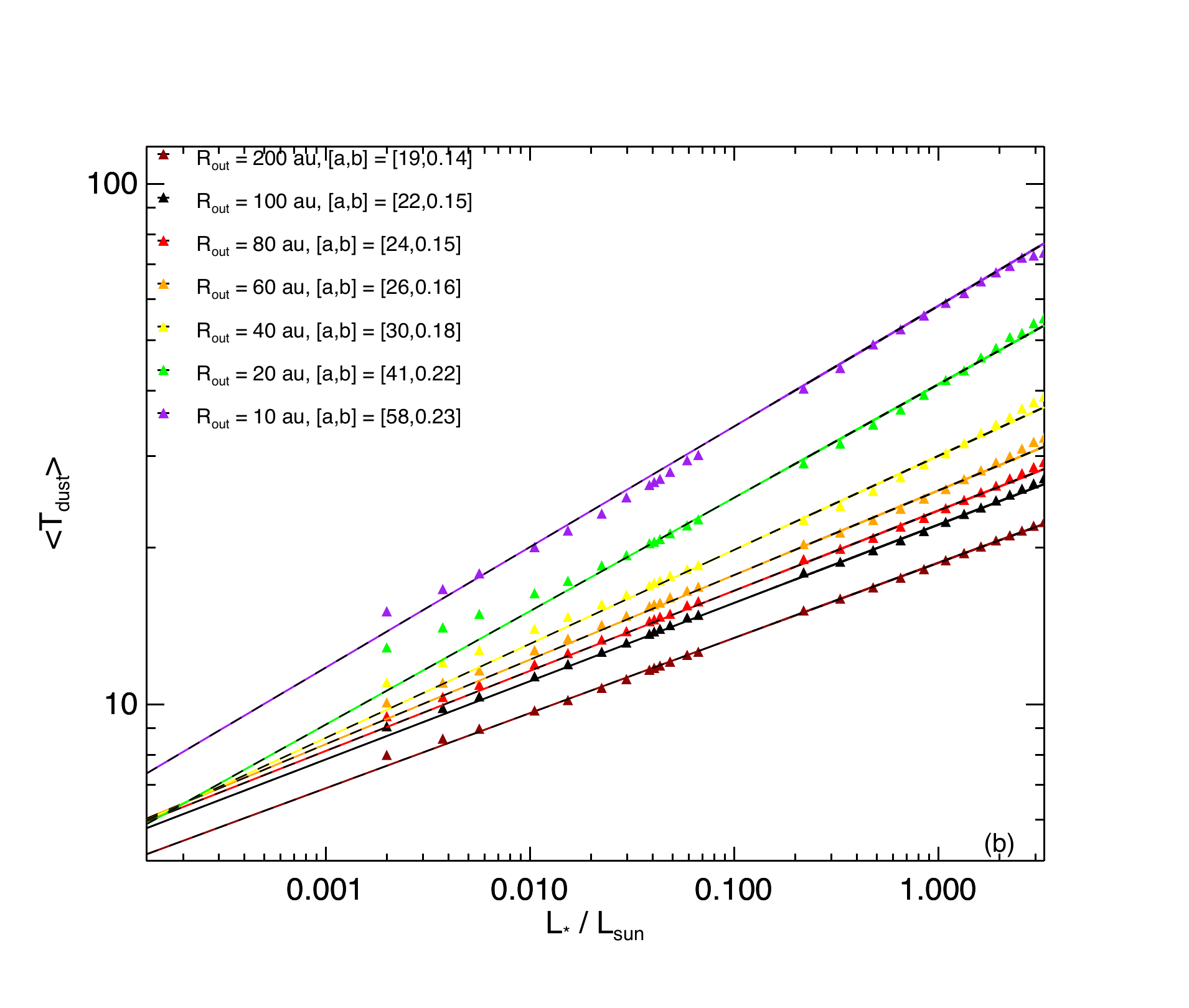}
    \caption{Dust temperature - luminosity scaling relations at 1 Myr, adopted for targets in the Taurus star forming region, and used in the estimation of dust temperatures for our TBOSS targets and those in \citet{andrews13}.}
    \label{fig:gvdp_1myr_relation}
\end{figure}

\begin{figure}
    \centering
    \includegraphics[scale=0.5]{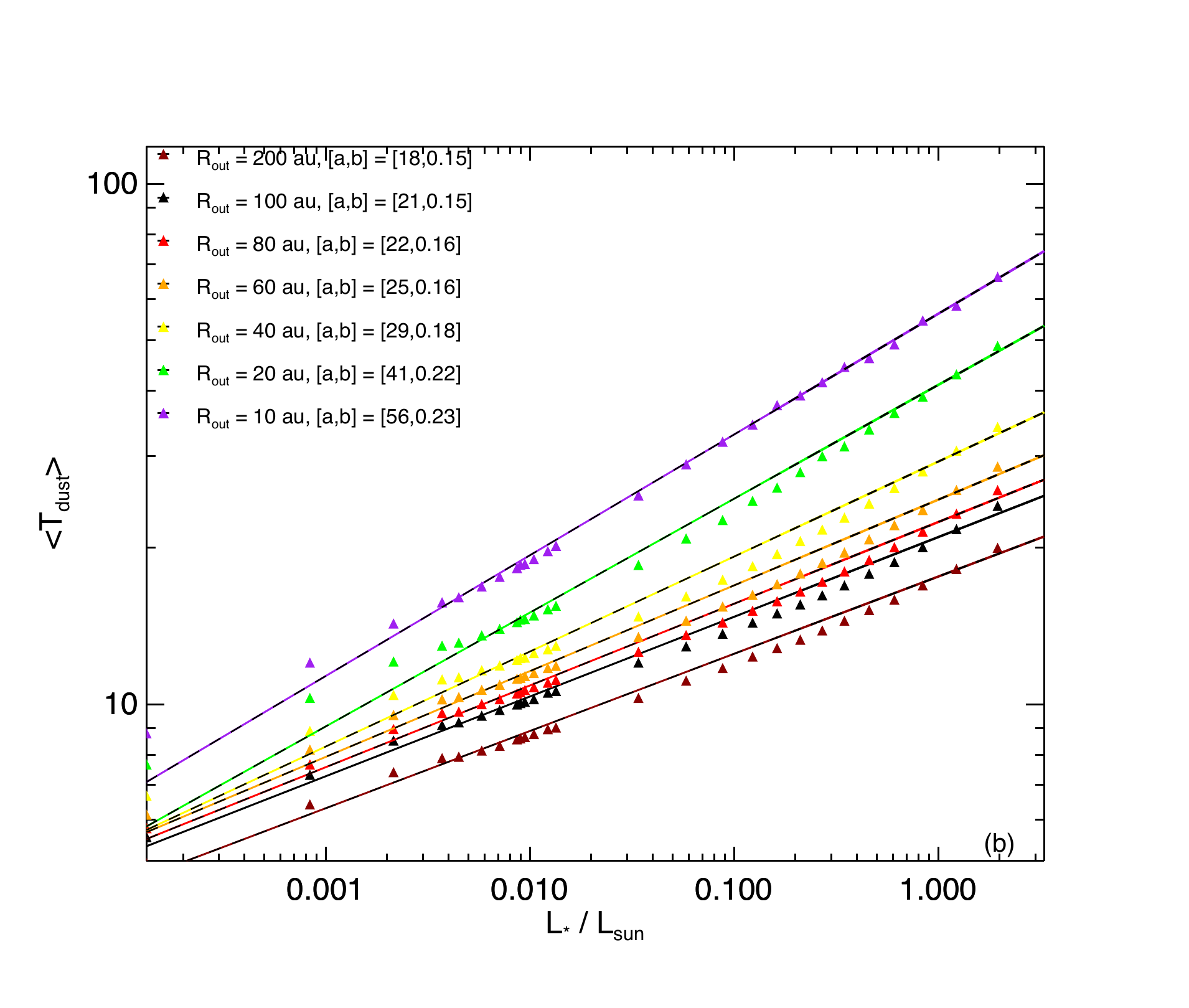}
    \caption{10 Myr scaling relations, used to enable comparison between our targets and older star forming regions; these relations were used for dust temperature estimation for targets in Upper Sco.}
    \label{fig:gvdp_10myr_relation}
\end{figure}

The power law coefficients, $A$ and $B$, are typically assumed to be $25$ and $1/4$, respectively, based upon stellar models of $L_{*} = 0.1 - 100 L_{\odot}$ and assuming disks of radius 100~au. Revised procedures to generate the dust temperature-luminosity scaling relations for lower-mass central objects, described in detail in \mbox{\citet{gvdp16}}, are outlined briefly here in their extension to the younger 1~Myr Taurus targets studied in this work. Relations are derived from grids of disk models which vary the parameters of luminosity and disk outer radius, while fixing scale height and profile, inclination, inner radius, surface density exponent, and disk mass ($\sim$1\% of the (sub)stellar mass). Figure~\ref{fig:gvdp_1myr_relation} shows the resulting power-law relation and coefficients for disks of different radii at 1~Myr, while Figure~\ref{fig:gvdp_10myr_relation} shows the same relations for the 10~Myr objects. Power-law coefficients are provided for the range of disk radii explored from 10~au to 200~au.

\begin{figure}
    \centering
    \includegraphics[scale=0.45]{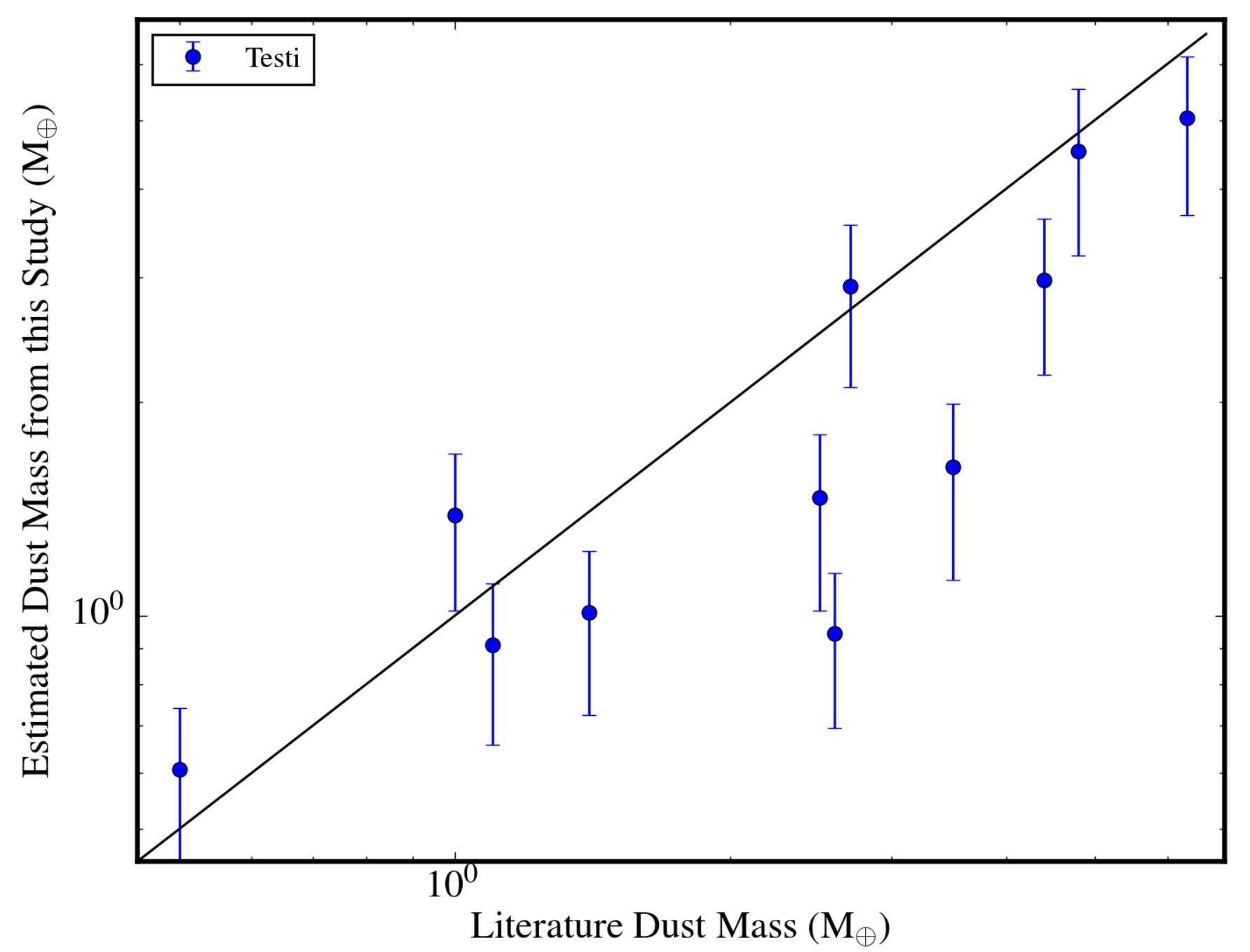}
    \caption{Comparison of dust mass measurements for a sample of objects from Rho Ophiuchus from \mbox{\citet{testi16}} and the uniform method of dust mass calculation presented in this study.}
    \label{fig:rho_oph_comparison}
\end{figure}

Following the approach outlined in Section~\ref{sec:analyticdust}, we then use the adopted dust temperatures from the revised power-law relations to infer the disk mass in sub-millimeter sized grains, using the flux density, dust opacity, and distance to the targets. To provide a uniform comparison with previous literature studies of disk-bearing objects in Upper Scorpius and Rho Ophiuchus, we also recalculated the dust masses for objects from previous studies using the same approach. Reported fluxes and spectral types from the previous surveys were used to re-estimate the stellar parameters of T$_\textnormal{eff}$ and $L_{\odot}$ (as described in Section~\ref{sec:starmassestimation}) using the \mbox{\citet{baraffe15}} evolutionary models.  The revised 1~Myr luminosity-dust temperature relations were applied to the Rho Oph population from \mbox{\citet{testi16}}, while the 10~Myr relations were applied to the Upper Sco population from \mbox{\citet{mathews12}}, \mbox{\citet{barenfeld16}}, and \mbox{\citet{gvdp16}}. For the Rho Oph population, the literature values are plotted against the new dust mass estimates in Figure~\ref{fig:rho_oph_comparison}, which show agreement at the 20 per cent level, with systematically higher previously-published values.

\begin{figure}
    \centering
    \includegraphics[width=0.65\textwidth]{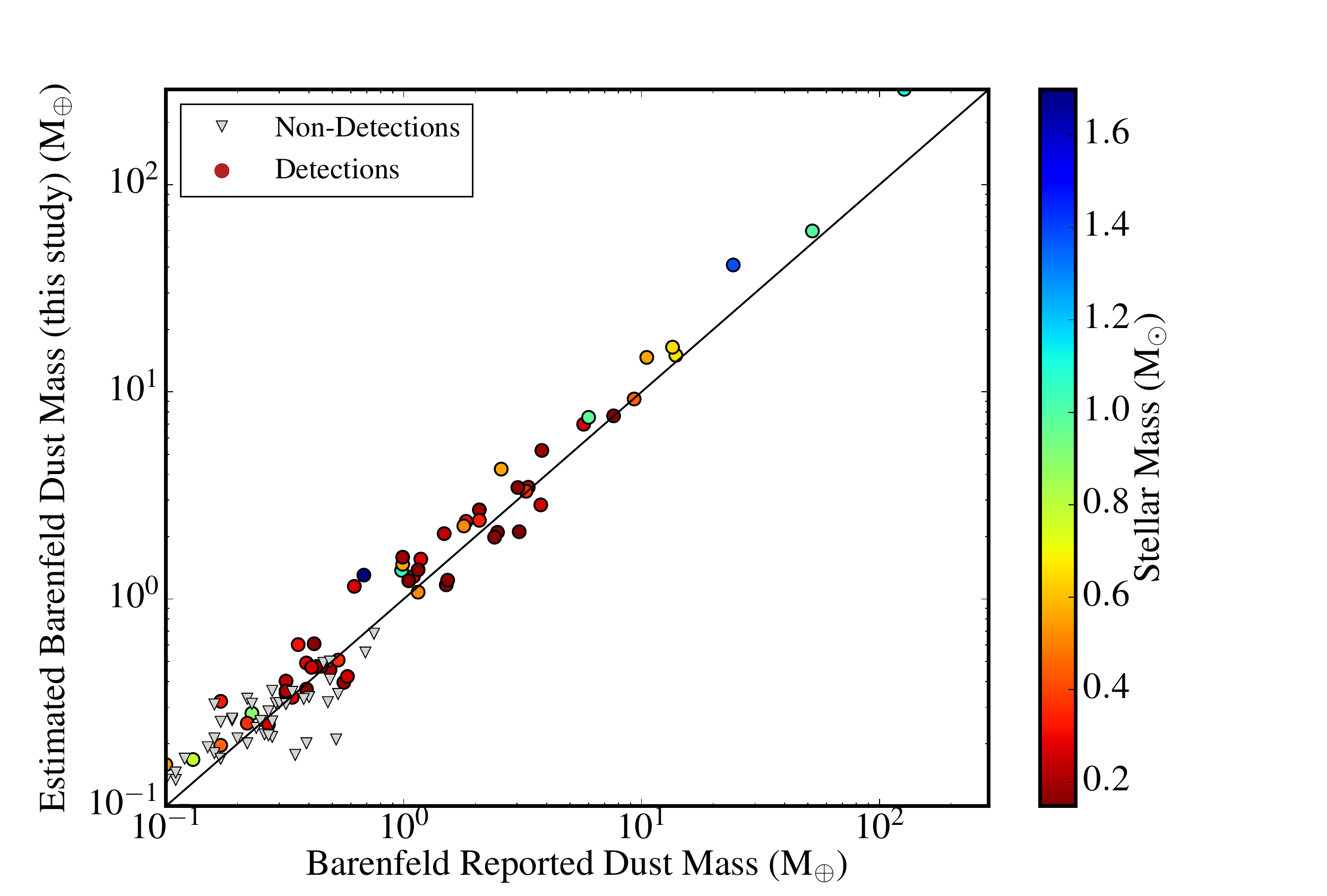}
    \caption{Comparison of the published Upper Sco dust masses from \citet{barenfeld16} and the dust masses re-derived for the same population, using the uniform method of dust mass calculation presented in this study.}
    \label{fig:dustcomp_barenfeld}
\end{figure}

\begin{figure}
    \centering
    \includegraphics[width=0.55\textwidth]{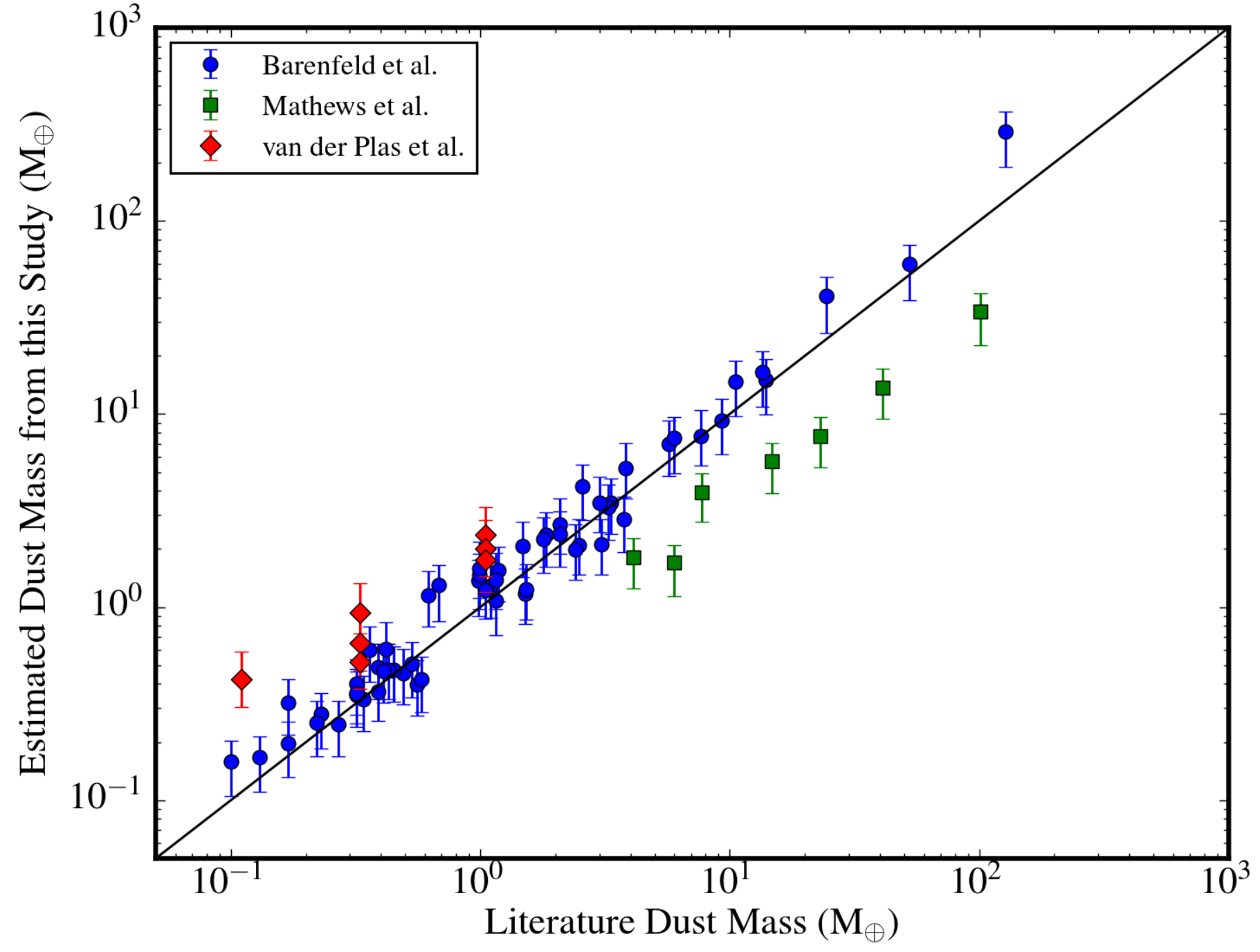}
    \caption{Dust mass measurements for the combined sample of Upper Sco objects from \mbox{\citet{mathews12}}, \mbox{\citet{barenfeld16}}, and \mbox{\citet{gvdp16}}, compared with the re-derived masses using the uniform method of dust mass calculation presented in this study.}
    \label{fig:uppersco_comparison}
\end{figure}

For the Upper Sco population, the literature and re-estimated dust mass comparison is shown for the \citet{barenfeld16} detections and upper limits in Figure~\ref{fig:dustcomp_barenfeld} and for the combined Upper Sco populations in Figure~\ref{fig:uppersco_comparison}. The literature dust masses and re-derived dust mass values using the approach described in this work are in close agreement, with no strong dependence on stellar mass, emphasizing the dependence of $M_{dust}$-$M_{*}$ relations on the choice of stellar model, as illustrated in Appendix~\ref{sec:starmassestimation}. The previous literature measurements and the re-derived dust masses from the uniform approach in this study agree at the 11 per cent level across all three Upper Sco surveys, with the higher-mass objects from \mbox{\citet{mathews12}} systematically higher and the \mbox{\citet{gvdp16}} population systematically lower. Offsets for the lowest and highest-mass members presented in \mbox{\citet{gvdp16}} and \mbox{\citet{mathews12}} may be explained by reporting of masses derived from radiative transfer modeling, similar to the differences between the analytic approach and model-estimated masses seen in this study (Section~\ref{sec:mcfost}). The dust mass values recalculated in this work for each previous literature study is given in full in Table~\ref{tab:recalc_dustmasses}.

\clearpage

\section{MCFOST Results}
\label{sec:mcfostseds}

\begin{figure*}
    \centering
    
    \includegraphics[width=1.0\textwidth]{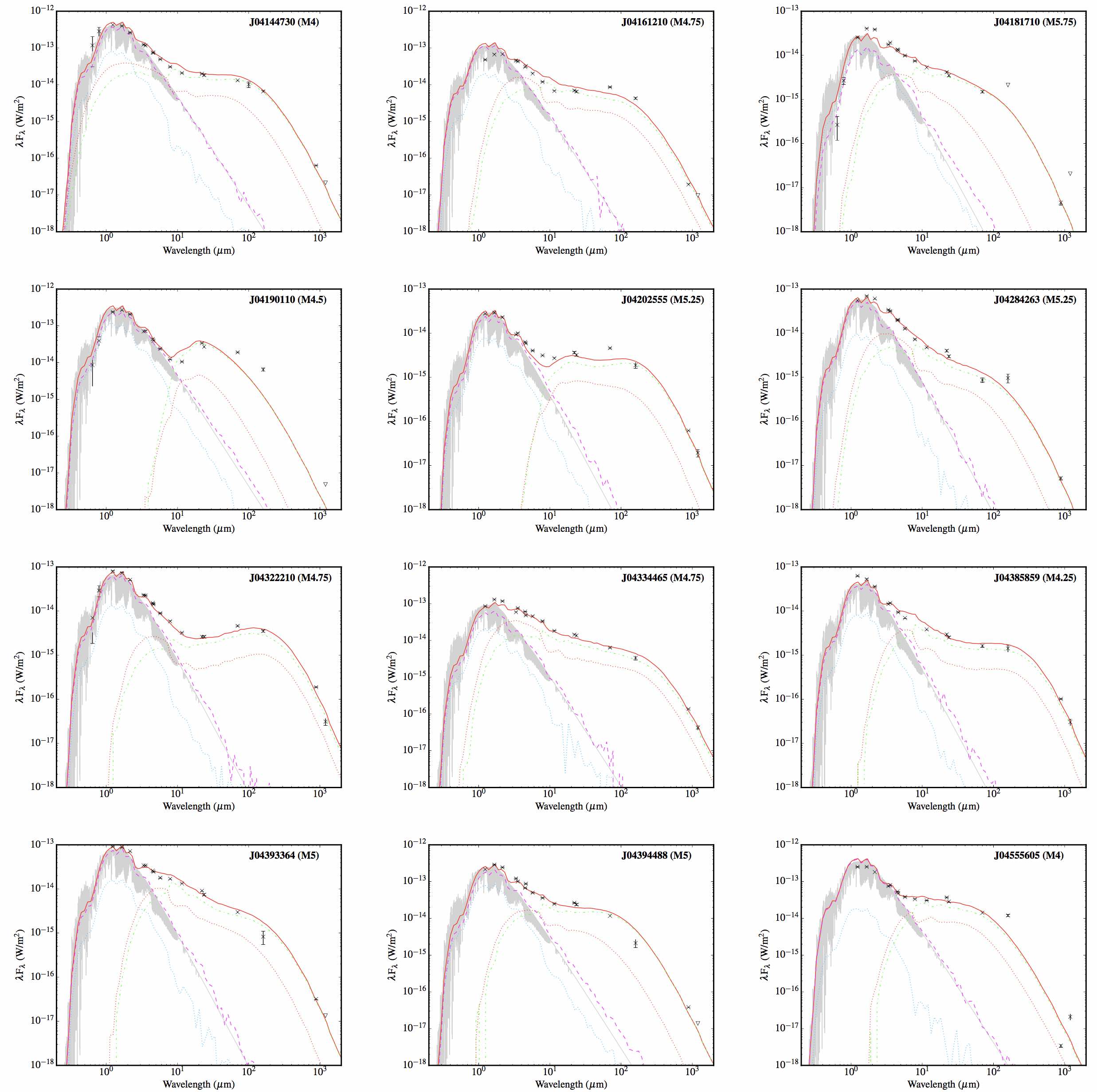}
    
    \caption{Spectral energy distributions (SEDs) with best-fit models from MCFOST, for the stellar targets within this survey (M4-M5.75) in order of increasing right ascension. Fluxes as a function of wavelength ($\mu$m; black symbols) are compiled from \mbox{\citet{bulger14}} with the new 885$\mu$m measurements from this study, and upper limits are denoted with downward triangles where applicable. In addition to the de-reddened stellar photosphere (gray) and best-fit full SED (red solid line), the contributing disk components are given by the following lines: scattered light (blue dotted); direct starlight (magenta dashed); thermal emission (red dot-dashed); and scattered thermal emission (green dot-dot dashed).}
    \label{fig:mcfost_seds}
\end{figure*}

\begin{figure*}
    \centering
    \includegraphics[width=1.0\textwidth]{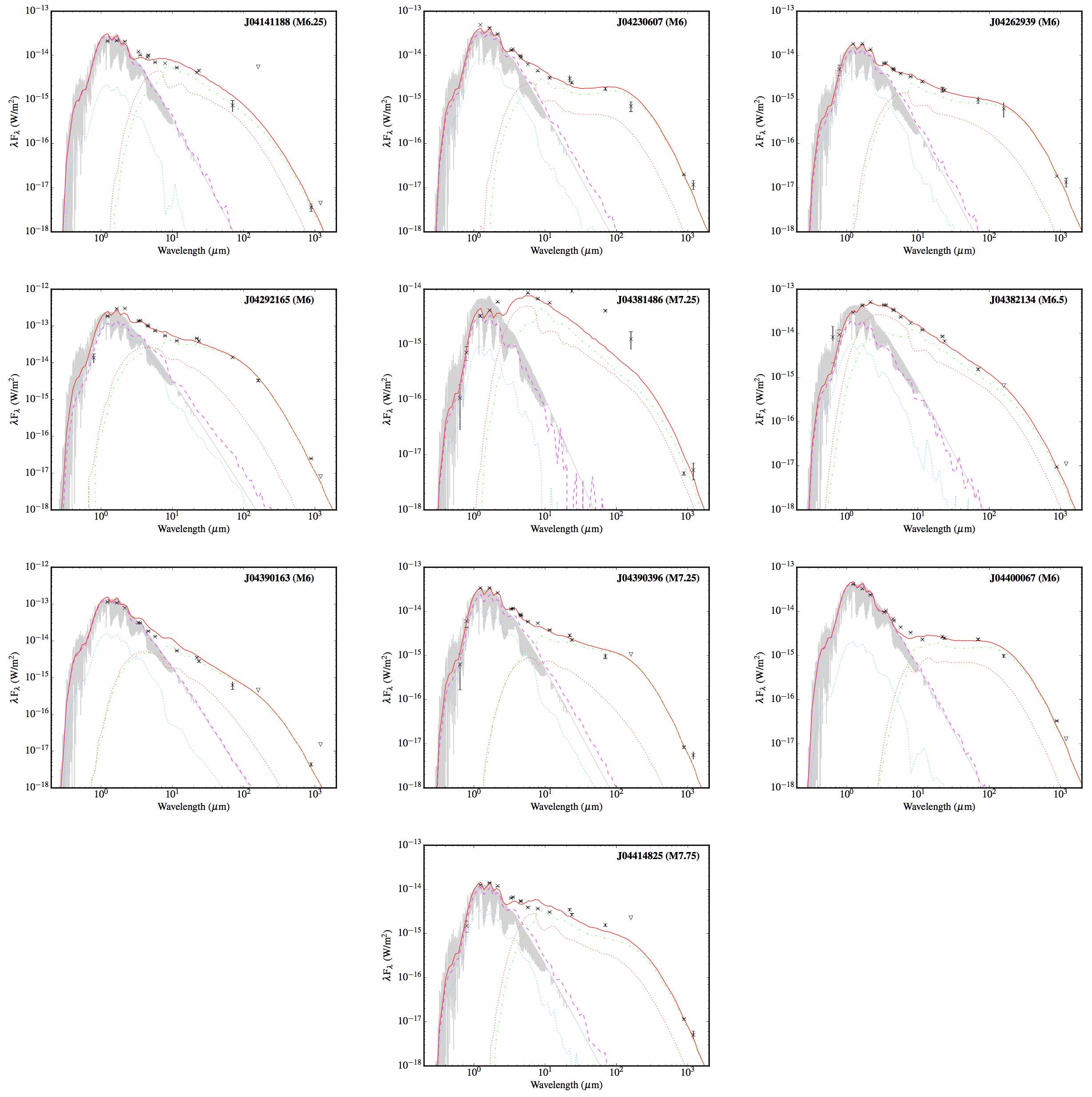}
    
    \caption{Spectral energy distributions (SEDs) with best-fit models from MCFOST, for the substellar targets within this survey (M6 and later) in order of increasing right ascension. Symbols and lines are as in Figure~\ref{fig:mcfost_seds}.}
    \label{fig:mcfost_seds_BDs}
\end{figure*}

This Appendix provides the spectral energy distributions for the targets within our survey, overlaid with the resulting best-fit MCFOST models as described in Section~\ref{sec:mcfost}. Data and fits for stellar hosts are shown in Figure~\ref{fig:mcfost_seds} and for brown dwarfs in Figure~\ref{fig:mcfost_seds_BDs}. 

\clearpage

\section{Table of Stellar Parameters and Disk Dust Masses From a Uniform Approach}
\label{sec:staranddiskparams}
In this Appendix, we provide the full set of stellar parameters and associated disk dust masses for the star-forming regions described in this paper, applying the uniform methodology described in Appendices~\ref{sec:starmassestimation} and \ref{sec:DustMassComparisons}. These results are summarized in Table~\ref{tab:recalc_dustmasses}. In Table~\ref{tab:dustmasses}, we provide the full table of analytic dust masses for the TBOSS ALMA sample, corresponding to dust temperatures for disks of radii ranging from 10 au -- 200 au.


\input{recalc_dustmass_wULs.txt}


\clearpage

\input{dustmass_updated.txt}

\end{document}

%% file: sample.txt
\begin{table*}
\caption{Sample table.}
\label{tab:sample}
\footnotesize
\begin{tabular}{llllllllll}
\hline
\hline
Target                & Other Name       & 2MASS RA     & 2MASS Dec    & SpTy  & F$_{24}$  & F$_{70}$  & F$_{160}$ & Notes                                       & Reference \\
                &        &    (J2000)   &  (J2000)   &   & (mJy) & (mJy) & (mJy) &                                        &  \\
\hline
J04144730+2646264  & FP Tau          & 04 14 47.309 & +26 46 26.44 & M4    & 143       & 307       & 351        & Transition (homologously depleted)     &      (1)     \\
J04555605+3036209  & XEST 26-062     & 04 55 56.055 & +30 36 20.96 & M4    & 226       & 330       & 639        &                                             &           \\
J05075496+2500156  & CIDA 12         & 05 07 54.966 & +25 00 15.61 & M4    & 0         & 51        & 44         &                                             &           \\
J04385859+2336351  &                 & 04 38 58.599 & +23 36 35.16 & M4.25 & 20        & 38        & 76         &                                             &           \\
J04190110+2819420  & V410 X-ray 6    & 04 19 01.106 & +28 19 42.05 & M4.5  & 213       & 445       & 342        & Transition (giant planet-forming)      &      (2)     \\
J04161210+2756385  &                 & 04 16 12.104 & +27 56 38.58 & M4.75 & 51        & 201       & 228        & Transition                             &    (3)       \\
J04322210+1827426  & MHO 6           & 04 32 22.109 & +18 27 42.64 & M4.75 & 20.7      & 107       & 188        & Transition                             &   (3)         \\
J04334465+2615005  &                 & 04 33 44.652 & +26 15 00.53 & M4.75 & 108       & 149       & 178        &                                             &           \\
J04393364+2359212  &                 & 04 39 33.645 & +23 59 21.23 & M5    & 59        & 70        & 44         &                                             &           \\
J04394488+2601527  & ITG 15          & 04 39 44.883 & +26 01 52.79 & M5    & 187       & 272       & 114        & Binary: $\rho \sim 3''$                     &     (4)      \\
J04202555+2700355  &                 & 04 20 25.554 & +27 00 35.55 & M5.25 & 25        & 107       & 100        & Transition (primordial disk)           &     (3), (1)      \\
J04284263+2714039  &                 & 04 28 42.635 & +27 14 03.91 & M5.25 & 24        & 20        & 51         & Transition, Binary: $\rho \sim 0''.63$ &      (2)     \\
J04213459+2701388  &                 & 04 21 34.599 & +27 01 38.85 & M5.5  & 9.6       & 37        & 101        & Transition                             &     (3)      \\
J04181710+2828419 & V410 Anon 13    & 04 18 17.106 & +28 28 41.92 & M5.75 & 28        & 35        & $<$113     &                                             &           \\
J04230607+2801194  &                 & 04 23 06.073 & +28 01 19.49 & M6    & 19        & 41        & 38         &                                             &           \\
J04262939+2624137  & KPNO 3          & 04 26 29.392 & +26 24 13.79 & M6    & 12.9      & 23        & 33         &                                             &           \\
J04292165+2701259  & IRAS 04263+2654 & 04 29 21.653 & +27 01 25.95 & M6    & 310       & 329       & 176        & Binary: $\rho \sim 0''.2$                   &   (5)        \\
J04390163+2336029  &                 & 04 39 01.631 & +23 36 02.99 & M6    & 22        & 15        & $<$24      &                                             &           \\
J04400067+2358211  &                 & 04 40 00.676 & +23 58 21.17 & M6    & 20        & 55        & 52         &                                             &           \\
J04141188+2811535  &                 & 04 14 11.881 & +28 11 53.51 & M6.25 & 36        & 17        & $<$293     & Truncated                              &     (3)      \\
J04382134+2609137  & GM Tau          & 04 38 21.340 & +26 09 13.74 & M6.5  & 53        & 36        & $<$35      &                                             &           \\
J04381486+2611399  &                 & 04 38 14.861 & +26 11 39.94 & M7.25 & 73        & 95        & 67         &                                             &           \\
J04390396+2544264  & CFHT 6          & 04 39 03.960 & +25 44 26.42 & M7.25 & 18        & 23        & $<$56      &                                             &           \\
J04414825+2534304  &                 & 04 41 48.250 & +25 34 30.50 & M7.75  & 21        & 37        & $<$122     &                                             &          \\

\hline
\multicolumn{9}{l}{\textbf{References for transition disks and binary system identifications.} (1) \citet{currie_siciliaaguilar11}; (2) \citet{cieza12}; (3) \citet{bulger14};}\\
\multicolumn{9}{l}{(4) \citet{itoh99}; (5) \citet{konopacky07}}
\end{tabular}
\end{table*}

%% file: observations.txt
\begin{table*}
\caption{Observations.}
\label{tab:obs}
\centering
\begin{tabular}{ccccccccc}
\hline
\hline
Group    & Obs. UT Dates   & Antennas   & Time on Target   & Baseline Lengths   & Median PWV   & \multicolumn{3}{c}{Calibrators:}          \\
         &                     &            & (min)                   & (m)                & (mm)         & Flux        & Bandpass     & Gain        \\
\hline
Taurus1  & 2013-11-05          & 31         & 22:57                   & 17.3 -- 1300       & 0.91         & J0238+166   & J0423-0120   & J0510+1800  \\
         & 2013-11-05          & 31         & 31:18                   & 17.3 -- 1300       & 1.13         & J0510+180   & J0423-0120   & J0510+1800  \\
         & 2014-07-26          & 30         & 25:59                   & 33.7 -- 820.2      & 0.36         & J0238+166   & J0510+1800   & J0510+1800  \\
         &                     &            &                         &                    &              &             &              &             \\
         & \multicolumn{8}{l}{\textit{Targets}: J04141188, J04230607, J04262939, J04292165, J04381486, J04382134, J04390163, J04390396, J04400067, J04414825} \\
         &                     &            &                         &                    &              &             &              &             \\
\hline
Taurus2a & 2013-11-19          & 28         & 41:43                   & 17.3 -- 1300       & 0.58         & J0510+180   & J0423-0120   & J0509+1806  \\
         & 2014-07-27          & 33         & 20:47                   & 24.2 -- 820.2      & 0.5          & J0510+180   & J0510+1800   & J0510+1800  \\
         &                     &            &                         &                    &              &             &              &             \\
         & \multicolumn{8}{l}{\textit{Targets}: J04144730, J04161210, J04181710, J04190110, J04202555, J04213459, J04284263, J04322210}                       \\
         &                     &            &                         &                    &              &             &              &             \\
\hline
Taurus2b & 2013-11-17          & 29         & 18:45                   & 17.3 -- 1300       & 0.77         & J0510+180   & J0423-0120   & J0509+1806  \\
         & 2014-07-27          & 33         & 15:35                   & 24.2 -- 820.2      & 0.36         & J0510+180   & J0510+1800   & J0510+1800  \\
         &                     &            &                         &                    &              &             &              &             \\
         & \multicolumn{8}{l}{\textit{Targets}: J04334465, J04385859, J04393364, J04394488, J04555605, J05075496}                                            \\
\hline
\end{tabular}
\end{table*}

%% file: positions.txt
\begin{table*}
\caption{Updated target positions from this study, and proper motions from \citet{zacharias15}.}
\label{tab:positions}
\centering
\begin{tabular}{llccccc}
\hline
\hline
\multicolumn{1}{c}{Target} & \multicolumn{1}{c}{J2000 Position (ALMA)} & \multicolumn{2}{c}{Offset from J2000 2MASS} & $\mu_\textnormal{RA}$   & $\mu_\textnormal{Dec}$   & Epoch \\
                           &                                           & RA (mas)             & Dec (mas)            & mas/yr       & mas/yr        &             \\
\hline
J04141188                  & 04 14 11.8872  +028 11 52.8848            & 81.963               & -625.2               & 8$\pm$5.3    & -26.7$\pm$5.3 & 2013.310    \\
J04144730                  & 04 14 47.3215  +026 46 26.1018            & 167.398              & -338.2               & 5.1$\pm$5.2  & -21.6$\pm$5.2 & 2013.424    \\
J04161210                  & 04 16 12.1253  +027 56 38.1025            & 282.248              & -477.5               & 11$\pm$5.2   & -29.5$\pm$5.2 & 2013.341    \\
J04181710                  & 04 18 17.1158  +028 28 41.6474            & 129.213              & -272.6               & 4.7$\pm$5.5  & -19.7$\pm$5.5 & 2013.108    \\
J04202555                  & 04 20 25.5760  +027 00 35.2819            & 294.006              & -268.1               & 14.3$\pm$5.3 & -19.6$\pm$5.3 & 2013.395    \\
J04230607                  & 04 23 06.0891  +028 01 19.1665            & 213.188              & -323.5               & 13.4$\pm$5.2 & -23.2$\pm$5.2 & 2013.335    \\
J04262939                  & 04 26 29.4038  +026 24 13.4991            & 158.536              & -290.9               & 9.3$\pm$5.3  & -20.4$\pm$5.3 & 2013.248    \\
J04284263                  & 04 28 42.6452  +027 14 03.3013            & 136.039              & -608.7               & -5.1$\pm$5.2 & -11.7$\pm$5.2 & 2013.326    \\
J04292165                  & 04 29 21.6580  +027 01 25.5845            & 66.811               & -365.5               & 5.5$\pm$5.2  & -22.7$\pm$5.2 & 2013.342    \\
J04322210                  & 04 32 22.1273  +018 27 42.4070            & 260.373              & -233                 & 13.8$\pm$6.3 & -16.8$\pm$6.3 & 2013.571    \\
J04334465                  & 04 33 44.6685  +026 15 00.1949            & 221.976              & -335.1               & 11.2$\pm$5.2 & -17.3$\pm$5.2 & 2013.499    \\
J04381486                  & 04 38 14.8866  +026 11 39.6288            & 344.564              & -311.2               & 7.8$\pm$10   & -17.8$\pm$10  & 2013.856    \\
J04382134                  & 04 38 21.3433  +026 09 13.4528            & 44.432               & -287.2               & 2.1$\pm$5.6  & -12.8$\pm$5.6 & 2013.309    \\
J04385859                  & 04 38 58.6108  +023 36 34.8674            & 162.184              & -292.6               & 11.5$\pm$5.5 & -19.8$\pm$5.5 & 2013.572    \\
J04390163                  & 04 39 01.6425  +023 36 02.6857            & 158.072              & -304.3               & 10.7$\pm$5.4 & -21.2$\pm$5.4 & 2013.681    \\
J04390396                  & 04 39 03.9673  +025 44 26.1032            & 98.634               & -316.8               & 4.7$\pm$5.5  & -19.9$\pm$5.5 & 2013.598    \\
J04393364                  & 04 39 33.6491  +023 59 20.9331            & 56.188               & -296.9               & 4.3$\pm$5.4  & -20.1$\pm$5.4 & 2013.831    \\
J04394488                  & 04 39 44.8920  +026 01 52.3806            & 121.305              & -409.4               & 2.9$\pm$5.5  & -21$\pm$5.5   & 2013.518    \\
J04400067                  & 04 40 00.6799  +023 58 20.7921            & 53.454               & -377.9               & 3.2$\pm$5.5  & -23.5$\pm$5.5 & 2013.579    \\
J04414825                  & 04 41 48.2591  +025 34 30.2815            & 123.126              & -218.5               & -1.8$\pm$6   & -9.7$\pm$6    & 2013.737    \\
J04555605                  & 04 55 56.0714  +030 36 20.4410            & 211.73               & -519                 & 6.3$\pm$5.1  & -30.5$\pm$5.1 & 2013.739    \\
J05075496                  & 05 07 54.9702  +025 00 15.3837            & 57.095               & -226.3               & 2.3$\pm$5.1  & -13.3$\pm$5.1 & 2013.609   \\
\hline
\end{tabular}
\end{table*}

%% file: fluxtable.txt
\begin{table*}
\caption{Measured flux density values for the 24 targets in this sample, with the spectral types and corresponding estimated effective temperatures, luminosities, and masses for the central objects. SC corresponds to sources for which self-calibration has been performed, and upper limits denote 3 $\times$ the RMS in the residual image.}
\label{tab:fluxes}
\centering
\begin{tabular}{llcccccccc}
\hline
\hline
&       & T$_\textnormal{eff}$ & log$L_{*}$  & $M_{*}$(B15) & Natural Weighting & Briggs     Weighting & Uniform Weighting & \textit{uvmodelfit}$^{\dagger}$ & Note \\
Target    & SpTy  & (K)       & ($L_{\odot}$) & ($M_{\odot}$)  & Flux (mJy)        & Flux (mJy)           & Flux (mJy)        & Flux (mJy)  &         \\
\hline
J04292165 & M6    & 2858      & -1.566         & 0.058       & 7.35$\pm$0.19     & 7.21$\pm$0.34    & 6.98$\pm$0.38     & 7.28$\pm$0.22       &          \\
J04141188 & M6.25 & 2836      & -1.628         & 0.053       & 1.06$\pm$0.21     & 0.71$\pm$0.20    & $\leq$0.55        & 1.25$\pm$0.30       &          \\
J04230607 & M6    & 2858      & -1.566         & 0.058       & 5.94$\pm$0.24     & 5.68$\pm$0.35    & 5.7$\pm$0.4       & 6.36$\pm$0.23       &          \\
J04262939 & M6    & 2858      & -1.566         & 0.058       & 5.4$\pm$0.13      & 5.7$\pm$0.22     & 5.58$\pm$0.25     & 5.61$\pm$0.15       &          \\
J04381486 & M7.25 & 2747      & -1.881         & 0.035       & 1.36$\pm$0.11     & 1.66$\pm$0.24    & 1.62$\pm$0.30     & 1.57$\pm$0.16       &          \\
J04382134 & M6.5  & 2814      & -1.689         & 0.048       & 2.8$\pm$0.12      & 2.62$\pm$0.18    & 2.62$\pm$0.21     & 2.75$\pm$0.15       &          \\
J04390163 & M6    & 2858      & -1.566         & 0.058       & 1.3$\pm$0.16      & 1.13$\pm$0.26    & 1.48$\pm$0.59     & 1.73$\pm$0.26       &          \\
J04390396 & M7.25 & 2747      & -1.881         & 0.035       & 2.46$\pm$0.13     & 2.33$\pm$0.21    & 2.23$\pm$0.23     & 2.28$\pm$0.15       &          \\
J04400067 & M6    & 2858      & -1.566         & 0.058       & 9.74$\pm$0.25     & 7.81$\pm$0.21    & 7.84$\pm$0.23     & 7.93$\pm$0.15       & SC       \\
J04414825 & M7.75 & 2696      & -2.02          & 0.028       & 3.41$\pm$0.14     & 3.34$\pm$0.22    & 3.45$\pm$0.26     & 3.52$\pm$0.16       &          \\
J04144730 & M4    & 3191      & -0.701         & 0.199       & 18.68$\pm$0.26    & 15.03$\pm$0.50   & 15.05$\pm$0.58    & 14.96$\pm$0.19      & SC       \\
J04161210 & M4.75 & 3027      & -0.959         & 0.135       & 5.71$\pm$0.17     & 5.47$\pm$0.30    & 5.38$\pm$0.38     & 5.84$\pm$0.19       &          \\
J04181710 & M5.75 & 2883      & -1.488         & 0.053       & 1.35$\pm$0.16     & 1.18$\pm$0.21    & 1.14$\pm$0.24     & 1.51$\pm$0.19       &          \\
J04202555 & M5.25 & 2943      & -1.15          & 0.101       & 18.31$\pm$0.27    & 14.43$\pm$0.40   & 14.21$\pm$0.44    & 15.32$\pm$0.19      & SC       \\
J04284263 & M5.25 & 2943      & -1.15          & 0.101       & 1.53$\pm$0.14     & 1.67$\pm$0.24    & 1.76$\pm$0.31     & 1.89$\pm$0.23       &          \\
J04322210 & M4.75 & 3027      & -0.959         & 0.135       & 55.65$\pm$0.54    & 48.28$\pm$0.75   & 48.4$\pm$0.85     & 47.77$\pm$0.23      & SC       \\
J04334465 & M4.75 & 3027      & -0.959         & 0.135       & 40.25$\pm$0.37    & 35.18$\pm$0.68   & 35.18$\pm$0.71    & 36.33$\pm$0.23      & SC       \\
J04385859 & M4.25 & 3133      & -0.777         & 0.177       & 30.00$\pm$0.30    & 26.75$\pm$0.45   & 26.56$\pm$0.49    & 28.29$\pm$0.22      & SC       \\
J04393364 & M5    & 2982      & -1.056         & 0.117       & 9.46$\pm$0.25     & 8.09$\pm$0.23    & 8.19$\pm$0.27     & 7.97$\pm$0.16       & SC       \\
J04394488 & M5    & 2982      & -1.056         & 0.117       & 11.26$\pm$0.27    & 9.01$\pm$0.25    & 8.7$\pm$0.26      & 9.44$\pm$0.16       & SC       \\
J04555605 & M4    & 3191      & -0.701         & 0.199       & 1.01$\pm$0.10     & 1.61$\pm$0.36    & 1.66$\pm$0.51     & 1.47$\pm$0.15       &          \\
J05075496 & M4    & 3191      & -0.701         & 0.199       & 2.88$\pm$0.12     & 2.9$\pm$0.23     & 2.93$\pm$0.28     & 3.06$\pm$0.17       &          \\
J04213459 & M5.5  & 2911      & -1.236         & 0.088       & $\leq$ 0.29       & $\leq$ 0.41      & $\leq$ 0.48       & --                  & Non-Det. \\
J04190110 & M4.5  & 3078      & -0.864         & 0.155       & $\leq$ 0.27       & $\leq$ 0.39      & $\leq$ 0.47       & --                  & Non-Det.\\
\hline
\multicolumn{5}{l}{$^{\dagger}$ Typical reduced $\chi^{2}$ value for fit is $\sim$1.45}
\end{tabular}
\end{table*}

%% file: binarytable.txt
\begin{table}
\caption{Binary companion candidates and their corresponding ALMA measurements within this study. Candidates denoted by $^{*}$ are not spatially resolved in the ALMA maps.}
\label{tab:binarytable}
\centering
\begin{tabular}{ccccc}
\hline
\hline
          & Cand. Flux & Sep. & Pos. Ang. &                             \\
System    & (mJy)                                & (asec)   & (deg)     & Ref.                        \\
\hline
J04181710 & 0.99 $\pm$ 0.16     & 9.6        & 77.6     & (1)     \\
J04394488 & 1.61 $\pm$ 0.18     & 3.1        & 324.8    & (1), (2) \\
J04202555 & $\leq$ 0.42         & 4.62       & 267.6    & (3)     \\
J04230607 & $\leq$ 0.42         & 6.44       & 291.6    & (3)     \\
J04284263 & $^{*}$        & 0.64       & 10       & (4)     \\
J04292165 & $^{*}$        & 0.22       & 268.6    & (5)     \\
\hline
\multicolumn{5}{l}{\textbf{Ref.} (1) This work; (2) \citet{itoh99}; (3) \citet{kraus12}};\\
\multicolumn{5}{l}{(4) \citet{cieza12};(5) \citet{konopacky07}.}
\end{tabular}
\end{table}

%% file: tdust_relations.txt
\begin{table}
\centering
\caption{1~Myr disk dust temperature power law coefficients for low-luminosity central objects (typically $L < 0.1 L_{\odot}$), from the relations provided in \citet{gvdp16}.}
\label{tab:tdust}
\begin{tabular}{ccc}
\hline
\hline
\multicolumn{3}{c}{$T_{\textnormal{dust}}=A(L_{*}/L_{\odot})^{B}$} \\
Disk Outer Radius       & Amplitude       & Index      \\
(au)                    & (A)             & (B)        \\
\hline
10                      & 58              & 0.23       \\
20                      & 41              & 0.22       \\
40                      & 30              & 0.18       \\
60                      & 26              & 0.16       \\
80                      & 24              & 0.15       \\
100                     & 22              & 0.15       \\
200                     & 19              & 0.14       \\
\hline
\end{tabular}
\end{table}

%% file: abridged_dustmass.txt
\begin{deluxetable}{ccccccc}


\tabletypesize{\footnotesize}


\tablecaption{Dust Masses \label{tab:abridged_dustmasses}}


\tablehead{
\colhead{Target} & \twocolhead{R=40 au} & \twocolhead{R=100 au} & \twocolhead{R=200 au} \\ 
 \colhead{} & \colhead{$T_{d}$ (K)} & \colhead{$M_{d} (\Earth)$} & \colhead{$T_{d}$ (K)} & \colhead{$M_{d} (\Earth)$} & \colhead{$T_{d}$ (K)} & \colhead{$M_{d} (\Earth)$} }

\startdata
J04292165 & 15.7 & 2.72  & 12.8 & 3.82  & 11.5 & 4.67  \\
J04141188 & 15.3 & 0.41  & 12.5 & 0.57  & 11.2 & 0.70  \\
J04230607 & 15.7 & 2.20  & 12.8 & 3.09  & 11.5 & 3.78  \\
J04262939 & 15.7 & 2.00  & 12.8 & 2.81  & 11.5 & 3.43  \\
J04381486 & 13.8 & 0.63  & 11.5 & 0.86  & 10.4 & 1.05  \\
J04382134 & 14.9 & 1.12  & 12.3 & 1.57  & 11.0 & 1.92  \\
J04390163 & 15.7 & 0.48  & 12.8 & 0.67  & 11.5 & 0.83  \\
J04390396 & 13.8 & 1.13  & 11.5 & 1.56  & 10.4 & 1.90  \\
J04400067 & 15.7 & 3.61  & 12.8 & 5.06  & 11.5 & 6.19  \\
J04414825 & 13.0 & 1.73  & 11.0 & 2.36  & 9.9 & 2.88  \\
J04144730 & 22.4 & 4.04  & 17.3 & 5.94  & 15.2 & 7.30  \\
J04161210 & 20.2 & 1.44  & 15.8 & 2.09  & 13.9 & 2.56  \\
J04181710 & 16.2 & 0.47  & 13.2 & 0.67  & 11.8 & 0.82  \\
J04202555 & 18.6 & 5.18  & 14.8 & 7.45  & 13.1 & 9.13  \\
J04284263 & 18.6 & 0.43  & 14.8 & 0.62  & 13.1 & 0.76  \\
J04322210 & 20.2 & 14.02 & 15.8 & 20.35 & 13.9 & 24.98 \\
J04334465 & 20.2 & 10.14 & 15.8 & 14.72 & 13.9 & 18.07 \\
J04385859 & 21.7 & 6.78  & 16.8 & 9.93  & 14.8 & 12.21 \\
J04393364 & 19.4 & 2.53  & 15.3 & 3.65  & 13.5 & 4.48  \\
J04394488 & 19.4 & 3.01  & 15.3 & 4.35  & 13.5 & 5.33  \\
J04555605 & 22.4 & 0.22  & 17.3 & 0.32  & 15.2 & 0.40  \\
J05075496 & 22.4 & 0.62  & 17.3 & 0.92  & 15.2 & 1.13  \\
\enddata




\end{deluxetable}

%% file: mcfost_stellarparams_new.txt
\begin{table}
\centering
\caption{Stellar parameters used in MCFOST models.}
\label{tab:mcfost_stellar}
\begin{tabular}{ccccc}
\hline
\hline
Target    & T$_\textnormal{eff}$ & log[L$_{*}$]    & R$_{*}$       & Av    \\
          & (K)                  &  (L$_{\odot}$)  & (R$_{\odot}$) & (mag) \\
\hline          
J04141188 & 2963 & -1.746 & 0.873 & 2.5 \\
J04144730 & 3270 & -0.49  & 1.734 & 0.7 \\
J04161210 & 3162 & -1.084 & 1.385 & 2   \\
J04181710 & 3023 & -0.987 & 0.422 & 2.8 \\
J04190110 & 3058 & -0.454 & 0.589 & 1.1 \\
J04202555 & 3091 & -1.343 & 0.487 & 1.6 \\
J04213459 & 3058 & -0.912 & 1.136 & 0.9 \\
J04230607 & 2990 & -1.332 & 0.942 & 1.5 \\
J04262939 & 2990 & -1.655 & 0.377 & 1.6 \\
J04284263 & 3091 & -1.258 & 1.217 & 1.3 \\
J04292165 & 3091 & -0.115 & 1.884 & 0.4 \\
J04322210 & 3162 & -1.134 & 1.385 & 1.4 \\
J04334465 & 3162 & -0.565 & 0.554 & 3.0 \\
J04381486 & 2837 & -2.358 & 0.579 & 1.0 \\
J04382134 & 2935 & -1.507 & 0.794 & 0.6 \\
J04385859 & 3234 & -1.148 & 0.622 & 1.5 \\
J04390163 & 2990 & -1.054 & 1.13  & 0.5 \\
J04390396 & 2837 & -1.336 & 0.552 & 0.5 \\
J04393364 & 3125 & -1.031 & 1.300 & 1.0 \\
J04394488 & 3125 & -0.295 & 2.600 & 0.5 \\
J04400067 & 2990 & -1.547 & 0.377 & 0.5 \\
J04414825 & 2752 & -1.683 & 0.634 & 1.3 \\
J04555605 & 3270 & -0.576 & 1.652 & 0.0 \\
J05075496 & 3270 & -1.095 & 1.652 & 1.2 \\
\hline
\end{tabular}
\end{table}

%% file: mcfost_results.txt
\begin{table}
\centering
\caption{Genetic algorithm results with SED fitting in MCFOST.}
\label{tab:mcfostresults}
\begin{tabular}{ccccccc}
\hline
\hline
Target    & M$_\textnormal{dust}$     & H$_{0}$ & $r_\textnormal{in}$ & $\beta$ & Surf. Dens. & $\chi^{2}$ \\
          & (M$_{\oplus}$) & (au)     & (au)     &         &             &            \\
\hline          
J04144730 & 3.00  & 18.5 & 0.01 & 1.2  & -0.65 & 45  \\
J04161210 & 2.16  & 23.5 & 0.05 & 1.2  & -1.4  & 100 \\
J04181710 & 0.33  & 23.5 & 0.02 & 1.08 & -0.5  & 10  \\
J04190110 & 0.10  & 18   & 1    & 1    & -1.05 & 500 \\
J04202555 & 6.66  & 14.5 & 0.4  & 1.2  & -0.35 & 55  \\
J04213459 & 0.11  & 17.5 & 0.04 & 1.25 & -1.25 & 125 \\
J04230607 & 2.00  & 16   & 0.08 & 1.2  & -0.75 & 15  \\
J04262939 & 4.16  & 18.5 & 0.04 & 1.08 & -0.3  & 13  \\
J04284263 & 0.67  & 18   & 0.02 & 1.08 & -0.75 & 58  \\
J04292165 & 0.67  & 22.5 & 0.05 & 1.08 & -0.4  & 42  \\
J04322210 & 23.31 & 11   & 0.05 & 1.27 & -0.8  & 35  \\
J04334465 & 16.65 & 14   & 0.05 & 1.08 & -0.9  & 16  \\
J04381486 & 2.66  & 25   & 0.03 & 1    & -1.4  & 600 \\
J04382134 & 1.50  & 20   & 0.02 & 1    & -1.2  & 20  \\
J04385859 & 18.31 & 10.5 & 0.08 & 1.09 & -0.4  & 9   \\
J04390163 & 0.33  & 12   & 0.02 & 1.07 & -0.55 & 10  \\
J04390396 & 0.92  & 19.5 & 0.04 & 1.1  & -0.25 & 33  \\
J04393364 & 4.16  & 15   & 0.08 & 1.06 & -0.55 & 17  \\
J04394488 & 1.33  & 20   & 0.08 & 1.16 & -0.5  & 45  \\
J04400067 & 3.33  & 10.5 & 0.11 & 1.23 & -0.8  & 60  \\
J04414825 & 1.66  & 20   & 0.9  & 1.13 & -1    & 35  \\
J04555605 & 0.58  & 18   & 0.3  & 1.3  & -1.4  & 500 \\
J05075496 & 0.75  & 16.5 & 0.07 & 1.14 & -0.4  & 10  \\
\hline
\end{tabular}
\end{table}


%% file: classIIIfrequency.txt
\begin{table*}
\centering
\caption{Disk detection and dust mass frequencies relative to Class II and III populations, with corresponding heavy element masses from the short-period \textit{Kepler} planet statistics, and the fraction of disks reaching minimum-mass solar nebula (MMSN) values.}
\label{tab:frequency}
\begin{tabular}{lcccccccc}
\hline
\hline
 &  Main Sequence Spectral Type:                        & F-stars                  & G-stars  & K-stars   & Early-M   & Mid-M     & Late-M    & Substellar \\
 \hline
 & Mass Range (M$_{\odot}$)             & 1.14--1.59                & 0.9--1.14 & 0.59--0.9 & 0.43--0.59 & 0.245--0.43 & 0.08--0.245 & $\leq$ 0.08   \\
 & Num. Class II Observed                       & 5             & 4        & 26        & 38        & 48        & 45        & 33         \\
 & \%Class II $>$Avg. Heavy Elem. Mass          & 80            & 75       & 69        & 57        & --        & --        & --          \\
 & \%Class II $>$ MMSN                          & 20            & 40       & 19        & 24        &  7        & 1         & 0    \\    
 \hline 
 & Num. Class III                       & 5             & 1        & 17        & 12        & 13        & 37        & 42         \\
 & \% Submm Det. in Class II+III        & 50            & 80       & 58        & 58        & 41        & 24        & 19         \\
\hline
\end{tabular}
\end{table*}

%% file: avgmedminmax_dustmasses.txt
\begin{table}
\caption{Disk dust mass values for various host mass regimes in Taurus, combining this study and \citet{andrews13}. Disk minimum values from upper limits denoted by UL, and \textit{Kepler} estimates from \citet{mulders15}.}
\label{tab:binned_values}
\centering
\begin{tabular}{lccccc}
\hline
\hline
\multicolumn{1}{l}{Object} & \multicolumn{4}{c}{Taurus Class II ($M_{\oplus}$)} &\multicolumn{1}{c}{\textit{Kepler}} \\
\hline            
$M_{*}$ ($M_{\odot}$)  & Min. (UL) & Min. (Det.)  & Max.  & Med. & Avg. \\
\hline
1.14--1.59          & --     & 2.0        & 94.5      & 26.5  & 3.6        \\
0.9--1.14           & --     & 3.0        & 102.6     & 21.3  & 5.0      \\
0.59--0.9           & --     & 1.8        & 303.8     & 13.1  & 5.4       \\
0.43--0.59          & 0.8    & 4.3        & 88.3      & 28.9  & 7.3        \\
0.245--0.43         & 1.4    & 2.3        & 147.8     & 10.4  & --        \\
0.08--0.245         & 0.1    & 0.3        & 107.7     & 10.9  & --         \\
$\leq$ 0.08         & 0.6    & 0.6        & 7.4        & 5.1  & --         \\
\hline
\end{tabular}
\end{table}

%% file: recalc_dustmass_wULs.txt
\startlongtable
\begin{deluxetable*}{ccccccccccc}
\tablecaption{Rederived stellar parameters and dust masses for our sample and comparison samples. \label{tab:recalc_dustmasses}}
\tabletypesize{\scriptsize}
\tablehead{
\colhead{Name} & 
\colhead{Source} & 
\colhead{Region} & 
\colhead{Age}  & 
\colhead{SpTy}  & 
\colhead{Teff} & 
\colhead{Stellar Mass}  & 
\colhead{Log L$_{*}$} & 
\colhead{M$_{dust40}$} & 
\colhead{M$_{dust100}$}  & 
\colhead{M$_{dust200}$}  \\
\colhead{} &        
\colhead{} &     
\colhead{} & 
\colhead{(Myr)}  &
\colhead{} &  
\colhead{(K)} &  
\colhead{(M$_{\odot}$)} & 
\colhead{(L$_{\odot}$)} & 
\colhead{(M$_{\oplus}$)} &  
\colhead{(M$_{\oplus}$)} &  
\colhead{(M$_{\oplus}$)} 
}

\startdata
J04292165                     & (1)        & Taurus    & 1               & M6    & 2858 & 0.06 \{0.05,0.09\}         & -1.57 \{-1.69,-1.24\}     & 2.72                             & 3.82                              & 4.67                              \\
J04141188                     & (1)        & Taurus    & 1               & M6.25 & 2836 & 0.05 \{0.04,0.05\}         & -1.63 \{-1.75,-1.49\}     & 0.41                             & 0.57                              & 0.70                              \\
J04230607                     & (1)        & Taurus    & 1               & M6    & 2858 & 0.06 \{0.05,0.09\}         & -1.57 \{-1.69,-1.24\}     & 2.20                             & 3.09                              & 3.78                              \\
J04262939                     & (1)        & Taurus    & 1               & M6    & 2858 & 0.06 \{0.05,0.09\}         & -1.57 \{-1.69,-1.24\}     & 2.00                             & 2.81                              & 3.43                              \\
J04381486                     & (1)        & Taurus    & 1               & M7.25 & 2747 & 0.04 \{0.03,0.04\}         & -1.88 \{-2.02,-1.75\}     & 0.63                             & 0.86                              & 1.05                              \\
J04382134                     & (1)        & Taurus    & 1               & M6.5  & 2814 & 0.05 \{0.04,0.06\}         & -1.69 \{-1.81,-1.57\}     & 1.12                             & 1.57                              & 1.92                              \\
J04390163                     & (1)        & Taurus    & 1               & M6    & 2858 & 0.06 \{0.05,0.09\}         & -1.57 \{-1.69,-1.24\}     & 0.48                             & 0.67                              & 0.83                              \\
\enddata

\tablecomments{\textbf{References.} (1) This study; (2) \citet{andrews13}; (3) \citet{testi16}; (4) \citet{barenfeld16}; (5) \citet{gvdp16}; (6) \citet{mathews13}. $^\mathrm{UL}$~denotes upper limits calculated from sub-mm/mm non-detection.}
\end{deluxetable*}

\clearpage

%% file: dustmass_updated.txt
\begin{longrotatetable}
\begin{table}
\caption{Dust Masses}
\label{tab:dustmasses}
\centering
\begin{tabular}{lcccccccccccccccc}
\hline
\hline
Target    & \multicolumn{2}{l}{$T_\textnormal{dust}=25(L_{*}/L_{\odot})^{0.25}$} & \multicolumn{2}{c}{R=10 au} & \multicolumn{2}{c}{R=20 au} & \multicolumn{2}{c}{R=40 au} & \multicolumn{2}{c}{R=60 au} & \multicolumn{2}{c}{R=80 au} & \multicolumn{2}{c}{R=100 au} & \multicolumn{2}{c}{R=200 au} \\
          & $T_\textnormal{dust}$(K)            & $M_\textnormal{dust}(\oplus)$            & $T_{d}$(K)        & $M_{d}(\oplus)$       & $T_{d}$(K)        & $M_{d}(\oplus)$       & $T_{d}$(K)        & $M_{d}(\oplus)$       & $T_{d}$(K)        & $M_{d}(\oplus)$       & $T_{d}$(K)        & $M_{d}(\oplus)$       & $T_{d}$(K)        & $M_{d}(\oplus)$        & $T_{d}$(K)        & $M_{d}(\oplus)$        \\
\hline
J04292165 & 10.1 & 5.92  & 25.3 & 1.35 & 18.5 & 2.10  & 15.68 & 2.72  & 14.6 & 3.05  & 14.0 & 3.29  & 12.8 & 3.82  & 11.5 & 4.67  \\
J04141188 & 9.8  & 0.92  & 24.5 & 0.20 & 18.0 & 0.32  & 15.28 & 0.41  & 14.3 & 0.46  & 13.7 & 0.49  & 12.5 & 0.57  & 11.2 & 0.70  \\
J04230607 & 10.1 & 4.78  & 25.3 & 1.09 & 18.5 & 1.69  & 15.68 & 2.20  & 14.6 & 2.47  & 14.0 & 2.66  & 12.8 & 3.09  & 11.5 & 3.78  \\
J04262939 & 10.1 & 4.35  & 25.3 & 0.99 & 18.5 & 1.54  & 15.68 & 2.00  & 14.6 & 2.24  & 14.0 & 2.42  & 12.8 & 2.81  & 11.5 & 3.43  \\
J04381486 & 8.5  & 1.61  & 21.4 & 0.31 & 15.8 & 0.50  & 13.76 & 0.62  & 13.0 & 0.69  & 12.5 & 0.74  & 11.5 & 0.86  & 10.4 & 1.05  \\
J04382134 & 9.5  & 2.61  & 23.7 & 0.56 & 17.4 & 0.88  & 14.90 & 1.13  & 14.0 & 1.26  & 13.4 & 1.35  & 12.3 & 1.57  & 11.0 & 1.92  \\
\hline
\end{tabular}
\end{table}
\end{longrotatetable}
